\documentclass[12pt,a4-paper]{article}

\usepackage{amsmath,amssymb} 
\usepackage{accents}
\usepackage{color}
\usepackage{xcolor}
\usepackage[bookmarks=true,hyperfigures=true]{hyperref}
\usepackage{graphicx}
\usepackage[nosort]{cite}
\usepackage[bulletsep]{collref}
\usepackage{tensor}
\usepackage{shuffle}
\usepackage{lscape}
\usepackage{empheq}

\allowdisplaybreaks

\usepackage[a4paper,text={173mm,216mm},centering]{geometry}

\usepackage[font=small,labelfont=bf,width=0.85\textwidth]{caption}

\let\oldbfseries=\bfseries
\let\oldmdseries=\mdseries
\let\oldnormalfont=\normalfont
\renewcommand{\bfseries}{\oldbfseries\boldmath}
\renewcommand{\mdseries}{\oldmdseries\unboldmath}
\renewcommand{\normalfont}{\oldnormalfont\unboldmath}

\makeatletter
\newlength{\apb@width}
\newcommand{\autoparbox}[2][c]{\settowidth{\apb@width}{#2}\parbox[#1]{\apb@width}{#2}}

\makeatother

\graphicspath{{./Images/}}

\numberwithin{equation}{section}

\makeatletter 
\renewcommand{\maketitle} 
{ \begingroup \begin{center} \large {\bf \@title}
		\vskip 5pt \large \@author \\ \vskip 5pt \@date \end{center}
	\vskip 5pt \endgroup \setcounter{footnote}{0} }
\makeatother 

\newcommand{\pic}[2]{\vcenter{\hbox{\includegraphics[scale=#1]{#2}}}}

\newcommand{\comments}[1]{}

\newcommand{\tl}{\widetilde\lambda}

\newcommand{\ft}[2]{{\textstyle\frac{#1}{#2}}}

\newcommand{\vev}[1]{\langle#1\rangle}
\newcommand{\bev}[1]{ [#1]}
\newcommand{\eqn}[1]{(\ref{#1})}
\newcommand\be{\begin{equation}}
\newcommand\ee{\end{equation}}
\newcommand\bea{\begin{align}}
\newcommand\eea{\end{align}}

\def\beqa{\begin{eqnarray}}
\def\eeqa{\end{eqnarray}}
\def\beq{\begin{equation}}
\def\eeq{\end{equation}}

\def\one{\mbox{1 \kern-.59em {\rm l}}}

\def\cA{{\cal A}}  
  
\def\cG{{\cal G}}  
\def\cJ{{\cal J}}  
  \def\cO{{\cal O}}

\def\uno{\mbox{1 \kern-.59em {\rm l}}}

\def\lan{\langle}
\def\ran{\rangle}

\def\one{1\!\!1\,\,}

\def\bcomment#1{}

\def\eps{\epsilon}

\def\bra(#1,#2){[#1\,#2]}
\def\ket(#1,#2){\langle #1\,#2\rangle}
\def\ketbra(#1,#2,#3){\langle #1\vert #2 \vert #3]}

\usepackage{slashed}
\usepackage{caption}

\usepackage[nosort]{cite}
\usepackage{color}

\usepackage{parskip}

\graphicspath{{./Images/}}

\long\def\symbolfootnote[#1]#2{\begingroup%
	\def\thefootnote{\fnsymbol{footnote}}\footnote[#1]{#2}\endgroup}

\begin{document}

\begingroup\raggedleft\footnotesize\ttfamily
HU-EP-18/04, CERN-TH-2018-048,
QMUL-PH-18-03
\vspace{15mm}
\endgroup

\begin{center}
{\LARGE\bfseries All   rational one-loop Einstein-Yang-Mills amplitudes at four points

\par}%

\vspace{15mm}

\begingroup\scshape\large 
Dhritiman~Nandan${}^{e}$, Jan~Plefka${}^{y}$  and  Gabriele~Travaglini$^{m}$
\endgroup
\vspace{5mm}

\textit{${}^{e}$Higgs Centre for Theoretical Physics, School of Physics and Astronomy \\
					The University of Edinburgh,
					Edinburgh EH9 3JZ, Scotland, United Kingdom } \\[0.25cm]
					
\textit{${}^{y}$Institut f\"ur Physik und IRIS Adlershof, Humboldt-Universit\"at zu Berlin, \phantom{$^\S$}\\
  Zum Gro{\ss}en Windkanal 6, D-12489 Berlin, Germany} \\[0.25cm]
  \textit{${}^{y}$ Theoretical Physics Department, CERN, 1211 Geneva 23, Switzerland}\\[0.25cm]
  
\textit{${}^{m}$Centre for Research in String Theory\\
						School of Physics and Astronomy,
						Queen Mary University of London\\
						Mile End Road, London E1 4NS, United Kingdom } \\[0.1cm]

\bigskip
  
\texttt{\small\{dhritiman.nandan@ed.ac.uk, jan.plefka@physik.hu-berlin.de, g.travaglini@qmul.ac.uk\}} 

\vspace{8mm}


\textbf{Abstract}\vspace{5mm}\par
\begin{minipage}{14.7cm}
All four-point mixed gluon-graviton amplitudes in pure Einstein-Yang-Mills theory 
with at most one state of negative helicity are computed at one-loop order and maximal powers of the gauge coupling, using $D$-dimensional generalized unitarity. The resulting
purely rational expressions take very compact forms. 
We comment on
the color-kinematics duality and a relation to collinear limits of pure gluon amplitudes. 
 \end{minipage}\par

\end{center}
\setcounter{page}{0}
\thispagestyle{empty}
\newpage

\hypersetup{
colorlinks=true,
linktoc=page,
citecolor=Blue,
linkcolor=Blue,
urlcolor=Blue}


	\setcounter{tocdepth}{4}
	\hrule height 0.75pt
	\tableofcontents
	\vspace{0.8cm}
	\hrule height 0.75pt
	\vspace{1cm}
	
	\setcounter{tocdepth}{2}

	\newpage

\section{Introduction and conclusions}\label{sec:Introduction}
	
It is a classic result in the field of scattering amplitudes that supersymmetric Ward identities
force gluon and graviton tree-level amplitudes  to vanish if all particles carry the same helicities or at most 
one state of opposite helicity \cite{Grisaru:1977px},
\begin{equation}
A_{n}(\pm , +,+, \ldots , +)=M_{n}(\pm , +,+, \ldots , +)= 0\, .
\label{intro1}
\end{equation}	
While this result holds at tree level in \emph{any} quantum field theory, in the presence of 
supersymmetry  the vanishing persists to all loops. In non-supersymmetric field theories,
in particular in the ``pure" Yang-Mills and gravity theories, the above amplitudes are very interesting as
they receive their leading  contributions at   one loop and are 
remarkably simple -- resembling tree-level expressions, although with more subtle factorization properties \cite{Bern:2005hs}.   
Their unitarity cuts vanish in four dimensions since the helicity configuration of any two-particle cut 
of the one-loop expressions in \eqn{intro1} implies that there is  at least one vanishing tree-level piece.
 Hence, these one-loop amplitudes are finite rational functions of the momentum invariants.

In the case of pure Yang-Mills theory they were efficiently constructed through their analytic properties
and even the all-multiplicity expression has been established in the all-plus case 
\cite{Bern:1993qk,Mahlon:1993fe}, resulting in a remarkably compact formula
\be
A^{\text{1-loop}}_{n}(1^{+}, \ldots , n^{+}) = \frac{i N_{p}}{96\pi^{2}}\, \sum_{1\leq k_{1}<k_{2}<k_{3}<k_{4} \leq n}
\frac{\vev{k_{1}k_{2}}\bev{k_{2}k_{3}}\vev{k_{3}k_{4}}\bev{k_{4}k_{1}}}{\vev{12}\vev{23}\cdots\vev{n1}}
\,, 
\ee
using spinor helicity variables\footnote{$N_{p}$ is the color weighted number of bosonic minus
fermionic states circling in the loop.}\footnote{See 
\cite{Dixon:1996wi,Elvang:2015rqa,Henn:2014yza} for comprehensive reviews.}.
These one-loop amplitudes are also generated by 
the self-dual Yang-Mills theory and represent their only non-vanishing  amplitudes \cite{Bardeen:1995gk,Chalmers:1996rq, Cangemi:1996rx}. 
The single-minus gluon  amplitudes at one loop are also known for all multiplicities and have been
constructed using Berends-Giele type \cite{Mahlon:1993si}, as well as BCFW-type recursion relations \cite{Bern:2005hs}. Their form is considerably more involved.

All-plus and single-minus helicity amplitudes have also been constructed in pure gravity. A conjecture for the all-plus graviton amplitude at any multiplicity exists \cite{Bern:1998sv} and agrees with explicit constructions at $n\!\leq\!7$ points. Again, this infinite series of graviton 
amplitudes is identical to one-loop self-dual gravity. For the single-minus amplitudes,
an explicit, yet not very compact expression has been recently derived \cite{Alston:2015gea}
using a spin-off of the BCFW method known as augmented recursion \cite{Dunbar:2010xk}, following earlier work in \cite{Bern:1993wt,Dunbar:1994bn,Brandhuber:2007up}. As is often the case, the analytic structure, in particular  consistency of soft and collinear limits,  helped to constrain the ansatz.

In this work we focus on explicit $S$-matrix elements for mixed graviton and gluon scattering in 
Einstein gravity minimally coupled to Yang-Mills theory, or EYM for short. In the 1990s EYM amplitudes in four dimensions for the maximally-helicity violating (MHV) case, i.e.~two negative-helicity states, were given at tree level
in \cite{Selivanov:1997aq,Bern:1999bx}. Only rather recently modern approaches to scattering
amplitudes based on the scattering equation formalism of CHY \cite{Cachazo:2013hca,Cachazo:2014nsa}, or the  color-kinematic duality relations \cite{Bern:2008qj,Bern:2010ue}, were applied to the realm of EYM amplitudes, leading to a number of explicit results. Double-copy constructions for gluon-graviton
scattering in supergravity theories were given in \cite{Chiodaroli:2014xia,Chiodaroli:2015rdg,Chiodaroli:2016jqw}. However, the most efficient way of establishing EYM amplitudes 
is by expanding them in a basis of pure gluon amplitudes multiplied by
 kinematic numerators to be determined  (also featuring in color-kinematic duality):
\be
A^{\text{tree}}_{\text{EYM}}(1,2,\ldots, n;h_{1},\ldots, h_{m}) = \sum_{\beta\in \text{Perm}(2,\ldots,n-1;h_{1},\ldots,h_{m})}n(1,\{\beta\}, n)\, A^{\text{tree}}_{\text{YM}}(1,\{\beta\}, n)
\, .
\ee
This form was initially presented by a string-based construction for one graviton and $n$-gluon scattering in \cite{Stieberger:2016lng}, the field theory proof followed shortly thereafter
\cite{Nandan:2016pya,delaCruz:2016gnm} and was further lifted to the sector of three gravitons in
\cite{Nandan:2016pya} employing the CHY formalism. A color-kinematic duality based construction extended this to
amplitudes involving up to five gravitons \cite{Chiodaroli:2017ngp}.
The complete recursive solution for the numerators $n(1,\{\beta\}, n)$ has recently been constructed in the single-trace sector in \cite{Teng:2017tbo} and for multi-traces in \cite{Du:2017gnh}. This, together with the
existing result for all tree-level color-ordered gluon amplitudes \cite{Drummond:2008cr,Dixon:2010ik,Bourjaily:2010wh}, constitutes the complete solution for the EYM $S$-matrix at tree level.

This state of affairs sets the stage for the investigation of the present paper. Here we compute the remaining
rational amplitudes of the EYM theory at the leading one-loop level at multiplicity four. These are the three all-plus helicity
amplitudes involving one, two or three gravitons, as well as the six single-minus amplitudes involving
one, two or three gravitons. 
An elegant way to determine such amplitudes consists in  employing  two-particle unitarity cuts in $D = 4 - 2 \epsilon$ dimensions  
\cite{Bern:1995db} (see also \cite{Brandhuber:2005jw} for the first uses of $D$-dimensional generalized unitarity). 
The main idea   is that a rational term in four dimensions,  $\cal{R}$, will in $D$ dimensions acquire a discontinuity, but to a higher order in the dimensional regularization parameter $\epsilon$. Schematically, 
\beq
\mathcal{R} \ \to \ \mathcal{R}  (- s) ^{ - \eps} \, = \, \mathcal{R} \, \big[1 - \eps \log ( -s)\big] + \cdots \ .  
\eeq
Technically, the calculation is greatly simplified by using 
the general supersymmetric Ward identity of \eqn{intro1} at the one-loop order,  which implies that
the contribution of an arbitrary state in the loop is proportional to that from   a scalar circulating in the
loop, 
\be
A_{n+m}^{\text{any state in loop}}(1,2,\ldots,n;h_{1},\ldots,h_{m}) = N_{p}\,  A_{n+m}^{\text{scalar in loop}}(1,2,\ldots,n;h_{1},\ldots,h_{m}) \, .
\label{keyeqn}
\ee
\begin{figure}[t]
\centering
\includegraphics[width=0.7\linewidth]{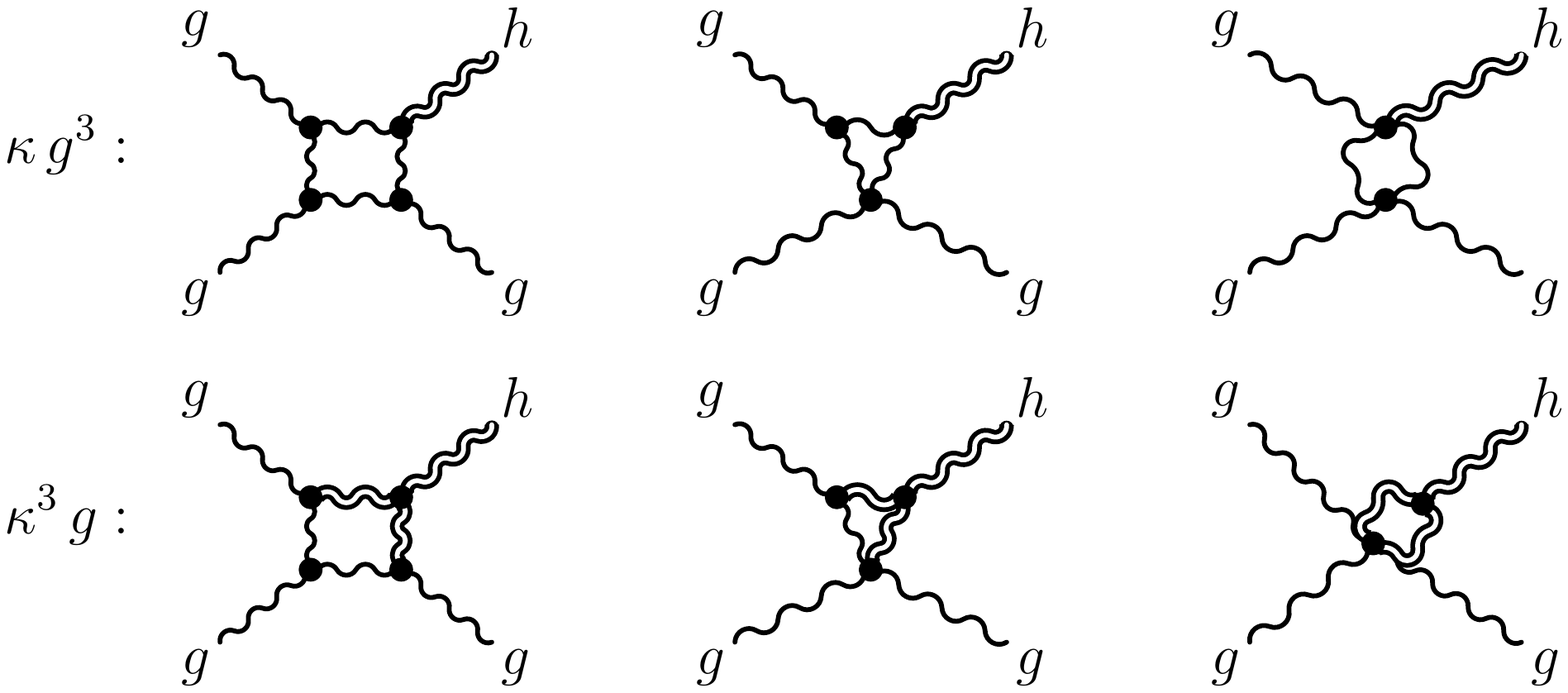} 
\caption{\it Contributions to the one-loop  EYM amplitude 
$A_{3+1}(1,2,3;h)$ at different orders in $\kappa$ and $g$. Only the top line of the
$\cO(\kappa\, g^{3})$
contribution of the amplitude is constructible by replacing the gluon by a (massive)  scalar from \eqn{keyeqn} in the rational case.}
\label{fig1}
\end{figure}
It is important to realize that ``any state in loop'' refers to a ``pure'' contribution
of a definite quantum field excitation (e.g.~graviton or gluon) propagating in the loop.
This relation may therefore be straightforwardly applied to the EYM situation of a gluon circulating inside the loop of a mixed gluon-graviton amplitude, see Figure~\ref{fig1} for a four-point example: A one-loop single-graviton three-gluon
amplitude will have one-loop contributions of order $\kappa g^{3}$ and $\kappa^{3} g$. 
A generic one-loop $m$-graviton and $n$-gluon amplitude will have $g$-leading contributions of
order $g^{n}\kappa^{m}$ representing only gluons in the loop, whereas the $g$-subleading
contributions $g^{n-2k}\kappa^{m+2k}$ reflect contributions where $2k$ gluon propagators
are turned into graviton propagators. Note that there is no single-gluon $l$-graviton vertex.

For the contributions to the amplitude
maximizing the powers of the gauge coupling constant, i.e.~the contributions to
$A_{n+m}(1,2,\ldots,n;h_{1},\ldots,h_{m})$ at order $g^{n}\kappa^{m}$,  we only have gluons 
running in the loop, and the relation \eqn{keyeqn} applies with $N_{p}=1$, i.e.~this contribution may
be computed upon replacing the gluon inside the loop by a scalar. 
The cuts are performed in $D$ dimensions, where a generic loop momentum $L$ satisfies 
$
L^2\!=\!0=\!l^2_{(- 2 \eps)}\!-\!l^2_{(4)}\!=\!0
$, 
where $l_{(- 2 \eps)}$ and $ l_{(4)}$ represent the $(-2\epsilon)$- and four-dimensional part of $L$. 
Because the external kinematics is four-dimensional, at one loop there is just one $l_{(- 2 \eps)}$. Setting 
$l^2_{(- 2 \eps)}\!:=\!\mu^2$, one then has $l^2_{(4)}\!=\!\mu^2$, i.e.~all  internal $D$-dimensional scalar can effectively be treated as  four-dimensional massive scalar with uniform mass $\mu^2$, over which one 
integrates at the end \cite{Bern:1995db}.

The ``non-pure'' 
contributions  of order $g^{n-2k}\kappa^{m+2k}$, however, have a mixture of gluons and
gravitons running inside the loop. Here the situation is less clear, as 
 \eqn{keyeqn} does not hold.  A simple dimensional analysis also reveals that
the mixed graviton-gluon contributions in the loop are not represented by \eqn{keyeqn}.

Hence in this work we only aim at finding the maximal $g$ contributions to the
one-loop rational amplitudes in EYM theory.
Here we find intriguingly simple results,
to wit\footnote{In our conventions we have $s=\vev{12}\bev{21}$, $t=\vev{23}\bev{32}$ and $u=\vev{13}\bev{31}$.} 
\begin{align}
 A^{(1)}(1^{+},2^{+},3^{+}; 4^{++})\Bigr |_{\kappa g^{3}} &= 0\,  ,  \nonumber \\
 A^{(1)}(1^{+},2^{+},3^{+}; 4^{--})\Bigr |_{\kappa g^{3}} &= \,  - {i \over (4 \pi)^2}\,
 \   \frac{[12][34]}{\lan 12 \ran \lan 34\ran} \,  {(\lan42\ran [23]\lan 34\ran)^3}\,  
{s^2 + t^2 +   u^2  \over  \, 6\,  s^2\, t^2\, u^2}
 ,  \nonumber  \\
 A^{(1)}(1^{-},2^{+},3^{+};4^{++})\Bigr |_{\kappa g^{3}} &= {i \over (4 \pi)^2} \,
    \frac{[24][34]}{\lan 24\ran \lan 34\ran}\,  
   {1\over\lan23\ran [21][31]}\,  {1\over 6} \, (s^2 + u^2)\, ,  \\
   A^{(1)}(1^+, 2^+; 3^{++}, 4^{++}) \Bigr |_{\kappa^{2} g^{2}} &= \   {i \over (4 \pi)^2} \,  {[12]\over \langle 12\rangle}\,  {[34]^2\over \langle 34\rangle^2}\,  {s\over 6}\, , 
    \nonumber  \\
   A^{(1)}(1^-, 2^+; 3^{++}, 4^{++})\Bigr |_{\kappa^{2} g^{2}}  &= \   {i \over (4 \pi)^2} \,
   \frac{[24]^2[34]^2\langle 14\rangle^2}{\langle 34\rangle^2}\,  
{s\over 6\,t\, u}\,  ,  \nonumber \\
   A^{(1)}(1^+, 2^+; 3^{++}, 4^{--})\Bigr |_{\kappa^{2} g^{2}}  &= \   {i \over (4 \pi)^2} \,
   \frac{\bra(1,2)\bra(1,3)^4\ket(1,4)^4}{\ket(1,2)}\,  
{t^2+u^2 \over 6\,s \,  t^2\, u^2}\,  ,  \nonumber \\
 A^{(1)}( 1^{\pm};2^{++},3^{++},4^{++})\Bigr |_{\kappa^{3} g} &= 0\,  ,   \nonumber \\
 A^{(1)}( 1^{+};2^{++},3^{++},4^{--})\Bigr |_{\kappa^{3} g} &= 0\,   . \nonumber 
\end{align}
It would be interesting to also construct the missing ``non-pure'' pieces at higher orders
in $\kappa$ as well, even though they will be numerically subleading at energies well
below the Planck mass. This should be possible using the double-copy techniques initiated in
\cite{Chiodaroli:2017ngp}.

The rest of our paper is organized as follows. In the next section we collect all relevant tree-level amplitudes involving gluons, gravitons and massive scalars entering the cuts needed to compute the rational amplitudes we are interested in. Sections \ref{allplus}--\ref{sec:5} are devoted to the calculation of all one-loop amplitudes with one graviton and three gluons. A particularly interesting case is that of Section \ref{allplus}, where we find that the all-plus amplitude $\langle 1^+ 2^+ 3^+ 4^{++}\rangle$, although non-vanishing in $D$ dimensions, actually vanishes in the four-dimensional limit.  Sections \ref{sec:6}--\ref{sec:8} discuss the derivation of the  amplitudes with two gravitons and two gluons, while 
Sections \ref{sec:3plusG1pmg}--\ref{sec:10} contain the (vanishing) amplitudes with three gravitons and one gluon. Finally in Section \ref{sec:11} we rederive the curiously vanishing
single-graviton all-plus amplitude from a double-copy construction.
Two appendices complete the paper. In Appendix \ref{Integrals}  we list the $D$-dimensional expressions of the relevant integrals and the appropriate limits contributing to the amplitudes of interest, while in Appendix \ref{recrel} we derive all the four-point  tree-level amplitudes with two massive scalars and gluons/gravitons using  recursion relations. 

\section{Relevant tree-level amplitudes}
\label{relevanttrees}

In this section we collect all  the tree-level amplitudes entering our calculation. The basic building blocks are the three-point amplitudes involving a gluon or graviton and two massive scalars. The color-ordered gluon-scalar-scalar amplitudes are \cite{Badger:2005zh}
\beq 
\label{gssb}
A(1^+, 2_{\phi}, 3_{\bar\phi}) \ = \ 
 i\,  {\lan q|3|1] \over \lan q1\ran} \ , \qquad 
A(1^-, 2_{\phi}, 3_{\bar\phi}) \ = \
i\,  {\lan 1|3|q] \over [ 1q]} \ , 
\eeq
where $p_2^2 = p_3^2 = \mu^2$, and   $\mu$ is the mass of the scalar particles. In these formulae, $\lambda_q$ and $\tl_q$ are reference spinors, and the amplitudes themselves are independent of their choice. 
The amplitudes involving a graviton are similarly given by the square of the previous amplitudes%
\footnote{We have confirmed this calculation also from Feynman rules, for which  a good
source  is \cite{TheoThesis}.
} 
\beq 
\label{hssb}
A(1^{++}; 2_{\phi}, 3_{\bar\phi}) \ = \ i\, \big[A(1^+, 2_{\phi}, 3_{\bar\phi})\big]^2 \ , \qquad 
A(1^{--};  2_{\phi}, 3_{\bar\phi}) \ = \ i\,\big[A(1^-,  2_{\phi}, 3_{\bar\phi})\big]^2 \ .
\eeq
 We will also need four-point amplitudes involving two gluons/gravitons and two scalars. The amplitudes involving gluons have been derived in \cite{Badger:2005zh} 
using BCFW recursion relations \cite{Britto:2004ap, Britto:2005fq} applied to massive scalars, and  
the relevant amplitudes with  gravitons can be obtained similarly (see Appendix 
\ref{recrel} for details). We quote here 
the  expression of the relevant  Yang-Mills amplitudes with two gluons and two scalars:
\beqa
\label{1p2p}
A(1^+, 2^+, 3_{\phi}, 4_{\bar\phi}) & = & \mu^2 {[12]\over \langle 12\rangle} {i\over (p_4 + p_1)^2 - \mu^2} \, , 
\\
\label{1m2p}
A(1^-, 2^+, 3_{\phi}, 4_{\bar\phi}) &= & 
{
\lan 1 |4|2]^2 \over s_{12} 
} {i \over \big[ (p_4 + p_1)^2 - \mu^2\big]}\, , 
\eeqa
while for the amplitudes involving a graviton, a gluon and two scalars we have:%
\footnote{The derivation of \eqref{4pp1p}, \eqref{4pp1m}  and \eqref{4g1gpp} is presented in Appendix \ref{recrel}.}
\beqa
\label{4pp1p}
A(1^+, 2_{\phi}, 3_{\bar\phi};4^{++}) &=& -   \mu^2 {[14]\over \langle 14\rangle^2}   {\lan 1 | 3 | 4]} \, 
\Big[ {i\over (p_3 + p_4)^2 - \mu^2} + {i\over (p_2 + p_4)^2 - \mu^2} \Big]\,\ , 
\\
\label{4pp1m}
A( 1^-, 2_{\phi}, 3_{\bar\phi}; 4^{++}) &=&  \, - {\lan1 | 3 |4]^3 \over s_{14} \, \lan 14\ran } \, 
\Big[ 
{i\over   (p_3 + p_4)^2 - \mu^2}  +{i\over   (p_2 + p_4)^2 - \mu^2} \Big]\,, 
\\
\label{4mm1p}
A( 1^{+}, 2_{\phi}, 3_{\bar\phi}; 4^{--}) &=& -  {\lan4 | 3 |1]^3 \over s_{14}\, [41]  } \, 
\Big[ 
{i\over   (p_3 + p_4)^2 - \mu^2}  +{i\over   (p_2 + p_4)^2 - \mu^2} \Big]\,. 
\eeqa
We have also double-checked these amplitudes through a direct Feynman diagrammatic calculation.
The two-graviton/two-scalar amplitudes in turn read
\beqa
\label{4g1gpp}
A(2_{\phi}, 3_{\bar\phi};4^{++}, 1^{++} ) &=& \,- \mu^4 \,  {[41]^{2}\over \lan 41\ran^2 } \, 
\Big[ 
{i\over   (p_3 + p_4)^2 - \mu^2}  +{i\over   (p_2 + p_4)^2 - \mu^2} \Big] \, , \\
A(2_{\phi}, 3_{\bar\phi};4^{++}, 1^{--} ) &=& \, - {\lan 1 | 3 |4]^4 \over s_{14}^2 } \, 
\Big[ 
{i\over   (p_3 + p_4)^2 - \mu^2}  +{i\over   (p_2 + p_4)^2 - \mu^2} \Big] 
\ .
\label{4g1gpm}
\eeqa
Note that \eqref{1p2p},  \eqref{4pp1p} and  \eqref{4g1gpp} manifestly vanish if the scalars are massless. \

For later convenience we shall  split up \eqref{4pp1p}--\eqref{4g1gpm} into a sum of  two
partial amplitudes which treat the single graviton effectively as if it were color ordered, in the sense that
\be
\label{1g1h}
A(1^\pm, 2_{\phi}, 3_{\bar\phi};4^{++}) := 
\cA(4^{++}, 1^\pm, 2_{\phi}, 3_{\bar\phi})  + \cA(1^\pm,4^{++}, 2_{\phi}, 3_{\bar\phi}) \ ,
\ee
with 
\begin{align}
\cA(4^{++}, 1^+, 2_{\phi}, 3_{\bar\phi}) &=   \mu^2 {[41]\over \langle 41\rangle^2}   {\lan 1 | 3 | 4]} \, 
{i\over (p_3 + p_4)^2 - \mu^2} \, , 
\\
\cA(1^+, 4^{++},2_{\phi}, 3_{\bar\phi}) &=    \mu^2 {[41]\over \langle 41\rangle^2}   {\lan 1 | 3 | 4]} \, 
{i\over (p_3 + p_1)^2 - \mu^2} \,  , \\
\cA(4^{++}, 1^-, 2_{\phi}, 3_{\bar\phi}) &=   {\lan1 | 2 |4]^3 \over \lan 14\ran \, s_{14} }\, 
{i\over (p_3 + p_4)^2 - \mu^2}\, , \\
\cA(1^-, 4^{++},2_{\phi}, 3_{\bar\phi}) &=   {\lan1 | 2 |4]^3 \over \lan 14\ran \, s_{14} } \, 
{i\over (p_3 + p_1)^2 - \mu^2} \, ,
\end{align}
and similarly for the other amplitudes. 
In the unitarity-based construction of the one-loop amplitudes to be discussed, we then
symmetrize explicitly in the graviton leg(s) attached.
	
\section{The {\normalfont $\langle 1^+\, 2^+\, 3^+\, 4^{++} \rangle$} amplitude} 
\label{allplus}

We begin our investigation with the four-point same-helicity amplitude with one graviton and three gluons. We will derive the integrand of this amplitude, as well as its  four-dimensional limit. We anticipate the interesting outcome  of this  computation, namely that this amplitude is zero in the four-dimensional limit -- a result that we will also confirm from the double-copy perspective in Section \ref{sec:11}.%
\footnote{We thank Henrik Johansson and Radu Roiban for confirming the vanishing of this amplitude in four dimensions  from  the double-copy approach implemented in   \cite{Chiodaroli:2017ngp}.}

\begin{figure}[t]
\centering
\includegraphics[width=0.7\linewidth]{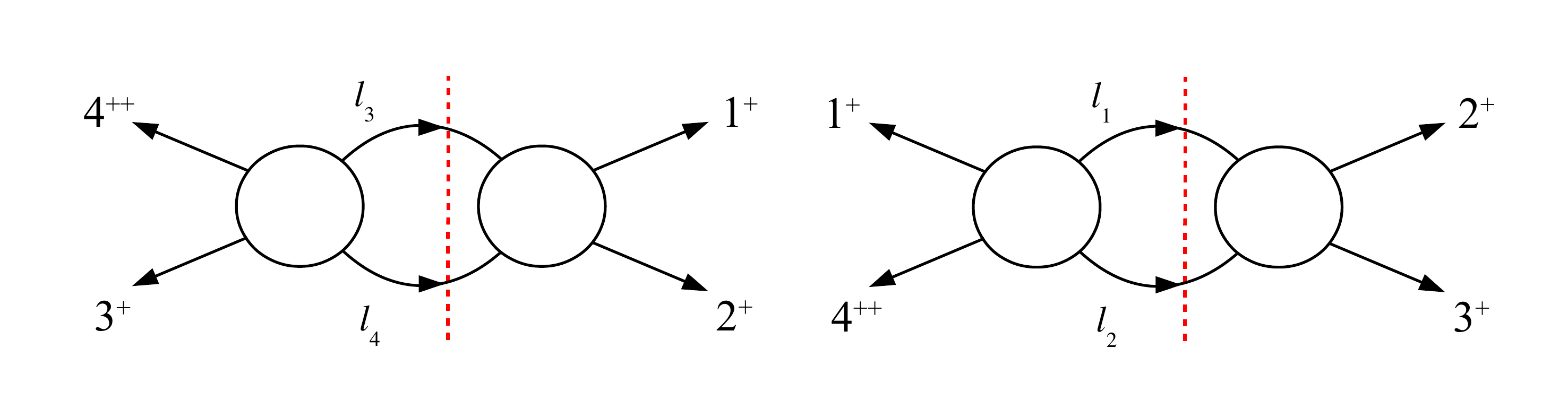} 
\caption{\it The $s$- and $t$-channel cuts of the all-plus single-graviton amplitude. Cyclic permutations of the labels $(1,2,3)$ should also be added. }
\label{fig2.1}
\end{figure}

To organize the computation efficiently, we employ the effective ``color"-ordered graviton partial amplitudes
introduced in the previous section. The diagrams to be considered are shown in 
Figure \ref{fig2.1}. As all gluons carry the same helicity, we need only to evaluate the
first diagram in Figure \ref{fig2.1}; 
 the final result will then be obtained by adding the terms obtained by cycling $(1,2,3)$ in the  partial result. 

For the configuration $(1234)$ of Figure \ref{fig2.1} there are two two-particle cuts, in the $s_{12}=s$ and $s_{23}=t$ channels.  
We start with the $t$-channel cut which is given by the product of the two partial amplitudes: 
\beqa
\label{cut1}
\left. A^{(1)} ( 4^{++}
; 1^+, 2^+, 3^+)\right|_{(1234), t}  &=& \cA(4^{++}, 1^+, l_{1,\phi}, l_{2, \bar\phi})  A( 2^+,  3^+, - l_{2, \phi}, - l_{1, \bar\phi}) 
\\ \nonumber 
&=& 2 \mu^4 { [41] [23] \over \langle 41\rangle \langle 23 \rangle } { \langle 1 | l_2 | 4] \over \langle 14\rangle} {i\over (l_2 + p_4)^2 - \mu^2} { i \over (l_1 - p_2)^2 - \mu^2}\ , 
\eeqa
where the explicit expressions of the tree-level amplitudes entering the cut are given in \eqref{1p2p} and \eqref{4pp1p}, and the factor of two arises from summing of the possible assignment ($\phi$, $\bar\phi$ and  $\bar\phi$, $\phi$) for the internal scalar particles. 

For the $s$-channel cut of the (1234)-configuration, one similarly arrives at  an integrand
\beqa
\label{cutt1}
\left. A^{(1)} (4^{++}
; 1^+, 2^+,3^+)\right|_{(1234), s}  &=& \cA(3^{+},4^{++}, l_{3,\phi}, l_{4, \bar\phi})  A(1^{+}, 2^+,  - l_{4, \phi}, - l_{3, \bar\phi}) 
\\ \nonumber 
&=& 2 \mu^4 { [43] [12] \over \langle 43\rangle \langle 12 \rangle } { \langle 3 | l_4 | 4] \over \langle 12\rangle} {i\over (l_4 + p_3)^2 - \mu^2} { i \over (l_3 - p_1)^2 - \mu^2}\ .
\eeqa
The strategy to find the integrand is now to rewrite the $t$-channel expression in such a way as to
reproduce the $s$-channel expression modulo terms that vanish on the $s$-cut.
For this we first introduce a uniform parametrization of the $(1234)$ box diagram in terms
of a single loop momentum $l$:
\be
l_{1}=l-p_{1}\, , \qquad l_{2}=-l-p_{4}\, , \qquad l_{3}=l\, , \qquad  l_{4}=p_{1}+p_{2}-l\, , 
\ee
with 
\be
D_{i}=(l-q_{i})^{2}-\mu^{2}\, , \qquad q_{0}=0,\, \quad q_{1}=p_{1}\, , \quad q_{2}=p_{1}+p_{2}\, , \quad q_{3}=-p_{4}\, .
\ee
Using these the, $s$- and $t$-channel cuts take the compact forms
\begin{align}
A^{(1)} ( 4^{++}; 1^+, 2^+, 3^+)\Bigr|_{(1234), t} &= 2i\mu^{4}\frac{[12][34]}{\vev{12}\vev{34}}
\, \frac{\langle 1|l|4]}{\vev{14}}\, \frac{\big[ (2\pi) \delta(D_{1})\big]\, \big[(2\pi) \delta(D_{3})\big]}{D_{0}\, D_{2}}\, , \\
A^{(1)} ( 4^{++}; 1^+, 2^+, 3^+)\Bigr|_{(1234), s} &= 2i\mu^{4}\frac{[12][34]}{\vev{12}\vev{34}}
\, \frac{\langle 3|l|4]}{\vev{34}}\, \frac{\big[ (2\pi) \delta(D_{0})\big]\, \big[ (2\pi) \delta(D_{2})\big]}{D_{1}\, D_{3}}\, , 
\label{allplust}
\end{align}
where we have explicitly indicated the cut propagators. 
From this it is obvious that we need to relate $\langle 1 |l |4]$ to $\langle 3|l|4]$.
The trick to do this is to exploit the identity
\be
\langle 3 | l |4] = \frac{[12]}{[23]}\, \langle 1 | l |4] + \frac{[24]}{[32]} s_{l4}\, , 
\ee
where $s_{l4}=\lan 4|l|4]=2\,( l\cdot p_{4})$, which in turn may be written as
\be
s_{l4}=(l+p_{4})^{2}-\mu^{2} - (l^{2}-\mu^{2}) = D_{3}-D_{0} \mathrel{\hat{=}} D_{3}\Bigr |_{\text{on $s$-cut}} \, .
\ee
We also note  the identity
\be
\frac{\langle 3 | l |4]}{\vev{34}} = \frac{\langle 1 | l|4]}{\vev{14}} + \frac{[24]}{[32]\vev{34}}\, (D_{3}-D_{0})\, .
\ee
Inserting this into the $s$-cut expression \eqn{allplust} and dropping the $D_{0}$ term gives us
an integrand which may be lifted off the cuts (with the usual replacement $(2\pi) \delta (D) \to i / D$ for the cut propagators):%
\footnote{The factor of $-1$ in the following expression arises from reinstating two  (uncut) propagators.}

\be
\left.A^{(1)} ( 4^{++}; 1^+, 2^+, 3^+)\right|_{1234}= 
-2i\frac{[12][34]}{\vev{12}\vev{34}} \int\!\frac{d^{4}l}{(2\pi)^{4}}\, \frac{d^{-2\eps}\mu}{(2\pi)^{-2\eps}}\,
\frac{\mu^{4}}{D_{0}D_{1}D_{2}D_{3}}\,
\left[  \frac{\langle 1|l|4]}{\vev{14}} +\frac{[24]}{[32]\vev{34}}\, D_{3}\right]\, .
\ee
The partial one-loop amplitude is thus given by a linear box integral and a scalar triangle.

The final step is to now reduce the linear box integral.
Here we use the {\tt Mathematica} package {\tt FeynCalc} \cite{Mertig:1990an,Shtabovenko:2016sxi}, which efficiently implements the Passarino-Veltman reduction algorithm \cite{Passarino:1978jh}.
Doing this we  arrive at the final result%
\footnote{The integral functions appearing in \eqref{FinalAllPlus} and in the rest of the paper are defined in Appendix \ref{Integrals}, following the conventions of  \cite{Bern:1995db}
up to a minus sign for the $I_{3}$ integrals.}
\begin{align}
A^{(1)}(1^{+},2^{+},3^{+}; 4^{++}) =
 \frac{2}{(4\pi)^{2-\epsilon}}\,
{[41][42][43]\over \langle 41\rangle \langle 23\rangle} \frac{t}{u}
 \, 
  \Big[
\frac{1}{2} I_4[\mu^{4}; s,t] + \frac{1}{t}I_3[\mu^{4}; t] + \frac{1}{s}I_3[\mu^{4}; s]
\Big] \ + \ \mathrm{perms}\, ,
   \label{FinalAllPlus}
\end{align}
where by ``perms'' we indicate  the  two permutations $(2314)$ and $(3124)$ of (1234), which interchange the Mandelstam invariants as     $(s,t,u)\to (t,u,s)$	 and $(s,t,u)\to (u,s,t)$, 
respectively. 
However, we need not do this explicitly as
taking the four-dimensional limit using the relations  in \eqref{hdint} 
 we get a vanishing result:
\be
\label{zero!}
 A^{(1)}(1^{+},2^{+},3^{+}; 4^{++}) = 0\, .
\ee
It would be  desirable to understand the deeper reason for this curious vanishing.

 We also quote an alternative expression of the $D$-dimensional amplitude, which  is given by:
\beq
\label{allplus-bis}
A^{(1)}(1^{+},2^{+},3^{+}; 4^{++}) = {2\over (4 \pi)^{2-\eps}} {[12]\, [34]\over \langle 12\rangle \, \langle 34\rangle}{1\over \langle 41\rangle[12]\langle 42\rangle}
\Big[ {s\,  t\over 2}  I_4[\mu^4; s,t] \, - \, s I_3[\mu^4;s]  \, +\, {\rm perms}\Big]\ , 
\eeq
where the two permutations are the same as in \eqref{FinalAllPlus}.  The vanishing of \eqref{allplus-bis} is of course obtained again upon using the formulae of Appendix \ref{Integrals}.
We also comment that this  integrand is  manifestly  odd  under the  exchange of any two same-helicity gluons. In color space this means that this amplitude is proportional to $f^{a_1 a_2 a_3}$, with no $d^{a_1 a_2 a_3}$ contribution. We will see that the same property is shared by all amplitudes involving three gluons computed in this paper -- they only come with an $f^{a_1 a_2 a_3}$ color factor.


\section{The  {\normalfont$\langle 1^-\, 2^+ \,3^+ \,4^{++}\rangle$} amplitude} 
Constructing this amplitude is a  slightly harder task, hence as an introduction we will first re-derive the four-point gluon amplitude with a single negative-helicity gluon of \cite{Bern:1995db}  and then apply a similar procedure to the more complicated EYM case. The form of the four-gluon integrand is also of use for a
double-copy based construction of the EYM amplitudes.

\begin{figure}[t]
\centering
\includegraphics[width=0.7\linewidth]{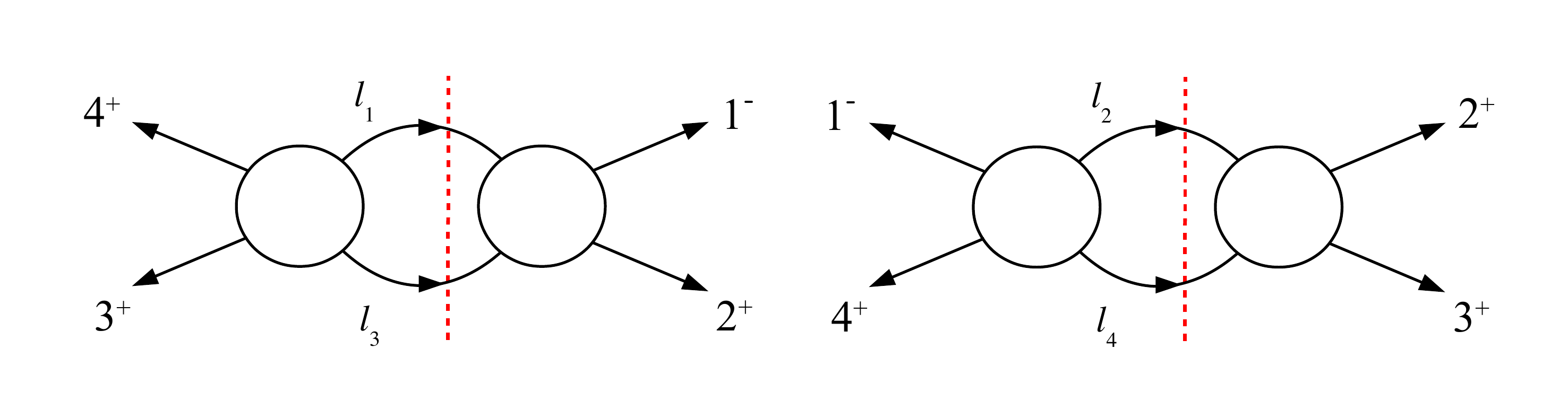}
\caption{\it The $s$- and $t$-channel cuts of the $A^{(1)}(1^-,2^+,3^+,4^+)$  
amplitude in pure Yang-Mills. }
\label{fig4.1}
\end{figure}

\noindent{\bf Warmup.}
As for the case of the all-plus amplitude derived in the previous section, we work with  two-particle cuts. Because only gluons are involved, color ordering leaves us with only  two channels to consider, see Figure \ref{fig4.1}. For the $s$-channel  
we  have
\begin{align}
A_{4}^{(1)}(1^-,2^+,3^+,4^+)\Bigr |_{s} &= 
A(3^{+},4^{+},l_{1,\phi},l_{3,\bar\phi})\,A(1^{-},2^{+},-l_{3,\phi},-l_{1,\bar\phi})\nonumber\\
& = 
\mu^{2}\frac{[34]}{\vev{34}\, (l_{4}^{2}-\mu^{2})}\,
\, {{-\langle 1 | l_{1}|2]^{2}}\over{\vev{12}\bev{21}\, (l_{2}^{2}-\mu^{2})}}  \ , 
\label{AYM2t}
\end{align}
whereas the $t$-channel cut reads 
\begin{align}
A_{4}^{(1)}(1^-,2^+,3^+,4^+)\Bigr |_{t} &= 
A(4^{+},1^{-},l_{2,\phi},l_{4,\bar\phi})\,A(2^{+},3^{+},-l_{4,\phi},-l_{2,\bar\phi})\nonumber\\
& = - \frac{\langle 1 | l_{4}|4]^{2}}{\vev{41}\bev{14}\, (l_{1}^{2}-\mu^{2})}\, 
\mu^{2}\frac{[23]}{\vev{23}\, (l_{3}^{2}-\mu^{2})}\, . 
\label{AYM2s}
\end{align}
\noindent
The strategy to find the integrand is now to rewrite the $t$-channel expression in such a way to
reproduce the $s$-channel one modulo terms that vanish on the $s$-cut. For this, we will make use of the following identity 
to rewrite the numerator in  \eqref{AYM2s}: 
\be
\langle 1 | l_{1} |4] = \frac{1}{\vev{34}}\, \Bigl
[ \vev{13}\, s_{l_{1}1} + \langle 1 | l_{1} |2]\, \vev{23}\, \Bigr ] \, , 
\label{1l14id}
\ee
where $s_{l_{1}1}=\lan 1|l_{1}|1]=2\, l_{1}\cdot p_{1}$, which in turn may be written as
\be
s_{l1}=(l_{1}^{2}-\mu^{2}) - (l_{2}^{2}-\mu^{2}) \mathrel{\hat{=}} (l_{1}^{2}-\mu^{2})\Bigr |_{\text{on $t$-cut}} \, .
\ee
This last expression holds on the $t$-channel cut. Inserting the expression \eqn{1l14id} for
$\lan 1|l_{1}|4]$ into the $t$-channel cut amplitude $A_{4}|_{t}$ of \eqn{AYM2s} then yields an expression
which may straightforwardly be lifted off the cut. Thus we get an integrand%
\footnote{Again, the minus sign in front of the following expression arises from two cut propagators.}
\begin{align}
A^{(1)}(1^{-},2^{+},3^{+},4^{+}) 
&  
= -\int\!\frac{d^{4}l}{(2\pi)^{4}}\, \frac{d^{-2\eps}\mu}{(2\pi)^{-2\eps}}\,
\left (-\frac{\mu^{2}}{\vev{34}^{2}}\right ) \, \Bigl [
\lan 1|l|2]^{2}  \nonumber \\ & 
+2 \frac{\vev{13}}{\vev{23}}\, D_{0}\, \lan 1|l|2]
+\frac{\vev{13}^{2}}{\vev{23}^{2}}\, D_{0}\, s_{l1}\Bigr ]{1\over D_{0} D_{1} D_{2} D_{3}} \, , \quad 
\label{YMoffcuts}
\end{align}
where we have chosen the loop momentum parametrization as $l=l_{1}$, and
\be
D_{0}=l^{2}-\mu^{2}\, , \quad 
D_{1}=(l-p_{1})^{2}-\mu^{2}\, , \quad 
D_{2}=(l-p_{1}-p_{2})^{2}-\mu^{2}\, , \quad 
D_{3}=(l+p_{4})^{2}-\mu^{2}\, .
\ee
Note that there is an ambiguity in treating the last term in \eqn{YMoffcuts}.
By the logic laid out above we could have also replaced $s_{l_{1}1}$ by $D_{0}$ as
the resulting expression would agree with \eqn{AYM2s} and \eqn{AYM2t} on the respective
cuts. However, only the choice quoted above does reproduce the result in the literature.%
\footnote{It would be valuable to understand this seeming ambiguity better. Such an ambiguity does not appear in the procedure of merging cuts employed in later sections, which we have used to confirm all calculations of this paper. In the latter procedure,  vanishing integrals are omitted, which may obscure a  double-copy interpretation of the results.}
The final step is to now reduce the tensor integrals appearing in \eqn{YMoffcuts}, which we do again using 
 the {\tt Mathematica} package {\tt FeynCalc} \cite{Mertig:1990an,Shtabovenko:2016sxi}.
Doing this we find
\begin{align}
A^{(1)}(1^{-},2^{+},3^{+},4^{+}) &= {2i\over (4\pi)^{2-\eps}}\frac{\lan 1 | 4 |2]^{2}}{\vev{34}^{2}}\, \frac{s}{tu}\, 
\Bigl [ \, I_{4}[\mu^{4}; s,t]+\frac{st}{2u}\, I_{4}[\mu^{2}; s,t] +\frac{s(u-t)}{tu}\, I_{3}[\mu^{2};t]
\\ & 
+\frac{t(s-u)}{su}\, I_{3}[\mu^{2};s] + \frac{u-s}{t^{2}}\, I_{2}[\mu^{2};t] 
+ \frac{u-t}{s^{2}}\, I_{2}[\mu^{2};s]\, \Bigl ]\, . \nonumber 
\end{align}
This result 
agrees with the result in the literature \cite{Bern:1995db}.%
\footnote{Had we taken 
$D_{0}^{2}$ instead of $D_{0}\, s_{l_{1}}$ in the last term of \eqn{YMoffcuts} we would on top find a term proportional to $[u/ (st)]\,I_{3}[\mu^{4}]$ in the above, in disagreement with
 \cite{Bern:1995db}.}

\noindent{\bf Single graviton amplitude.} 
After this warmup let us now consider the EYM amplitude for a single graviton and three gluons with one negative-helicity state. Again we shall construct the integrand from two-particle 
cuts. Now, due to the presence of the graviton $4^{++}$ which we here include with the
effectively colored ordered tree-amplitudes $\mathcal{A}$ of \eqn{1g1h}, we will have to consider three distinct
type of two-particle cut diagrams. These follow from the particle
configurations $(1234)$, $(1243)$ and $(1423)$ pushing the graviton leg $4^{++}$ through the
color-ordered gluons.
The full amplitude is then divided into three parts,
\be
A^{(1)}(1^{-},2^{+},3^{+};4^{++})= A_{(1234)}+A_{(1243)}+A_{(1423)}\, ,
\ee
which we now construct  in turn from two-particle cuts.


\noindent{\bf Diagram (1234).} Here we encounter an $s$-channel and a $t$-channel cut. For the $s$-channel of the 
$(1234)$-configuration we find
\begin{align}
\begin{split}
A_{(1234)}|_{s}&= \pic{0.35}{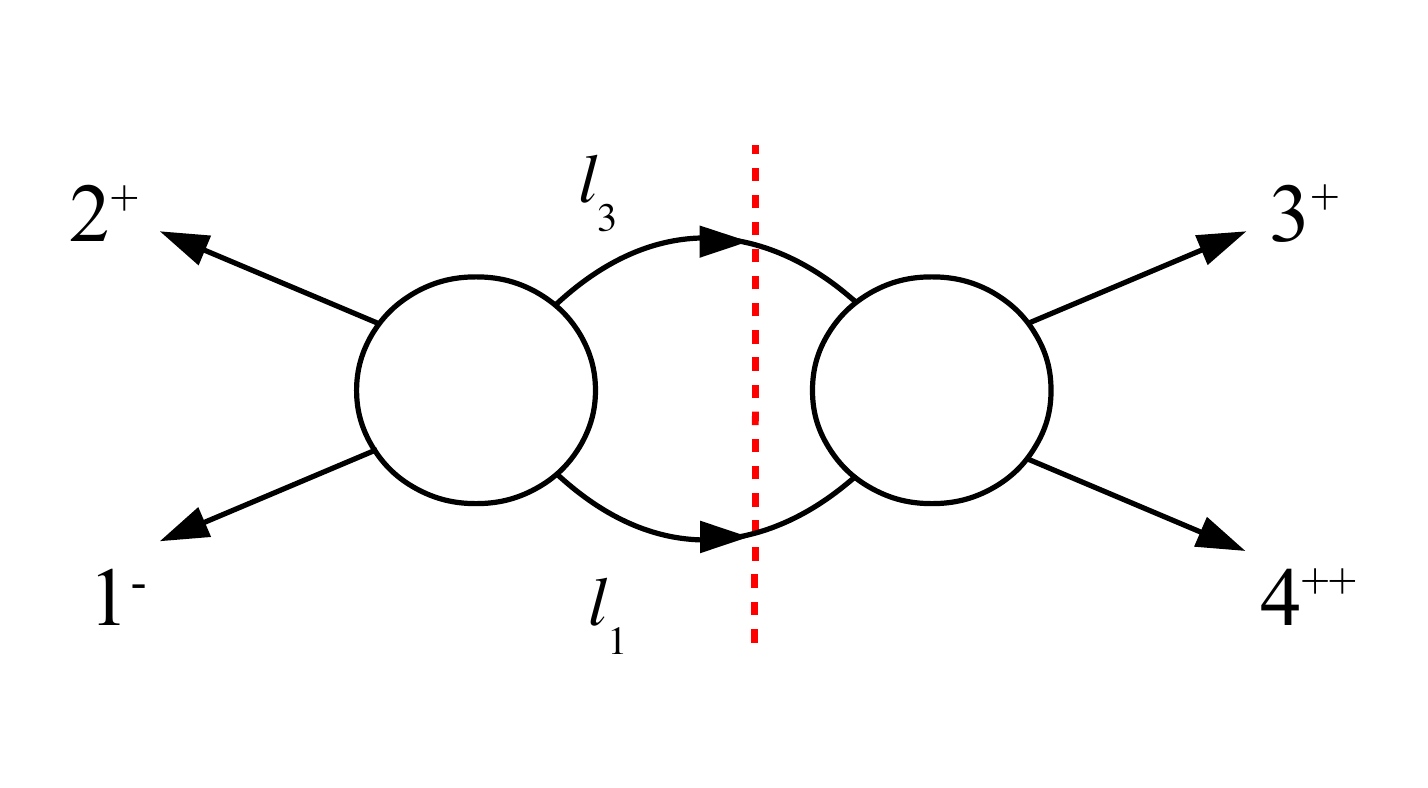} 
=
A(1^{-},2^{+},\phi_{l_{3}},\bar\phi_{l_{1}})\, \cA(3^{+},4^{++},\phi_{-l_{1}},\bar\phi_{-l_{3}}) 
\\ & =
2\mu^{2}i^2\, \frac{\bev{12}\bev{34}}{\vev{12}\vev{34}}\, \frac{\lan 3|l_{1}|4]\,
 \lan 1|l_{1}|2]^{2}}{\bev{12}^{2}\, \vev{34}}\, \frac{\big[ (2\pi)\delta({D_{0})\big]\,\big[ (2\pi)\delta(D_{2})\big]}}{D_{1}\, D_{3}} \ ,
 \label{3s}
 \end{split}
\end{align}
where for the diagram (1234) we use the following loop momentum assignments:
\begin{align}
D_{0}&=l_{1}^{2}-\mu^{2}=:l^{2}-\mu^{2}\, , \qquad &D_{1}=l_{2}^{2}-\mu^{2}=(l-p_{1})^{2}-\mu^{2}\, , \nonumber \\ 
 D_{2}&=l_{3}^{2}-\mu^{2}=(l-p_{1}-p_{2})^{2}-\mu^{2}\, ,
\quad &D_{3}=l_{3}^{2}-\mu^{2}=(l+p_{4})^{2}-\mu^{2}\, .
\label{LoopMomentaA}
\end{align}
Note that we have set $l_{1}=-l$.
The $t$-channel cut of the $(1234)$-configuration  on the other side takes the form  
\begin{align}
A_{(1234)}|_{t}&= 
\pic{0.35}{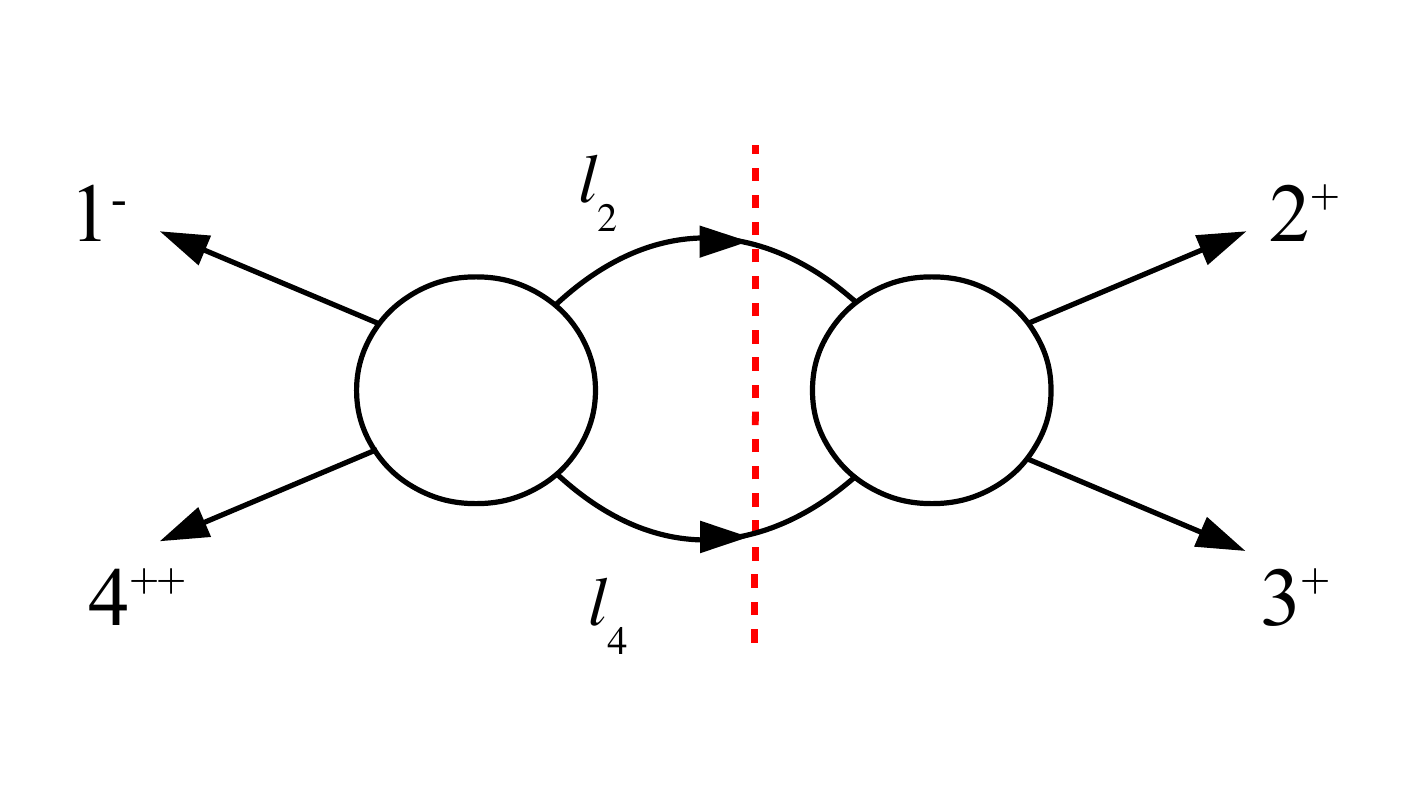} =
\cA(4^{++},1^{-},\phi_{l_{2}},\bar\phi_{l_{4}})\, A(2^{+},3^{+},\phi_{-l_{4}},\bar\phi_{-l_{2}}) \nonumber \\ & =
-2\mu^{2}i^2\, \frac{\bev{12}\bev{34}}{\vev{12}\vev{34}}\, \frac{\lan 1|l|4]^{3}\,}{\bev{41}^{2}\, \vev{14}}\, \frac{\big[ (2\pi)\delta({D_{1})\big]\,\big[ (2\pi)\delta(D_{3})\big]}}{D_{0}\, D_{2}}\, .
 \label{3t}
\end{align}
We now lift the two expressions \eqn{3s} and \eqn{3t} off the cuts by the same strategy that
was applied previously. We rewrite the two $l$-dependent spinorial expressions in $A_{(1234)}|_{s}$ as
\begin{align}
\lan 3 |l|4] &= \frac{\bev{12}}{\bev{23}}\, \lan 1 | l|4] + 
\frac{\bev{42}}{\bev{23}}\, s_{l4}\, , \qquad
\lan 1 | l|2] = \frac{\vev{34}}{\vev{23}}\, \lan 1 | l|4] + 
\frac{\vev{31}}{\vev{23}}\, s_{l1}\, .
\end{align}
Using these relations, we observe the identity
\begin{align}
\lan 3|l|4]\,
 \lan 1|l|2]^{2} &=
 \frac{\bev{12}\vev{34}^{2}}{s_{23}\,\vev{32}}\, \lan 1|l|4]^{3}+
\frac{\bev{24}\vev{23}}{s_{23}} s_{l4} \, \lan 1|l|4]^{2}\nonumber \\ & 
 +\frac{\bev{12}\vev{31}}{s_{23}\vev{32}}
s_{l1}\,  \lan 1 | l|4]\, \Bigl (\vev{34}\, \lan 1 | l|4] + \vev{23}\, \lan 1 | l|2]\Bigr ) \ .
\end{align}
Inserting this into the $s$-cut amplitude \eqn{3s}, and  rewriting  the Mandelstam invariants 
$s_{li}=2(l\cdot p_{i})$  as 
\be
s_{l4}=D_{3}-D_{0}\mathrel{\hat{=}} D_{3}\Bigr |_{\text{on $s$-cut}} 
\, , \qquad s_{l1}=D_{0}-D_{1}\mathrel{\hat{=}} -D_{1}\Bigr |_{\text{on $s$-cut}} \, , 
\ee
leads us to an expression for the $A_{(1234)}$ integrand manifestly agreeing with both cuts
\eqn{3s} and \eqn{3t}, 
\begin{align}
A_{(1234)}=2 i^2\frac{\bev{12}\bev{34}}{\vev{12}\vev{34}}\, \int& \frac{d^{4}l}{(2\pi)^{4}}\, \frac{d^{-2\eps}\mu}{(2\pi)^{-2\eps}}\,
\Bigl \{  \frac{\lan 1|l|4]^{3}\,}{\bev{41}^{2}\, \vev{14}}
 -\frac{\bev{42}}{\vev{14}\bev{12}^{3}}\, D_{3}\, \lan 1|l|2]^{2}\nonumber \\ & 
 -\frac{\vev{31}\, \lan 1 | l|4]}{\vev{14}\bev{41}^{2}\vev{34}^{2}}\, D_{1}\,
  \Bigl (\vev{34}\, \lan 1 | l|4] + \vev{23}\, \lan 1 | l|2]\Bigr )
\, \Bigr\}\, \frac{\mu^{2}}{D_{0}D_{1}D_{2}D_{3}}\, .
\label{Aa}
\end{align}
This expression may be straightforwardly reduced to scalar integrals using
e.g.~{\tt FeynCalc}. As a matter of fact, one quickly sees that the second term in the
above vanishes upon integration.

An alternative representation for $A_{(1234)}$ is obtained if one rewrites the $t$-cut expression
\eqn{3t} in terms of the $s$-cut one plus $D_{0}$ terms, arriving at 
\begin{align}
A_{(1234)}'\, =\, 2i^2 \frac{\bev{12}\bev{34}}{\vev{12}\vev{34}}\, \int& \frac{d^{4}l}{(2\pi)^{4}}\, \frac{d^{-2\eps}\mu}{(2\pi)^{-2\eps}}\,
\Bigl \{  \frac{\lan 3 |l |4]\lan 1|l|2]^{2}\,}{\bev{12}^{2}\, \vev{34}}
 +\frac{\bev{24}}{\bev{14}\bev{12} s_{23}}\, D_{0}\, \lan 1|l|4]^{2}\nonumber \\ & 
 +\frac{\vev{13}\, \bev{23}\, \lan 3 | l|4]}{\bev{12}\bev{41}\vev{34}^{2}\, s_{23}}\, D_{0}\,
  \Bigl (\vev{34}\, \lan 1 | l|4] + \vev{23}\, \lan 1 | l|2]\Bigr )
\, \Bigr\}\, \frac{\mu^{2}}{D_{0}D_{1}D_{2}D_{3}}\, , 
\end{align}
which upon Passarino-Veltman reduction indeed matches $A_{(1234)}$ of \eqn{Aa}.
The result after reduction reads:
\begin{align}
A_{(1234)}=&
{2i\over (4\pi)^{2-\eps}}\frac{[24][34]}{\lan 24 \ran \lan 34\ran} \frac{1}{[12]\lan23\ran[31]} 
 \Bigl[-\frac{3}{2}  s t I_{4}[\mu^{4}; s,t] -\frac{s^2 t^2}{2
   u} I_{4}[\mu^{2}; s,t] \nonumber \\&
   -\frac{s^2 (s+3 u)}{t^2}I_{3}[\mu^{4}; t]-\frac{s^2
   \left(s^2+3 s t+3 t^2\right)}{t u}I_{3}[\mu^{2}; t]+
   \frac{ t
   (u-s)}{s}I_{3}[\mu^{4}; s]+\frac{s^2 t}{u}I_{3}[\mu^{2}; s]
   \nonumber \\ & 
   +\frac{ s^2 (2 t-u)+u^3}{2 s t}I_{2}[\mu^{2}; s] -\frac{
   s (2 s-u) (s+3 u)}{2 t^2}I_{2}[\mu^{2}; t]\Bigr]
   \ .
\end{align}

\noindent
{\bf Diagram (1243).} 
For the (1243)-contribution we have a $u$-channel and a $s$-channel cut, 
which read
\begin{align}
A_{(1243)}|_{u}&= 
\pic{0.35}{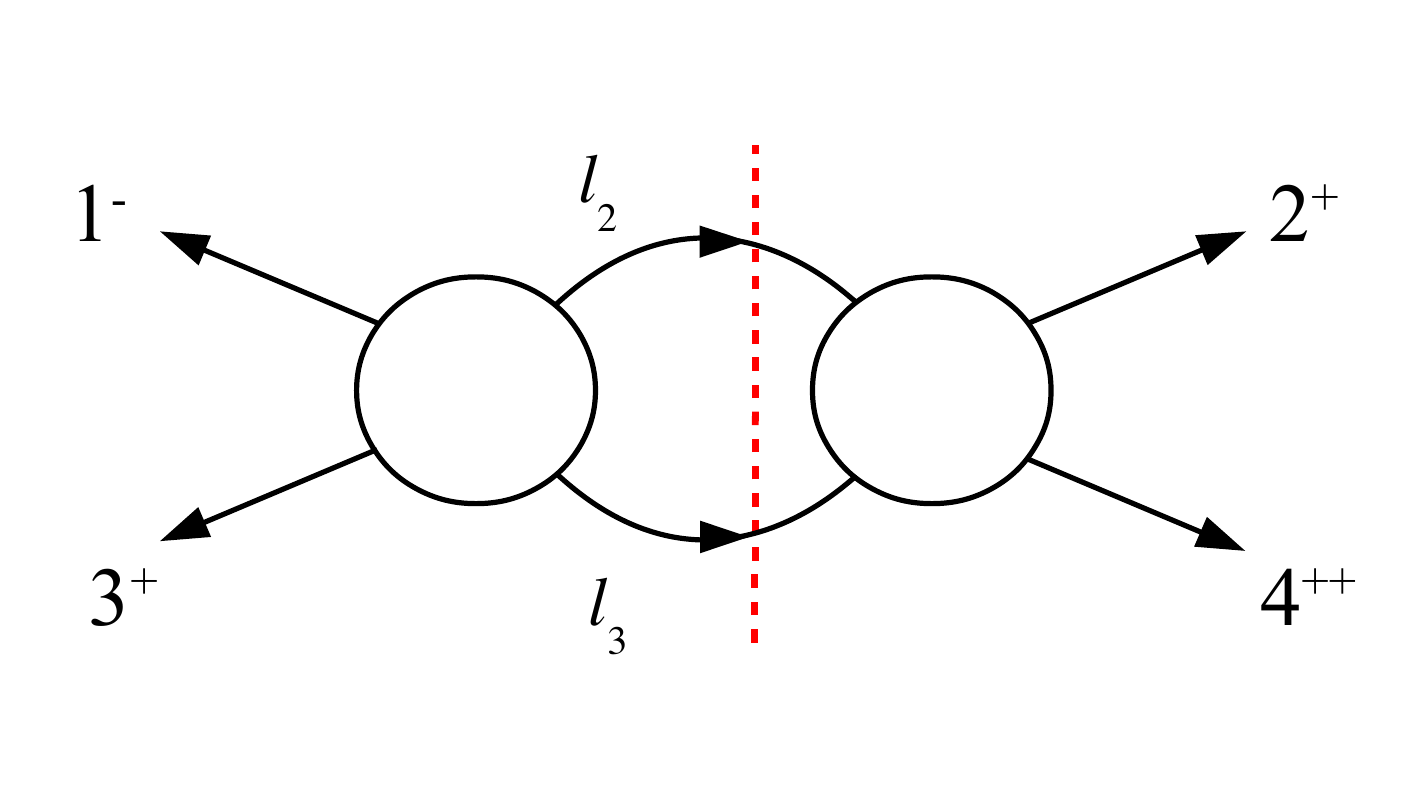} =
A(3^{+},1^{-}, \phi_{l_{2}},\bar{\phi}_{l_{3}})\, \cA(2^{+},4^{++},\phi_{-l_{3}}
, \bar{\phi}_{-l_{2}}) \nonumber \\
& =  2\mu^{2} i^2\frac{\bev{12}\bev{34}}{\vev{12}\vev{34}}\, \frac{\lan 2|l|4]\, 
\lan 1 | l | 3]^{2}}{\vev{24}\, \bev{31}^{2}}\, \frac{\big[ (2\pi)\delta(D_{0})\big]\, \big[ (2\pi)\delta(D_{2})\big]}{D_{1}D_{3}}\, ,
\end{align}
and
\begin{align}
A_{(1243)}|_{s}&= 
\pic{0.35}{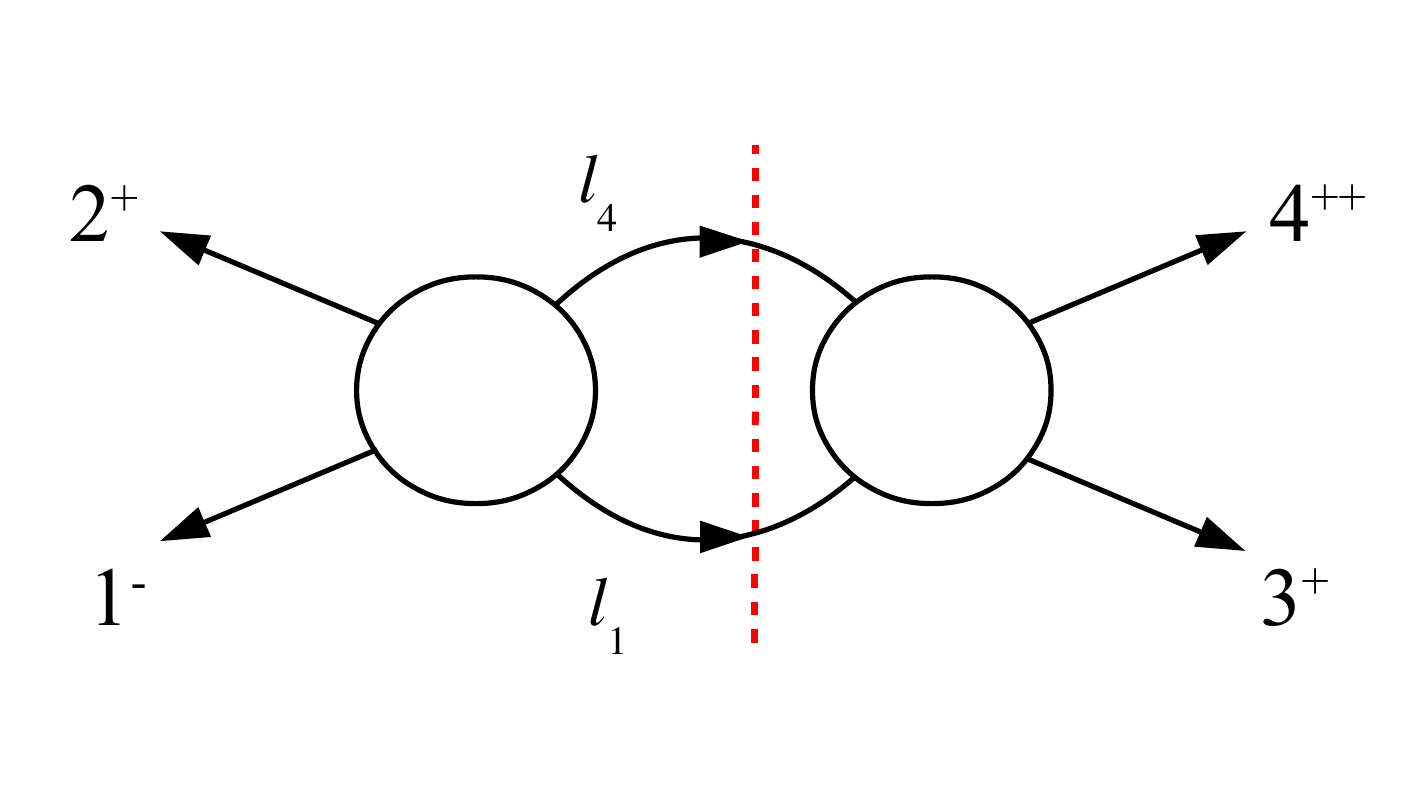} =
A(1^{-},2^{+}, \phi_{l_{4}},\bar{\phi}_{l_{1}})\, \cA(4^{++},3^{+},\phi_{-l_{1}}
, \bar{\phi}_{-l_{4}}) \nonumber \\
& = - 2\mu^{2} i^2\frac{\bev{12}\bev{34}}{\vev{12}\vev{34}}\, \frac{\lan 3|l|4]\, 
\lan 1 | l-p_{3} | 2]^{2}}{\vev{34}\, \bev{21}^{2}}\, \frac{\big[ (2\pi)\delta(D_{1})\big]\, \big[ (2\pi)\delta(D_{3})\big]}{D_{0}D_{2}}\, , 
\end{align}
where we have introduced the loop parametrization $l:=-l_{3}$ along with
\begin{align}
D_{0}&=l_{3}^{2}-\mu^{2}=:l^{2}-\mu^{2}\, , \qquad &D_{1}=l_{1}^{2}-\mu^{2}=(l-p_{3})^{2}-\mu^{2}\, , \nonumber \\ 
 D_{2}&=l_{2}^{2}-\mu^{2}=(l-p_{1}-p_{3})^{2}-\mu^{2}\, ,
\quad &D_{3}=l_{4}^{2}-\mu^{2}=(l+p_{4})^{2}-\mu^{2}\, .
\label{LoopMomentaB}
\end{align}
The $s$-cut expression may now be lifted off the cut by using the identities
\be
\bev{31}\, \lan 3 |l |4] = \bev{12}\, \lan 2| l |4] + \bev{14}\, s_{l4}\, , \qquad
\vev{42}\, \lan 1 |l-p_{3} |2] = \vev{34}\, \lan 1| l |3] + \vev{14}\, 2 (l-p_{3})\cdot p_{1}\, .\quad
\ee
On the $s$-cut (where $D_{1}=D_{3}=0$) we may replace $s_{l4}=D_{3}-D_{0}\mathrel{\hat{=}}-D_{0}$ as well as $2 (l-p_{3})\cdot p_{1}=D_{2}-D_{1}\mathrel{\hat{=}}D_{2}$. Using this we 
arrive at the integrand for the $(1243)$-type contribution, 
\begin{align}
\label{422}
A_{(1243)}&=2 i^2\frac{\bev{12}\bev{34}}{\vev{12}\vev{34}}\, \int\!\frac{d^{4}l}{(2\pi)^{4}}\, \frac{d^{-2\eps}\mu}{(2\pi)^{-2\eps}}\,
\Bigl \{ \frac{\lan 2|l|4]\, \lan 1 | l |3]^{2}}{\vev{24}\bev{31}^{2}}
  -\frac{\bev{14}}{\bev{12}\vev{24}\bev{31}^{2}}\, D_{0}\, \lan 1|l|3]^{2}\nonumber \\ & \qquad\qquad
 -\frac{\vev{14}}{\bev{12}^{2}\vev{24}^{2}\vev{34}}\, D_{2}\,
 \lan 3 | l|4]\, \Bigl (\vev{34}\, \lan 1 | l|3] + \vev{42}\, \lan 1 | l-p_{3}|2]\Bigr )
\, \Bigr\}\, \frac{\mu^{2}}{D_{0}D_{1}D_{2}D_{3}}\, .
\end{align}
Again we have an expression in terms of box and triangle tensor integrals amenable to
standard integral reduction techniques. An alternative and more compact expression may derived if one rewrites the $u$-cut in terms of the $s$-cut followed by a shift in
the integration variable $l\to l +p_{3}$. One then finds
\begin{align}
\begin{split}
\label{432}
A_{(1243)}' &=2i^2 \frac{\bev{12}\bev{34}}{\vev{12}\vev{34}}\,  
\int\!\frac{d^{4}l}{(2\pi)^{4}}\, \frac{d^{-2\eps}\mu}{(2\pi)^{-2\eps}}\,
\Bigl \{ \frac{\lan 3|l|4]\, \lan 1 | l |2]^{2}}{\vev{34}\bev{12}^{2}}   \\ &
   -\frac{\vev{14}}{\bev{12}^{2}\vev{24}^{2}\vev{34}}\, D_{1}\,
 \lan 3 | l|4]\, \Bigl (\vev{34}\, \lan 1 | l|3] + \vev{42}\, \lan 1 | l|2]\Bigr )
\, \Bigr\}\, \frac{\mu^{2}}{D_{0}D_{1}D_{2}D_{3}}\, , 
\end{split}
\end{align}
where now 
\be
D_{0}=(l+p_{3})^{2}-\mu^{2}\, , \qquad
D_{1}=l^{2}-\mu^{2}\, , \qquad
D_{0}=(l-p_{1})^{2}-\mu^{2}\, , \qquad
D_{0}=(l+p_{3}+p_{4})^{2}-\mu^{2}\, .
\ee
Passarino-Veltman reducing  \eqref{422} or \eqref{432}, one arrives at
\begin{align}
A_{(1243)}=
-{2i\over (4\pi)^{2-\eps}}\frac{[24][34]}{\lan 24 \ran \lan 34\ran} \frac{1}{[12]\lan23\ran[31]} 
 \Bigl[&
 -\frac{us}{2}  I_{4}[\mu^{4}; u,s]
 +\frac{s^2}{u} I_{3}[\mu^{4}; u] \nonumber\\ 
 &+\frac{u^2}{s} I_{3}[\mu^{4}; s]
 -\frac{t u}{2 s} I_{2}[\mu^{2}; s]-\frac{s t}{2 u}I_{2}[\mu^{2}; u]
 \Bigr]\, .
\end{align}

\noindent
{\bf Diagram (1423).} 
The remaining $(1423)$-contribution carries a $u$-channel and a $t$-channel cut.
These read
\begin{align}
A_{(1423)}|_{t}&= 
\pic{0.35}{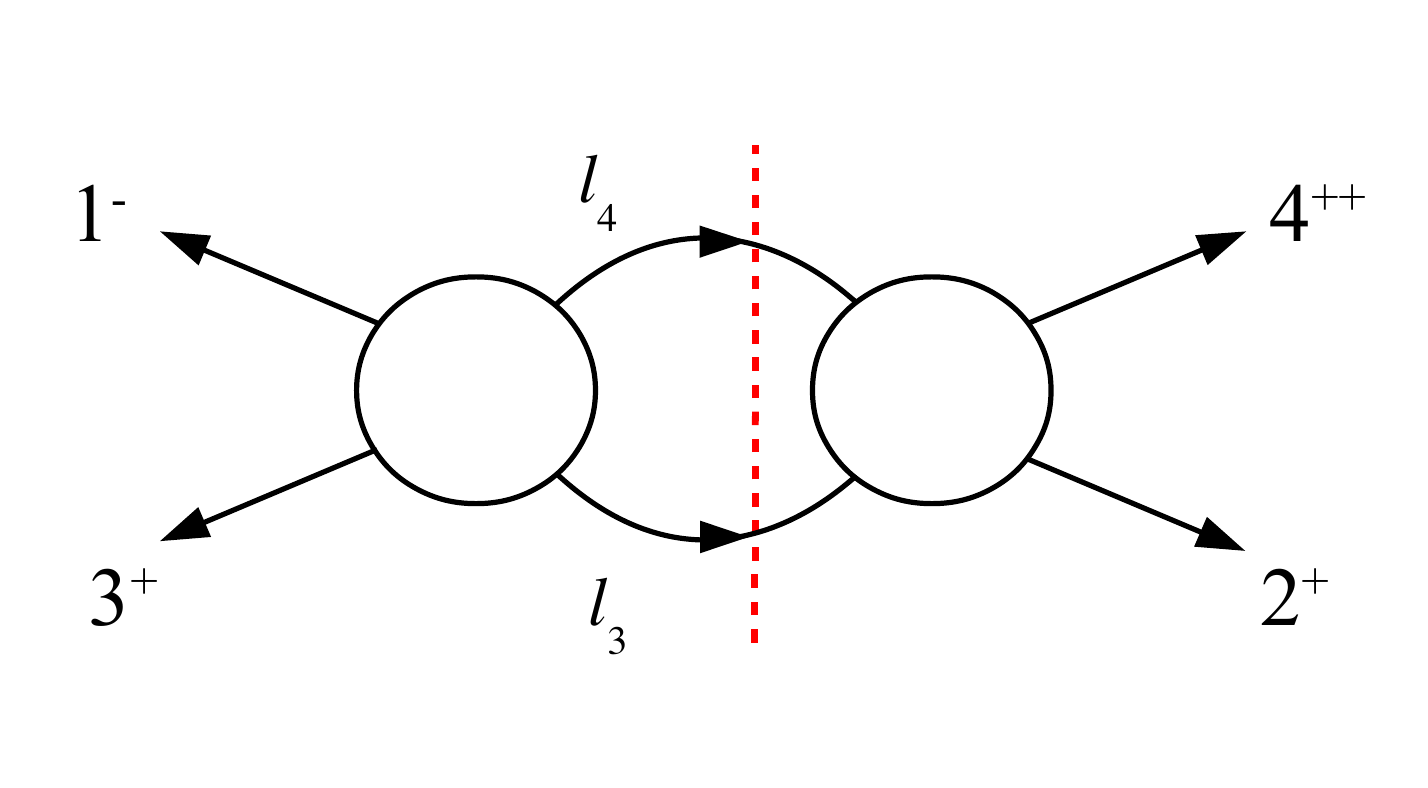} =
A(3^{+},1^{-}, \phi_{l_{4}},\bar{\phi}_{l_{3}})\, \cA(4^{++},2^{+},\phi_{-l_{3}}
, \bar{\phi}_{-l_{4}}) \nonumber \\
& = -2i^2 \mu^{2} \, \frac{\lan 1 |l|4]^{3}}{\vev{14}\vev{23}^{2}}\, \frac{\big[ (2\pi)\delta(D_{0})\big]\,
\big[ (2\pi)\delta(D_{2})\big]}{D_{1}\, D_{3}}\, ,
\end{align}
and
\begin{align}
A_{(1423)}|_{u}&= 
\pic{0.35}{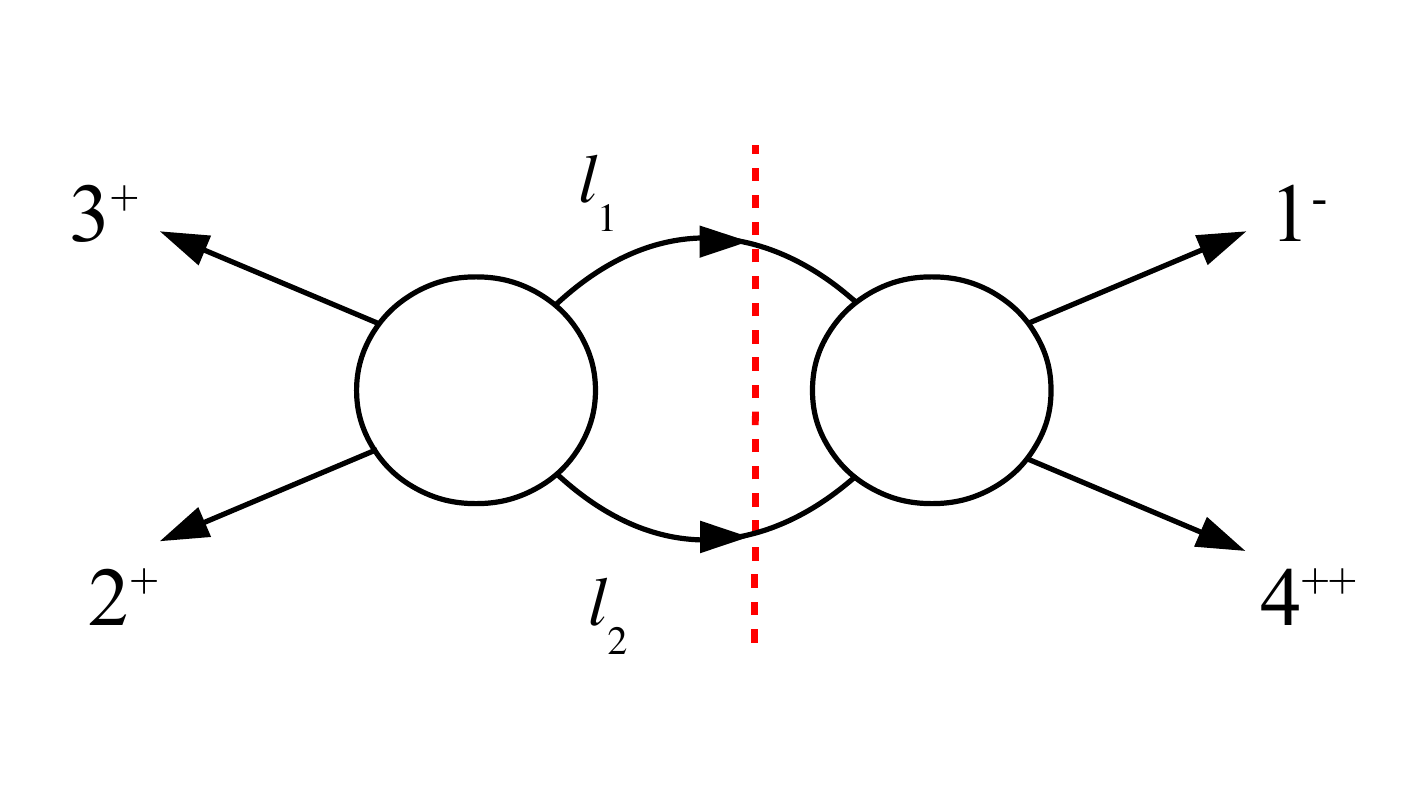} =
A(2^{+},3^{+}, \phi_{l_{1}},\bar{\phi}_{l_{2}})\, \cA(1^{-},4^{++},\phi_{-l_{2}}
, \bar{\phi}_{-l_{1}}) \nonumber \\
& = -2i^2 \mu^{2}\, \frac{\lan 2| l|4]\, \lan 1|l-p_{2}|3]^{2}}{\vev{24}^{3}}\,
\frac{\big[ (2\pi)\delta(D_{1})\big]\, \big[ (2\pi)\delta(D_{3})\big]}{D_{0}\, D_{2}}\, ,
\label{boxC2}
\end{align}
where we identified the loop momentum as $l:=-l_{2}$ and used the inverse propagators 
suitable for  diagram (1423),
\begin{align}
D_{0}&=l_{2}^{2}-\mu^{2}=:l^{2}-\mu^{2}\, , \qquad &D_{1}=l_{3}^{2}-\mu^{2}=(l-p_{2})^{2}-\mu^{2}\, , \nonumber \\ 
 D_{2}&=l_{1}^{2}-\mu^{2}=(l-p_{2}-p_{3})^{2}-\mu^{2}\, ,
\quad &D_{3}=l_{4}^{2}-\mu^{2}=(l+p_{4})^{2}-\mu^{2}\, .
\label{LoopMomentaC}
\end{align}
However, by inspection we see that $A_{(1423)}$ may be obtained from the
$(1234)$-configuration by simply swapping $2\leftrightarrow 3$ (or $s\leftrightarrow u$).
Hence we conclude that
\begin{align}
A_{(1423)}&= A_{(1234)}\Bigr|_{2\leftrightarrow 3} \,  \nonumber \\ &= 
-{2i\over (4\pi)^{2-\eps}} \frac{[24][34]}{\lan 24 \ran \lan 34\ran} \frac{1}{[12]\lan23\ran[31]} 
 \Bigl[\frac{3}{2}  \, u\,  t I_{4}[\mu^{4}; u,t] +\frac{u^2 t^2}{2
   s} I_{4}[\mu^{2}; u,t] \nonumber \\&
   +\frac{u^2 (u+3 s)}{t^2}I_{3}[\mu^{4}; t]+\frac{u^2
   \left(u^2+3 s u+3 s^2\right)}{t s}I_{3}[\mu^{2}; t]+
   \frac{ t
   (u-s)}{u}I_{3}[\mu^{4}; u]-\frac{u^2 t}{s}I_{3}[\mu^{2}; u]
   \nonumber \\ & 
   -\frac{ u^2 (2 t-s)+s^3}{2 u t}I_{2}[\mu^{2}; u] +\frac{
   u (2u^2+5u s-3s^2)}{2 t^2}I_{2}[\mu^{2}; t]\Bigr]
   \ . 
\end{align}

\noindent
{\bf Final result. } 
Adding all the three contributions $A_{(1234)}+A_{(1243)}+A_{(1423)}$  leads to the final $D$-dimensional
result:
\begin{align}
A^{(1)}&(1^{-},2^{+},3^{+};4^{++}) =\, 
-{2i\over (4\pi)^{2-\eps}} \frac{[24][34]}{\lan 24 \ran \lan 34\ran} \frac{1}{[12]\lan23\ran[31]} 
 \Bigl\{
 -\frac{3}{2}  s t I_{4}[\mu^{4}; s,t] -\frac{s^2 t^2}{2 u}I_{4}[\mu^{2}; s,t]
 \nonumber \\ & +
 \frac{1}{2}  s u I_{4}[\mu^{4}; u,s] -\frac{3}{2} t u I_{4}[\mu^{4}; u,t]-\frac{t^2 u^2}{2 s}
 I_{4}[\mu^{2}; u,t]
 -\frac{ \left(-t^2 u-2 t u^2+u^3\right)}{s u}I_{3}[\mu^{4}; s] \nonumber \\
 & -\frac{ \left(t^4+3 t^3 u+3 t^2
   u^2+t u^3\right)}{s u}I_{3}[\mu^{2}; s]
 -\frac{\left(t^2 u+t u^2\right)}{s u}I_{3}[\mu^{4}; t]  \nonumber \\ &
 -\frac{ \left(-t^4-2 t^3 u-t^2 u^2+2 t
   u^3+u^4\right)}{s u}I_{3}[\mu^{2}; t] 
   -\frac{ \left(2 t^2+4 t u+u^2\right)}{u}I_{3}[\mu^{4}; u] 
   +\frac{t u^2}{s}I_{3}[\mu^{2}; u] \nonumber \\ &
   \frac{ t (t+2 u)}{s}I_{2}[\mu^{2}; s]
  + \frac{\left(t^2+2 t u+2 u^2\right)}{t}I_{2}[\mu^{2}; t]
  -\frac{t (t+2 u)}{u}I_{2}[\mu^{2}; u]
 \Bigr\}
   \ . 
\end{align}
Taking the four-dimensional limit yields the compact final expression 
\beq
A^{(1)}(1^{-},2^{+},3^{+};4^{++}) = {i\over (4 \pi)^2}\, 
   \frac{[24][34]}{\lan 24\ran \lan 34\ran}\, {1\over\lan23\ran [21][31]}\,  {1\over 6} \, (s^2 + u^2)
   \ . 
   \eeq

\section{The  {\normalfont$\langle 1^+\, 2^+ \,3^+ \,4^{--}\rangle$} amplitude} 
\label{sec:5}
We now consider the  rational one-loop   amplitude with a single negative-helicity graviton and three positive-helicity gluons $A^{(1)}(1^{+},2^{+},3^{+};4^{--}) $.  For amplitudes containing progressively more negative helicities, the procedure described in previous sections to construct the integrand becomes tedious. Hence, from now on, 
rather than constructing the integrand, we will  use the standard  approach of  \cite{Bern:1994zx,Bern:1994cg}
where we directly merge all two-particle cuts into a single function. The case at hand is particularly simple given the very symmetric helicity configuration chosen. Using the tree-level amplitudes  in Section \ref{relevanttrees}, we find that the   $s$-cut of the amplitude is given by
\begin{align}
\label{ttb}
\textrm{{\it s}-cut:}&\pic{0.35}{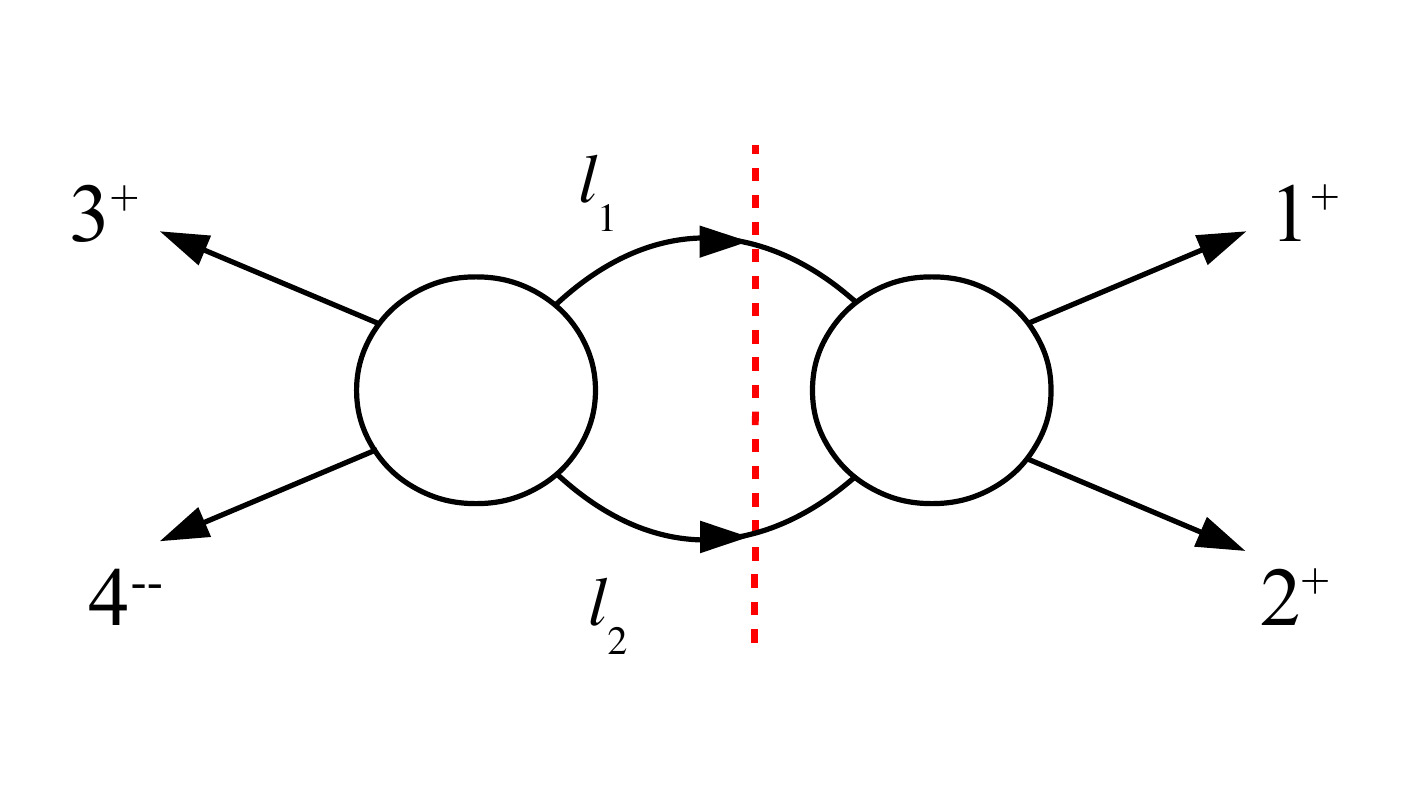} = \ 
-i^2  \mu^2 {[12]\over \langle 12 \rangle} {\langle 4|l_1|3]^3\over s[34] } \Big[
\pic{0.35}{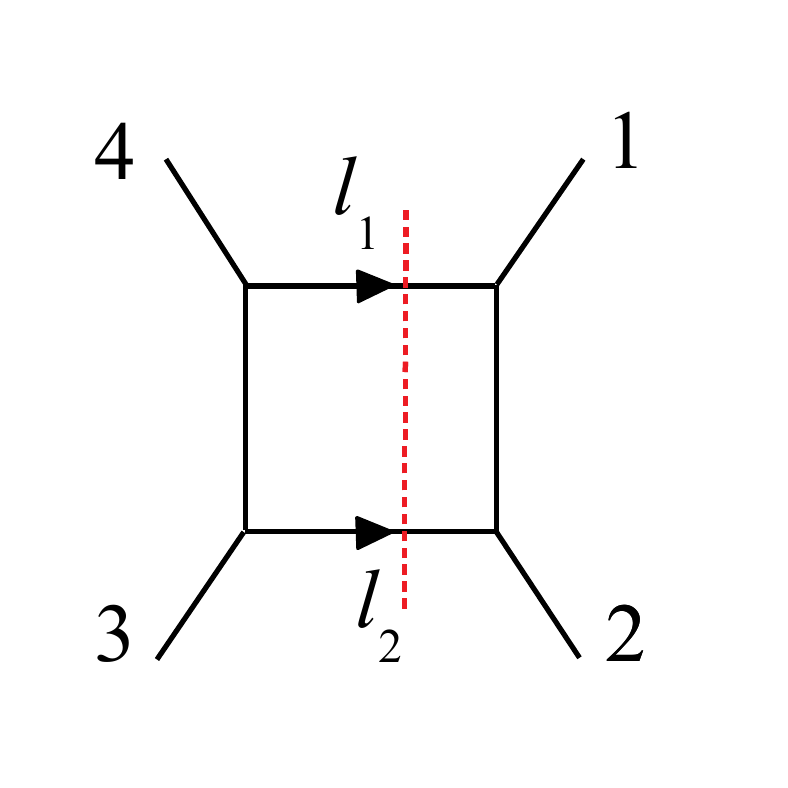} \ + 
\pic{0.35}{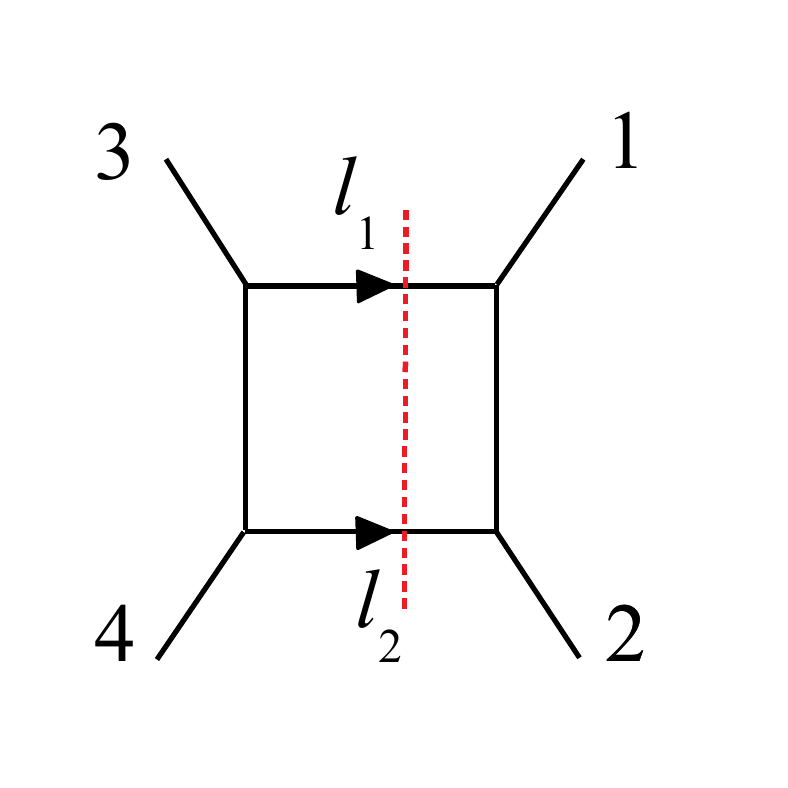} \Big]
\ .
\end{align}
This amplitude also has $t$- and $u$-cuts which are obtained by simply cycling  the labels $(312)\to (123)$ and $(312) \to (231)$, respectively. As in the previous sections, we use {\tt FeynCalc}  \cite{Mertig:1990an,Shtabovenko:2016sxi} to perform 
efficiently all relevant Passarino-Veltman reductions of the three-tensor box in \eqref{ttb} (and its permutations). We work first in the $s$-cut, and focus on the tensor box with particle ordering $(1234)$. We lift the integral off the cut, and perform a 
 Passarino-Veltman reduction. This will generate  scalar boxes with particle ordering $(1234)$ (and powers of the $(-2\eps)$-momentum $\mu$ in the numerator), whose coefficient(s) we will then confirm from the $t$-cut. It will also generate one-mass triangles and bubbles in the $s$-channel (again with  powers of  $\mu$ in the numerator), which we  keep, as well as spurious one-mass triangles and bubbles with a $t$-channel discontinuity, which we drop. We then repeat the same operation for the two other box topologies with particle orderings $(1243)$ and $(1324)$. 
Merging all contributions thus obtained, 
we  arrive at our final result:
\begin{align}
\label{1234m}
A^{(1)}(1^{+},2^{+},3^{+};4^{--})  \ = \ 2    i^2 \frac{[12][34]}{\lan 12 \ran \lan 34\ran}  (\lan42\ran [23]\lan 34\ran)^3  \Big[  {f(s,t,u) }  \  + \mathrm{perms}\Big]
\ , 
\end{align}
where 
\begin{align}
\begin{split}
f (s,t,u) & = \, {i\over (4\pi)^{2-\eps}}{1\over stu^2}\, \Big[ \, \frac{3 }{2}I_4[\mu^4; s,t] -\frac{t
   (s-2 u)}{s^3} I_3[\mu^4; s]
   -\frac{s 
   (t-2 u)}{t^3}I_3[\mu^4;  t]
   \\
   &+
   \frac{ s t}{2
   u}I_4[\mu^2; s,t] +\frac{ s\left(s^2-3  t
   u\right)}{t^2 u}I_3[\mu^2;  t] +\frac{t  \left(t^2-3 s
    u\right)}{s^2 u}I_3[\mu^2; t]
 \\
   & +  \frac{(s-2 u) (u-2 t)}{2
   s^3}I_2[\mu^2; s] +\frac{ (u-2 s) (t-2 u)}{2
   t^3}I_2[\mu^2; t] \Big] \ . 
   \end{split}
   \end{align}
   As in the case of the $\langle 4^{++} 1^+ 2^+ 3^+\rangle$ amplitude computed in Section \ref{allplus}, by ``perms'' we denote the 
  two permutations $2314$ and $3124$ of 1234, with the   the Mandelstam invariants interchanged as    
  $(s\to t, t \to u, u\to s)$ and $(s\to u, t \to s, u\to t)$.
  Performing the four-dimensional limit using the results of Appendix \ref{Integrals}, we find: 
   \beq
   f(s,t,u)\ \to  \ -{i\over (4\pi)^2}\frac{3 t^2+3 t u+2 u^2}{24\,  s^3 t^3}\ .
  \eeq
    Adding the permutations,   we arrive at a very compact final result: 
     \begin{align}
\label{eq:gravitononeminusfinalresult}
A^{(1)}(1^{+},2^{+},3^{+};4^{--})  \ =  \ 
 -{i \over (4\pi)^2} \, 
  \frac{[12][34]}{\lan 12 \ran \lan 34\ran} \,  {(\lan42\ran [23]\lan 34\ran)^3}\,  
{s^2 + t^2 + u^2  \over  \,  6\, s^2\, t^2\, u^2}
\ .
\end{align}
Note that 
the kinematic function in \eqref{eq:gravitononeminusfinalresult} is an odd function under any exchange of two gluons, and hence the complete amplitude is even under such an exchange  (including a minus sign from the colour factor $f^{abc}$), as it should.

\section{The  {\normalfont$\langle 1^+\, 2^+ \,3^{++} \,4^{++}\rangle$} amplitude}
\label{sec:6}

In this section we move on to  amplitudes which contain two gravitons and two gluons. The simplest case to consider occurs when all particles have the same helicity -- a particularly symmetric configuration. 

We briefly describe the outline of the derivation, similarly with previous calculations. 
As usual there are three cut diagrams to consider, in the $s$-, $t$- and $u$-channels. These cuts will give rise to tensor boxes with particle ordering $(1234)$, $(1243)$ and $(1324)$. These are given by: 
\begin{align}
\begin{split}
\textrm{{\it s}-cut:}\quad &
A\big( 3^{++}, 4^{++}, l_{1, \bar\phi},  l_{2,  \phi}\big) \, \big[ A\big(1^+,  2^+ , -l_{2, \bar\phi} ,  -l_{1, \phi}\big)\, + \, 1\leftrightarrow 2\big]\, , \\
\textrm{{\it t}-cut:}\quad &
A\big( 4^{++}, 1^{+},  l_{1, \bar\phi},  l_{2, \phi }\big) \ \ A\big(2^+,  3^{++} ,  -l_{2, \bar\phi} ,  -l_{1, \phi}\big)\, , \\
\textrm{{\it u}-cut:}\quad &
A\big( 3^{++}, 1^{+},  l_{1, \bar\phi},   l_{2, \phi}\big) \ \ A\big(2^+,  4^{++} ,  -l_{2, \bar\phi} ,  -l_{1, \phi}\big)\ .
\end{split}
\label{eq:1p2p3pp4ppAllcuts}
\end{align}
Note that  on the right-hand side of the the $s$-cut in  \eqref{eq:1p2p3pp4ppAllcuts} we have to include the sum of two color-ordered amplitudes,   $ A\big(1^+,  2^+ , -l_{2, \bar\phi} ,  -l_{1, \phi}\big)$ and $ A\big(2^+,  1^+ , -l_{2, \bar\phi} ,  -l_{1, \phi}\big)$. 
Indeed, since the left-hand side of the cut is an amplitude with a colorless (two-graviton) external state,   both terms contribute to the same color ordering. This will be a recurrent feature of all cuts where one side of the cut is colorless. 
Moreover,   there will be an additional contribution from the cut obtained by swapping $\phi$ with $\bar\phi$, which will double up the result of the previous cuts, as usual. 

Using the tree-level amplitudes given in Section \ref{relevanttrees}, we work out the expressions of these cuts, which give rise to  three tensor boxes with the different particle orderings $(1234)$, $(1243)$ and $(1324)$. Inspecting all cuts we can reconstruct the amplitude. We find the following results: 
\begin{align}
\begin{split} 
\textrm{{\it s}-cut:}&\pic{0.35}{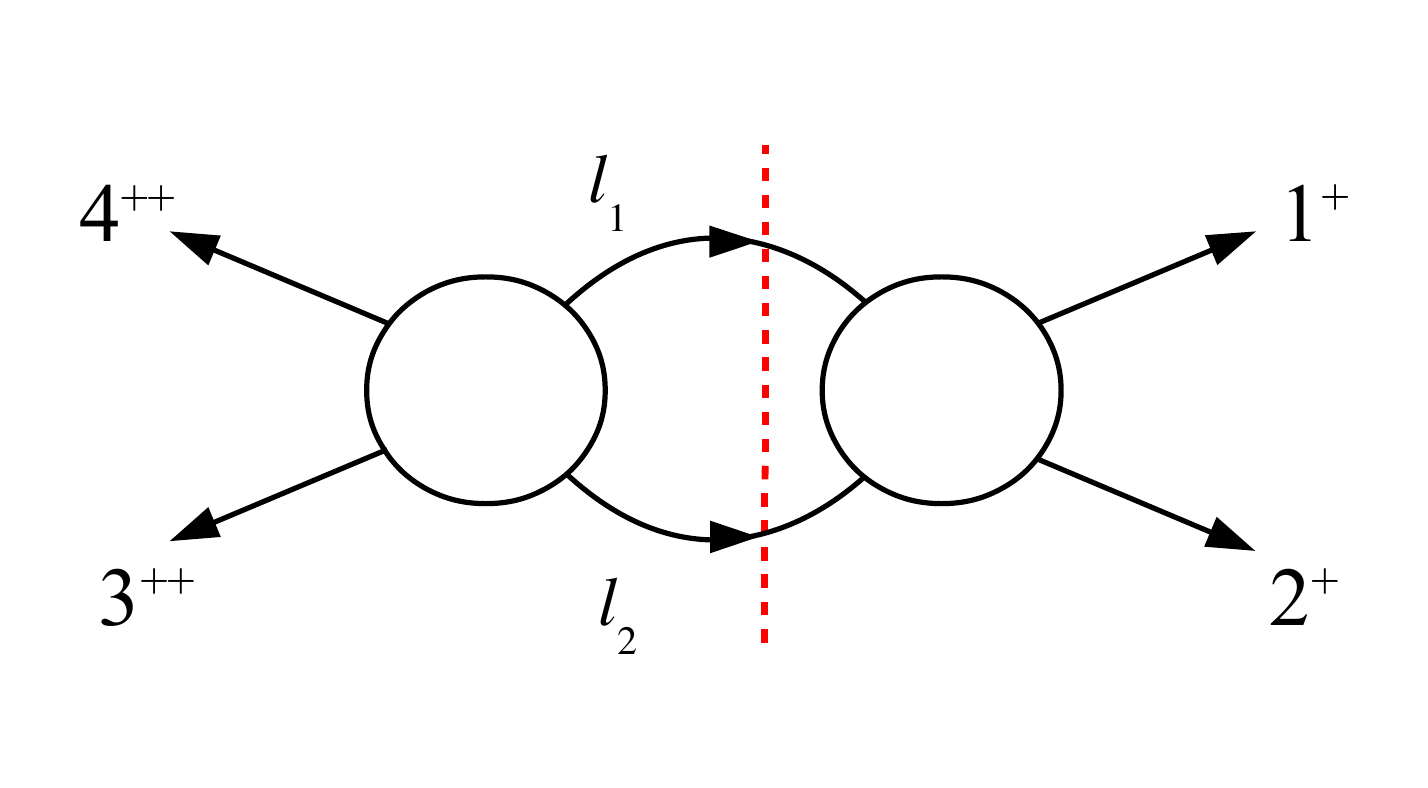} = \ 
   2 \mu^6 {[34]^2\over \langle 34 \rangle^2} {[12]\over \langle 12 \rangle} 
  \\
 & \Big[
\pic{0.35}{box1234-s-cut} \ + \  \pic{0.35}{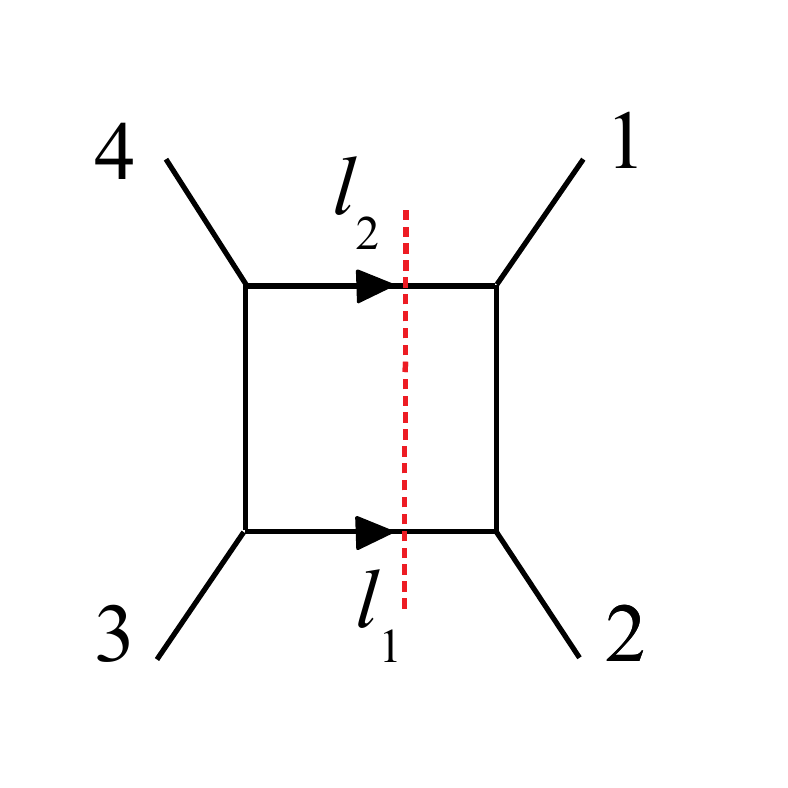}\, + \, 
\pic{0.35}{box1243-s-cut} \, + \, \pic{0.35}{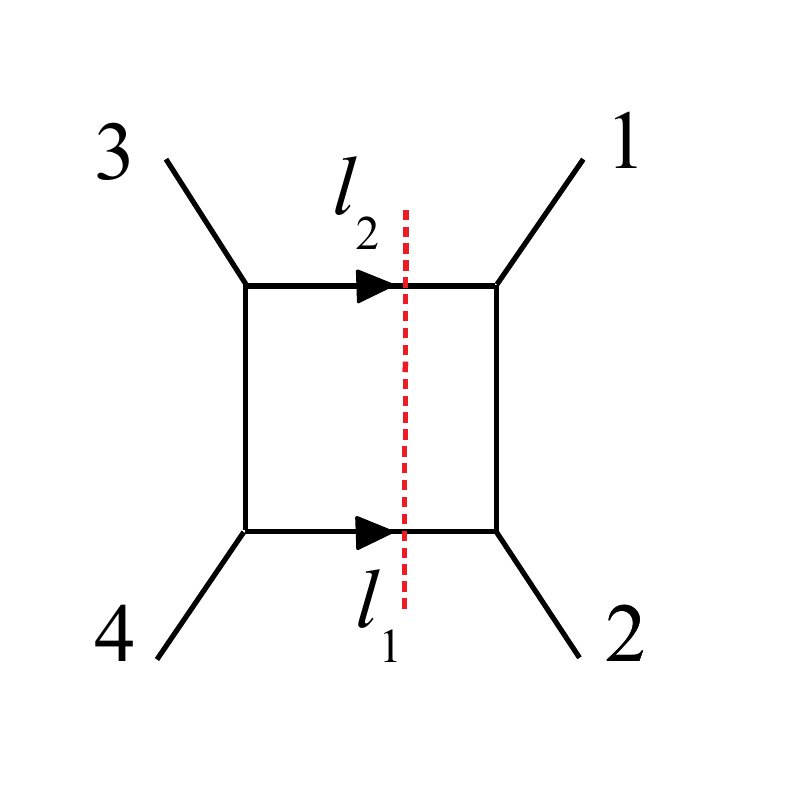} \Big]
\ , 
\label{pr}
\end{split}
\end{align}
\begin{align}
\begin{split}
\textrm{{\it t}-cut:}&\pic{0.35}{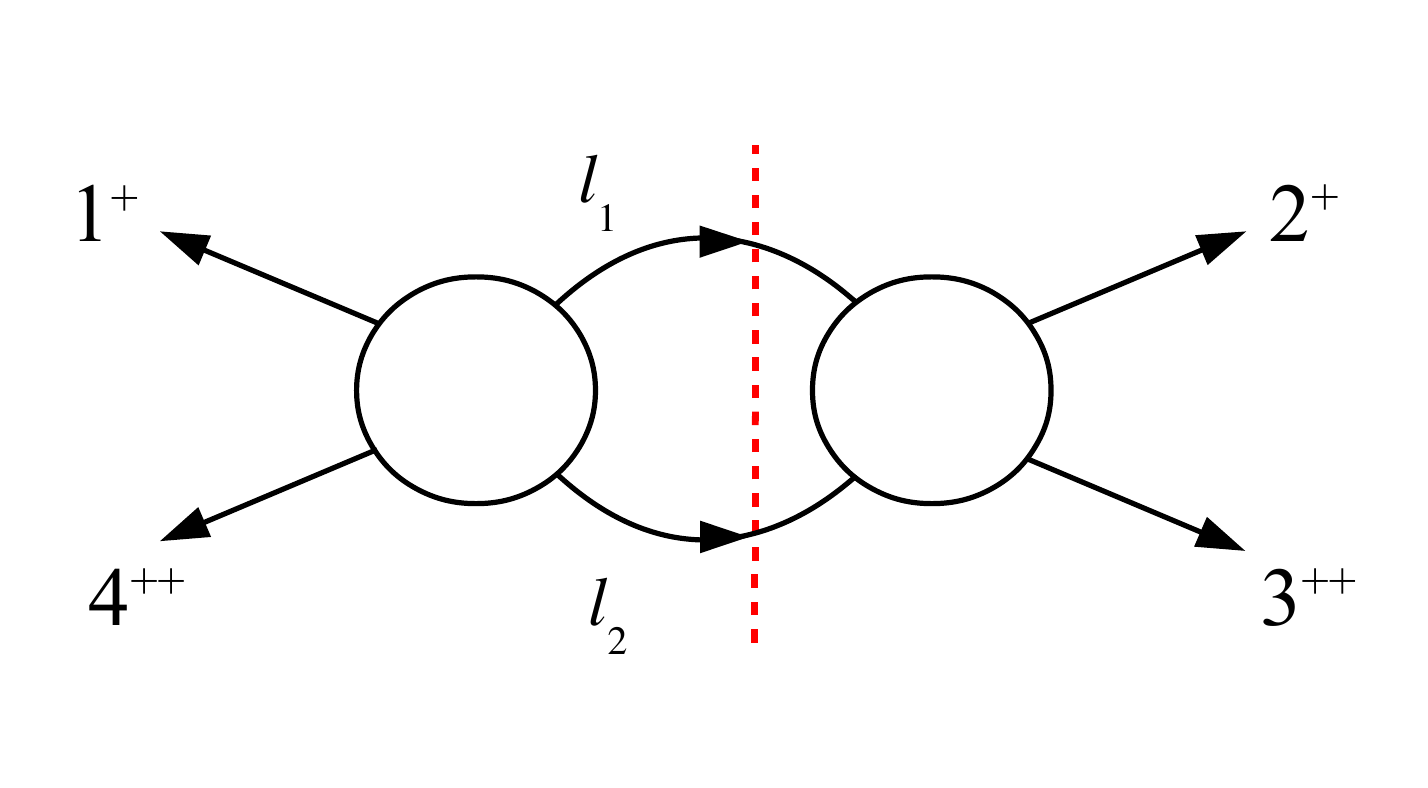} = \ 
 \ 2 \mu^4 {[41]\over \langle 41 \rangle^2} {[32]\over \langle 32 \rangle^2}
 \langle 1 |l_1 |4]
 \langle 2 |l_2 |3]
 \\ 
 & \cdot \Big[
\pic{0.35}{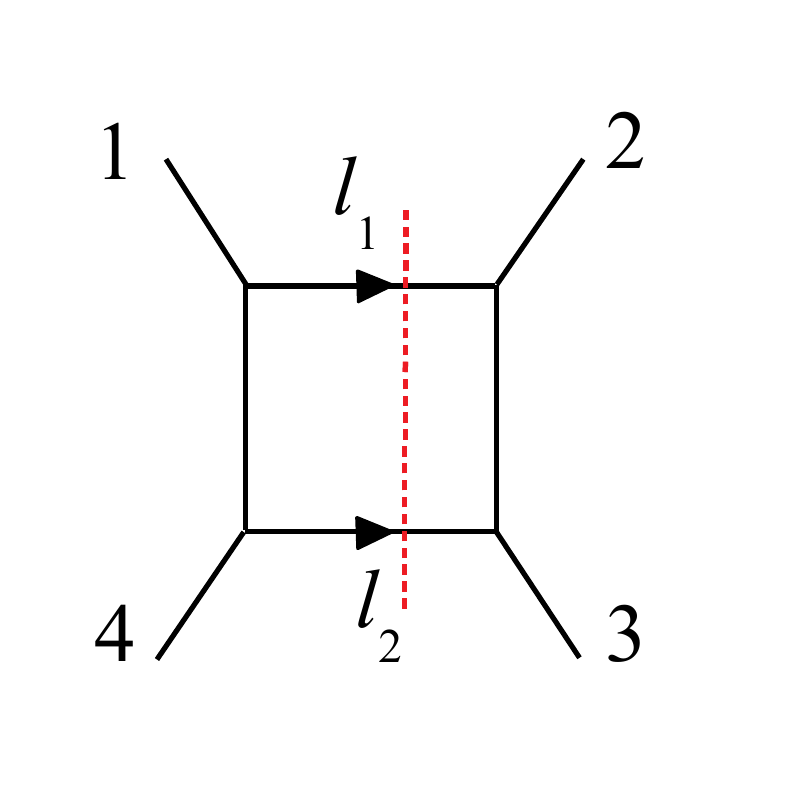} + 
\pic{0.35}{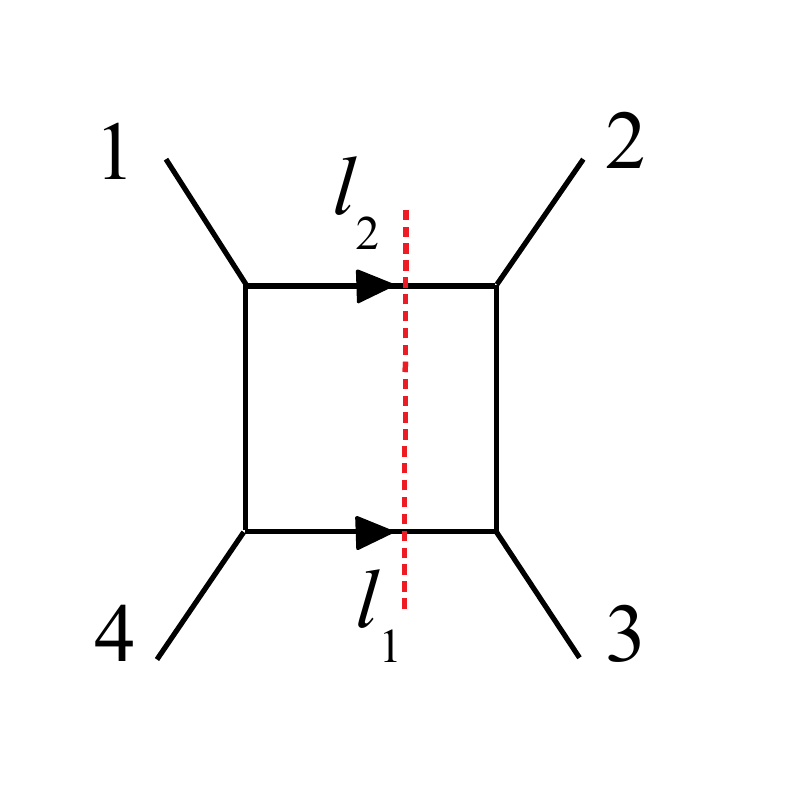} + 
\pic{0.35}{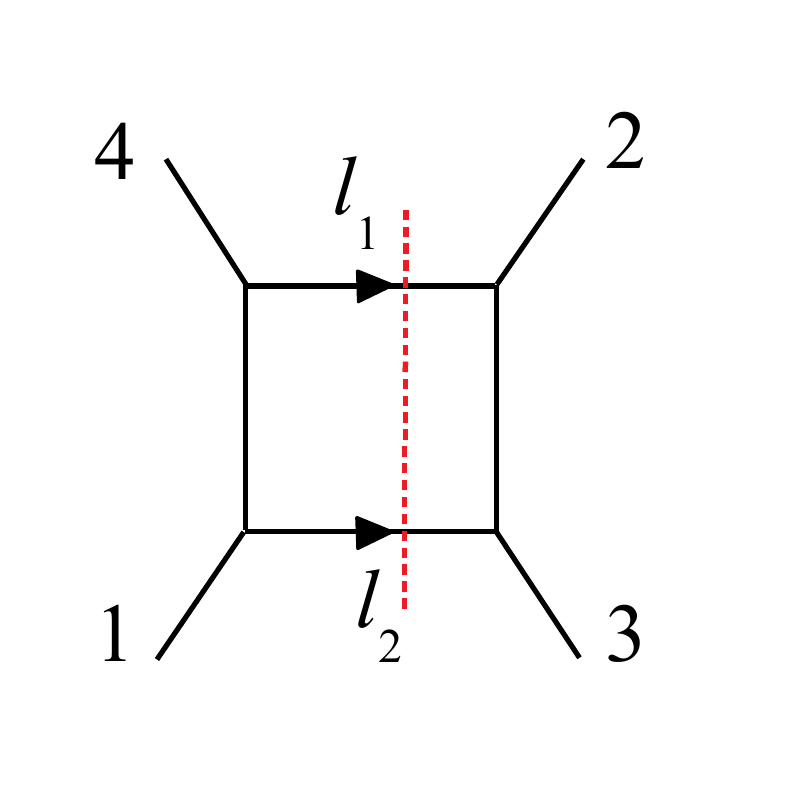}  + 
\pic{0.35}{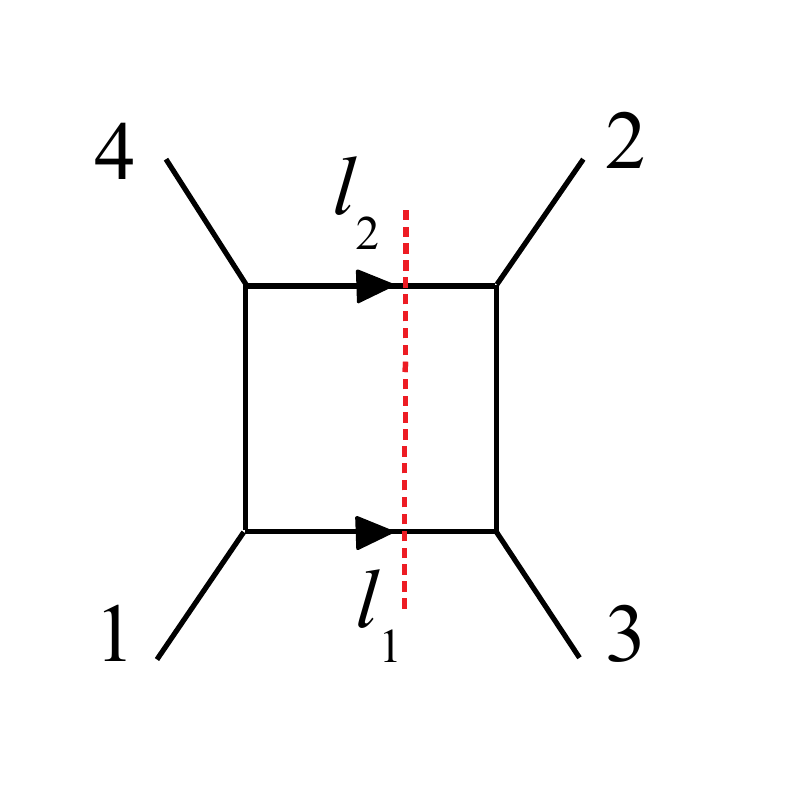} \Big]\ , 
\label{sec}
\end{split}
\end{align}
\begin{align}
\begin{split}
\textrm{{\it u}-cut:}&\pic{0.35}{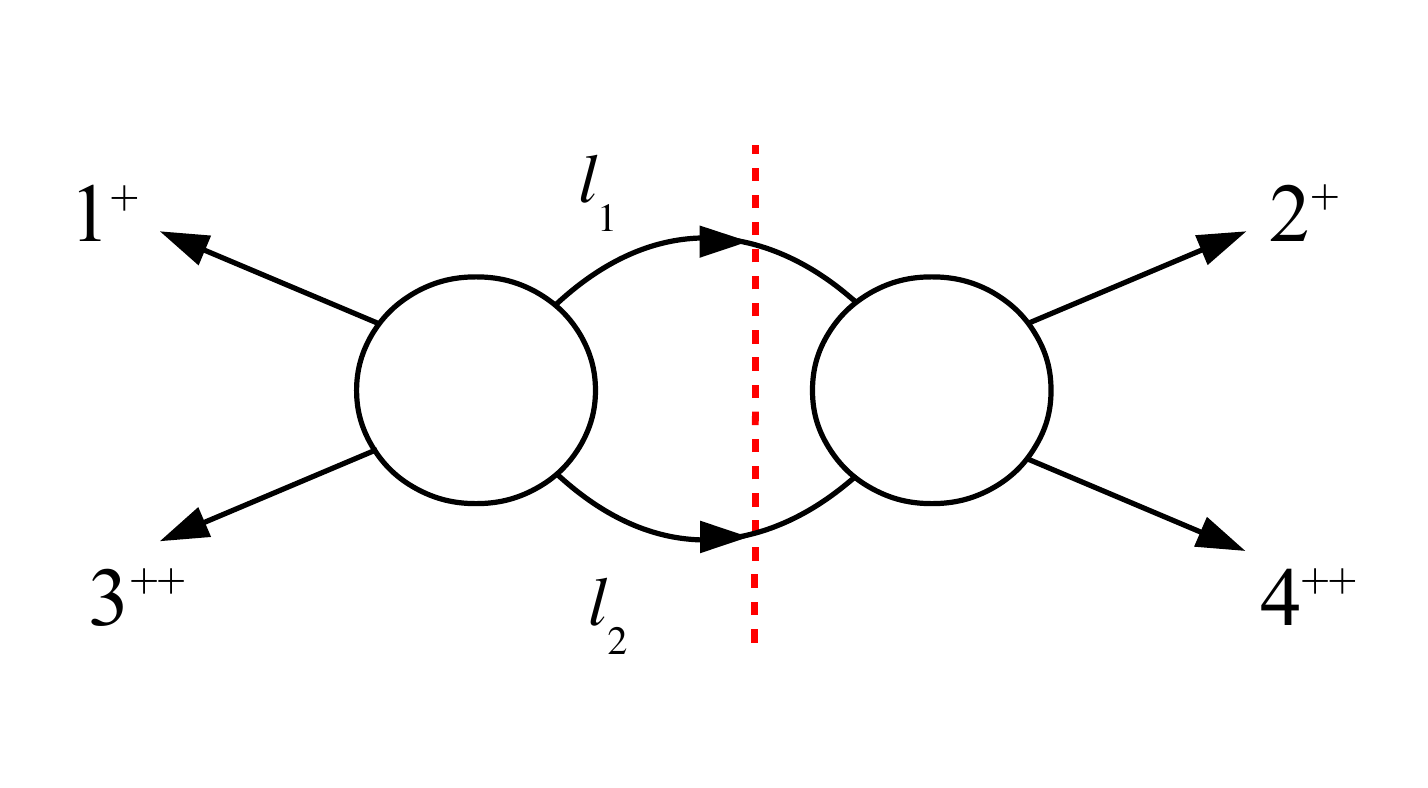} = \ 
 2 \mu^4 {[31]\over \langle 31 \rangle^2} {[42]\over \langle 42 \rangle^2} 
 \langle 1 |l_1 |3]
 \langle 2 |l_2 |4]
 \\
 &\cdot \Big[
\pic{0.35}{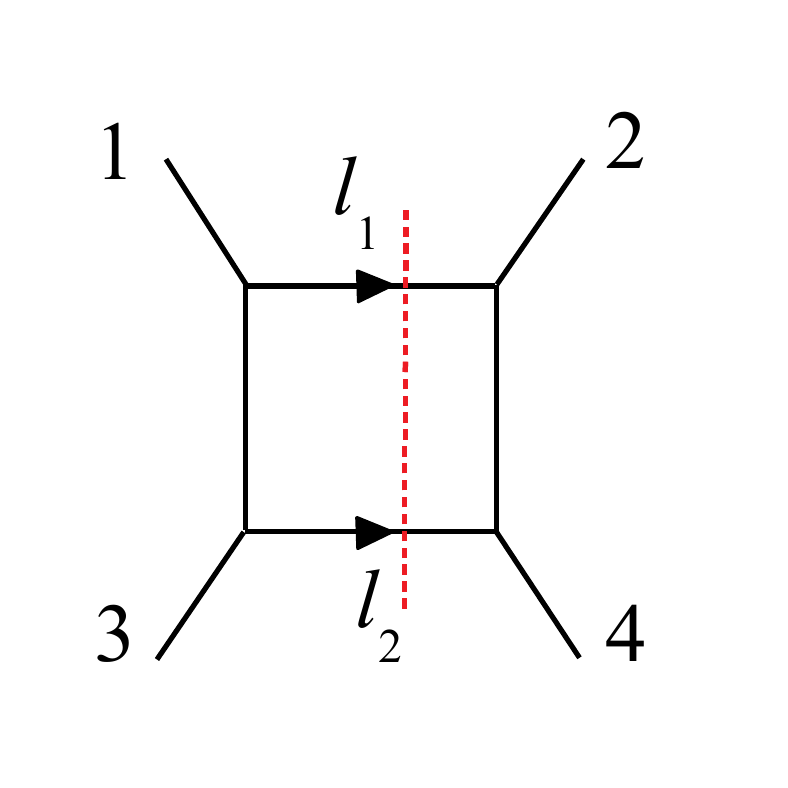} + 
\pic{0.35}{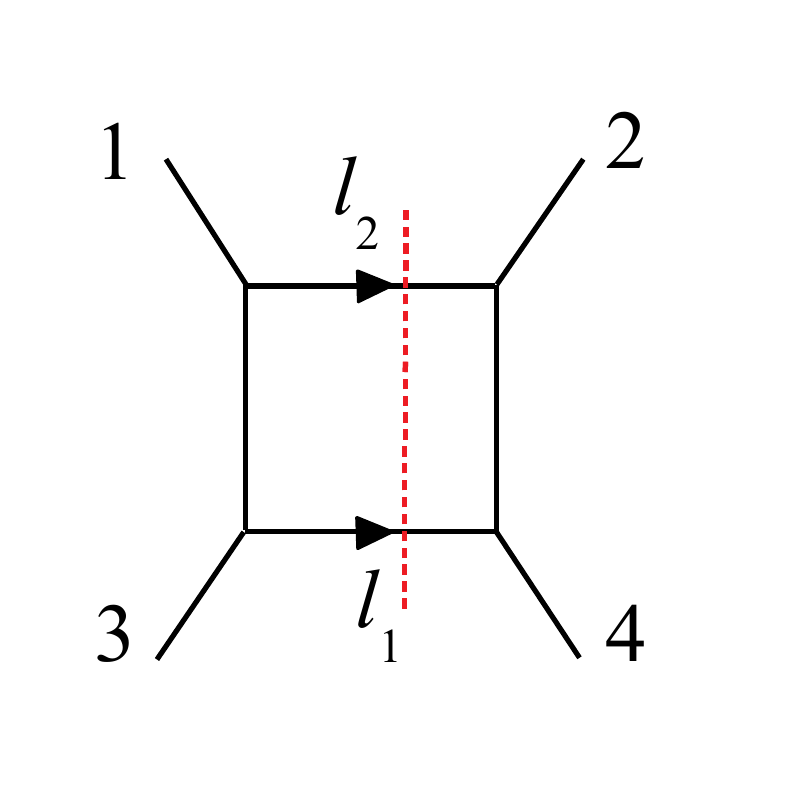} + 
\pic{0.35}{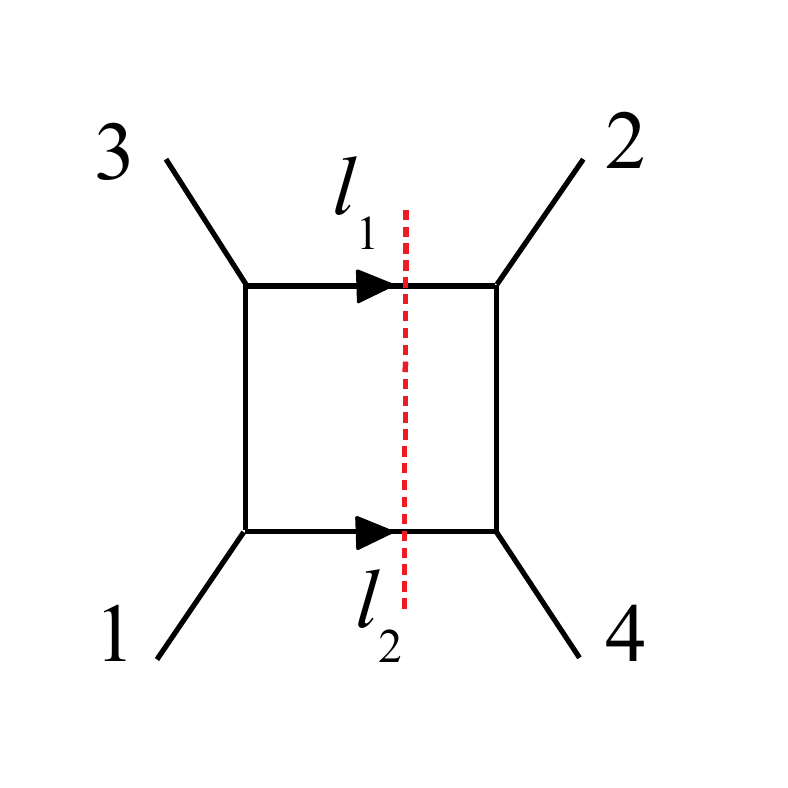}  + 
\pic{0.35}{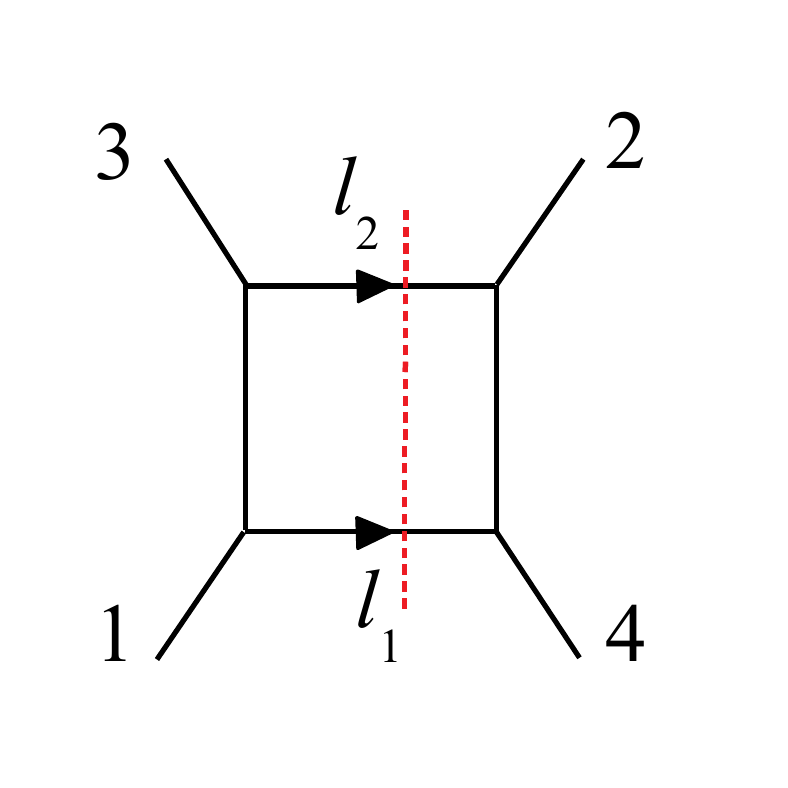} \Big]\ .
\label{ter}
\end{split} 
\end{align}
Note that  our cut integrand  contains 
tensor boxes with  cut momenta $l_1$ and $l_2$ as well as the same contribution but with  $l_1$ and $l_2$ flipped. At the level of the integral, this will be taken into account by doubling up the contribution of a single copy.

The next step consists in combining all cuts, which we will do for each box topology separately. 
Doing so,  we arrive at the following result 
for the topology $(1234)$: 
\beq
(1234):  \qquad -{i\over (4\pi)^{2-\eps}} \  {[12]\over \langle 12\rangle} {[34]^2\over \langle 34\rangle^2} \cdot 4 \Big( I_4[ \mu^6; s,t]- {1\over t} I_2[ \mu^4; t]\Big)
\ , 
\eeq
which is obtained from combining the relevant terms in the $s$-cut given in \eqref{pr} and the $t$-cut of \eqref{sec}. 
The topology (1243) is simply obtained by swapping $3\leftrightarrow 4$, or $s\to s, t\to u, u\to s$ in the previous result: 
\beq(1243):  \qquad -{i\over (4\pi)^{2-\eps}} \  {[12]\over \langle 12\rangle} {[34]^2\over \langle 34\rangle^2} \cdot 4 \Big( I_4[ \mu^6; s,u]- {1\over u} I_2[ \mu^4; u]\Big)
\ . 
\eeq
The last topology to consider is (1324), which  is obtained from combining the relevant terms from the $s$- and $u$-cuts, given in \eqref{pr} and \eqref{ter}.  Doing so we get: 
\begin{align}
\begin{split}
(1324):  \qquad  & -{i\over (4\pi)^{2-\eps}} \ {[12]\over \langle 12\rangle} {[34]^2\over \langle 34\rangle^2} \cdot 
\\
& 4  \Big( I_4[ \mu^6; u,t]+ {ut\over 2s}  I_4[ \mu^4; u,t]- {t\over s} I_3[ \mu^4; t]- {u\over s} I_3[ \mu^4; u]+ {I_2[\mu^4; t]\over t} +{I_2[\mu^4; u]\over u} \Big)
\ . 
\end{split}
\end{align}
Finally we take the four-dimensional limit: 
\begin{align}
4\Big( I_4[ \mu^6; s,t]- {1\over t} I_2[ \mu^4; t]\Big) \to - {s\over 15}
\ , 
\end{align}
while 
\beq
4  \Big( I_4[ \mu^6; u,t]+ {ut\over 2s}  I_4[ \mu^4; u,t]- {t\over s} I_3[ \mu^4; t]- {u\over s} I_3[ \mu^4; u]+ {I_2[\mu^4; t]\over t} +{I_2[\mu^4; u]\over u} \Big)
\to - {s\over 30}\ . 
\eeq
Combining all terms we arrive at a remarkably simple final result:
\beq
A^{(1)}(1^+, 2^+; 3^{++}, 4^{++}) \ = \   {i \over (4 \pi)^2} \, {[12]\over \langle 12\rangle}\,  {[34]^2\over \langle 34\rangle^2}\, {s\over 6}\,  
\ . 
\label{fr}
\eeq
We note that \eqref{fr} is symmetric under the exchange of the two gluons. This is consistent with the colour factor $\delta^{ab}$ of this amplitude --  indeed, the complete, color-dressed result should be symmetric under a   swapping of the two gluons.

We also quote the compact expression of the $D$-dimensional result: 
\begin{align}
A^{(1)}(1^+, 2^+; 3^{++}, 4^{++}) &= -{4 i \over  (4\pi)^{2-\eps}}
{[12]\over \langle 12\rangle}\,  {[34]^2\over \langle 34\rangle^2}
\Bigl \{ I_{4}[\mu^{6};s,t] + I_{4}[\mu^{6};s,u] + I_{4}[\mu^{6};u,t]\nonumber \\
& +\, \frac{tu}{2s} I_{4}[\mu^{4};u,t]  \, 
-\frac t s \,  I_{3}[\mu^{4}; t] -\frac u s \, I_{3}[\mu^{4}; u]
 \Bigr \}\, .
\end{align}

\section{The  {\normalfont$\langle 1^-\, 2^+ \,3^{++} \,4^{++}\rangle$} amplitude}
\label{sec:7}
Here we follow the same strategy as in the previous  section,  and derive the complete amplitude from merging   two-particle cuts. As we will see,  this procedure will now give rise to  three tensor boxes with different particle orderings as before, with numerators that are up to  quartic order in the loop momenta. These will then be Passarino-Veltman reduced as usual.

We now compute  the three possible two-particle cuts of the amplitude. We also include the usual factor of two  from  swapping $\phi$ and $\bar{\phi}$ in the loop. 
The $s$-cut is given by
\begin{align}
\begin{split}
\textrm{{\it s}-cut:}&\pic{0.35}{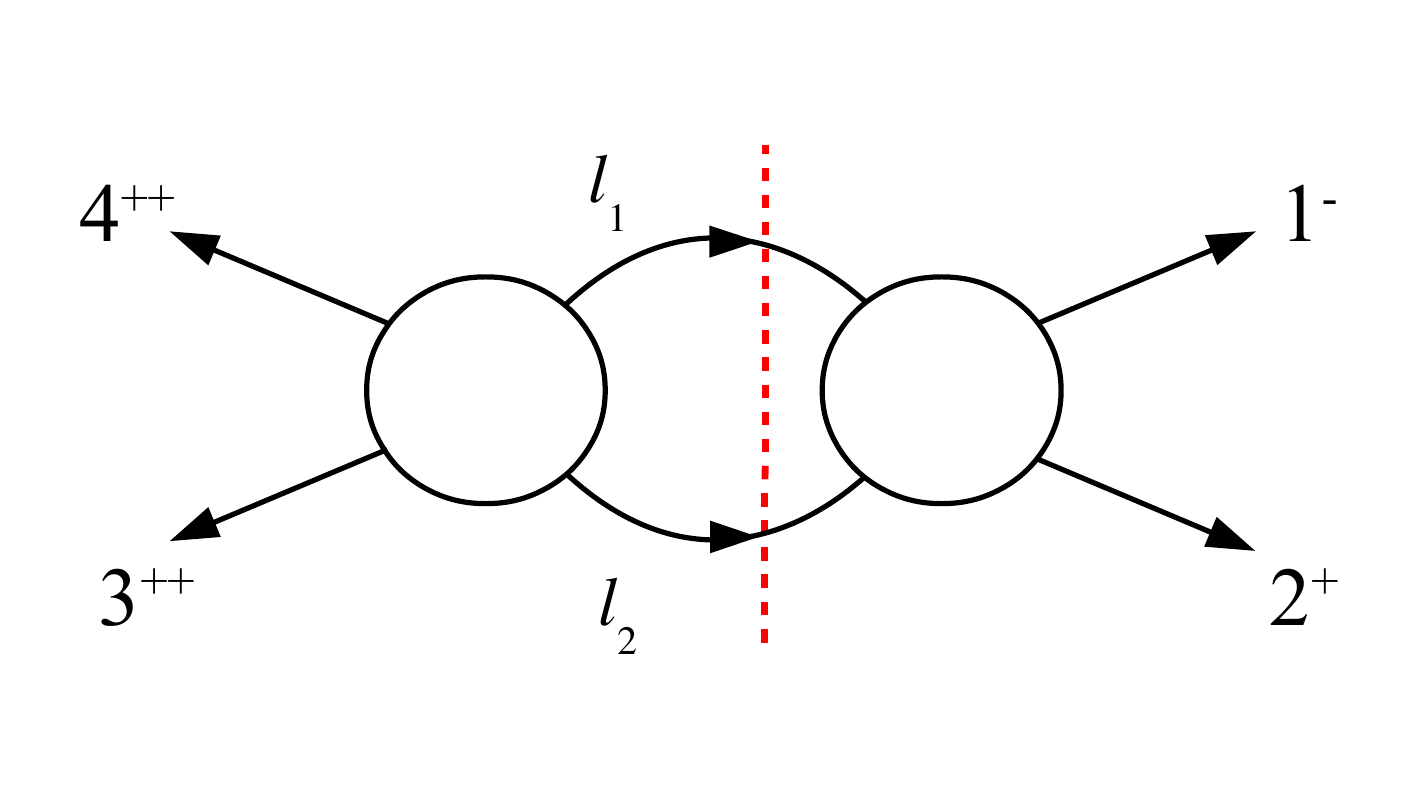} = \ 
2 \mu^4 {[34]^2\over \langle 34 \rangle^2} {\langle 1|l_1|2]^2\over s}
 \\
&\cdot  \Big[
\pic{0.35}{box1234-s-cut} +
\pic{0.35}{box1234-s-cut-swapped} + 
\pic{0.35}{box1243-s-cut} +
\pic{0.35}{box1243-s-cut-swapped} \Big] \ ,
\end{split}
\end{align}
arising from  $ A(3^{++},4^{++},l_{1,\phi},l_{2,\bar{\phi}})\big[ A(1^{-},2^{+},-l_{2,\phi},-l_{1,\bar{\phi}})+
A(2^{+},1^{-},-l_{2,\phi},-l_{1,\bar{\phi}})
\big]$. Again, the appearance of two terms on the right-hand side of the cut, with two different gluon orderings, is due to the fact that the amplitude on the left-hand side of the cut contains a colorless external state. 
The next cut to look at is: 
\begin{align}
\begin{split}
\textrm{{\it t}-cut:}&\pic{0.35}{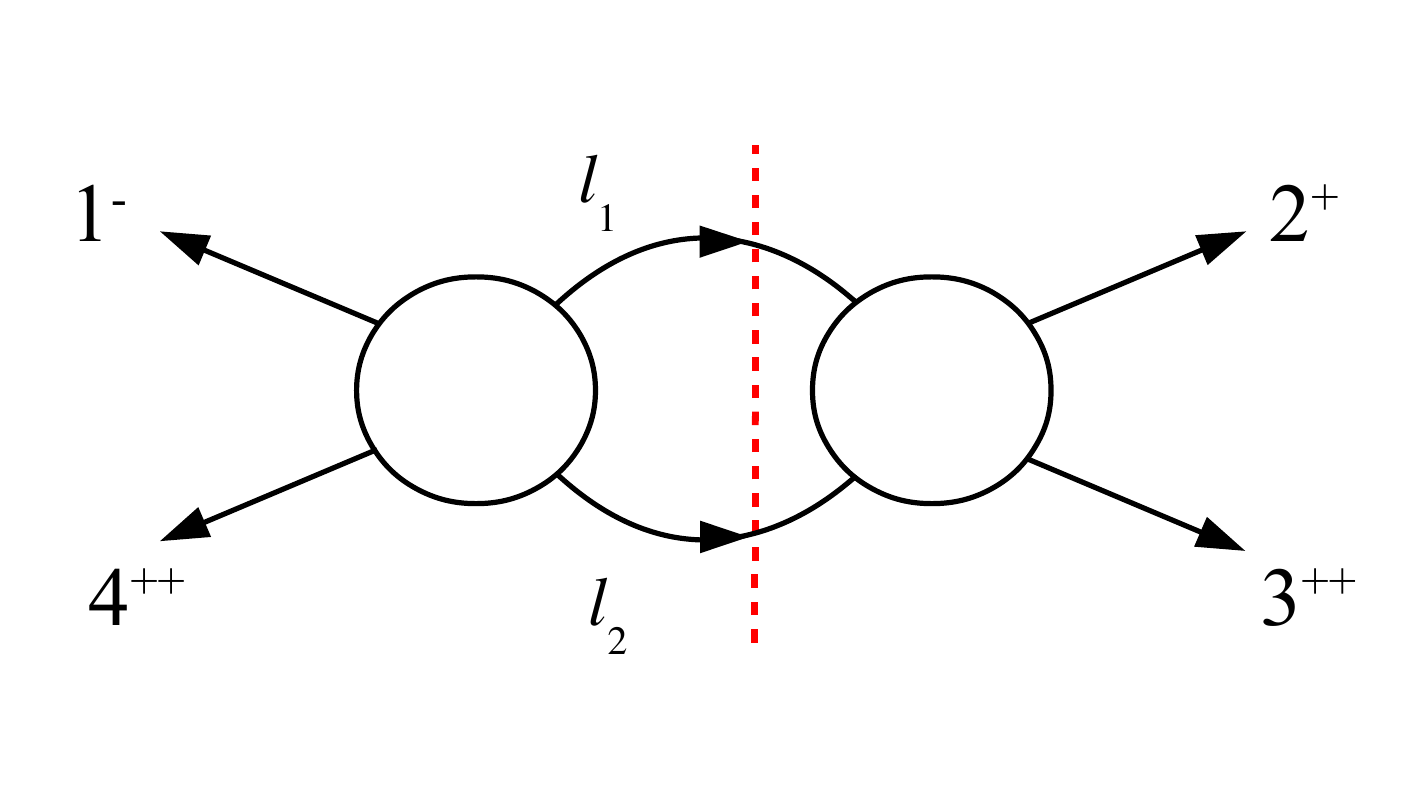} = \ 
\  2 \mu^2 {[32]\over \langle 32 \rangle^2 \langle 14 \rangle }
{\langle 1 |l_1 |4]^3
\langle 2 |l_1 |3] \over t}
\\ 
& \cdot \Big[
\pic{0.35}{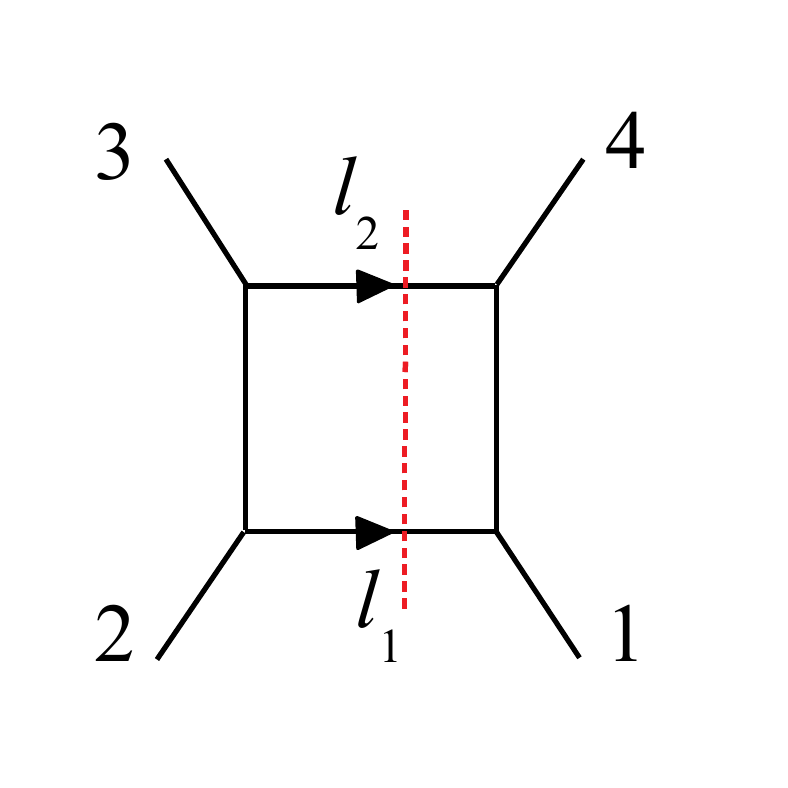} + 
\pic{0.35}{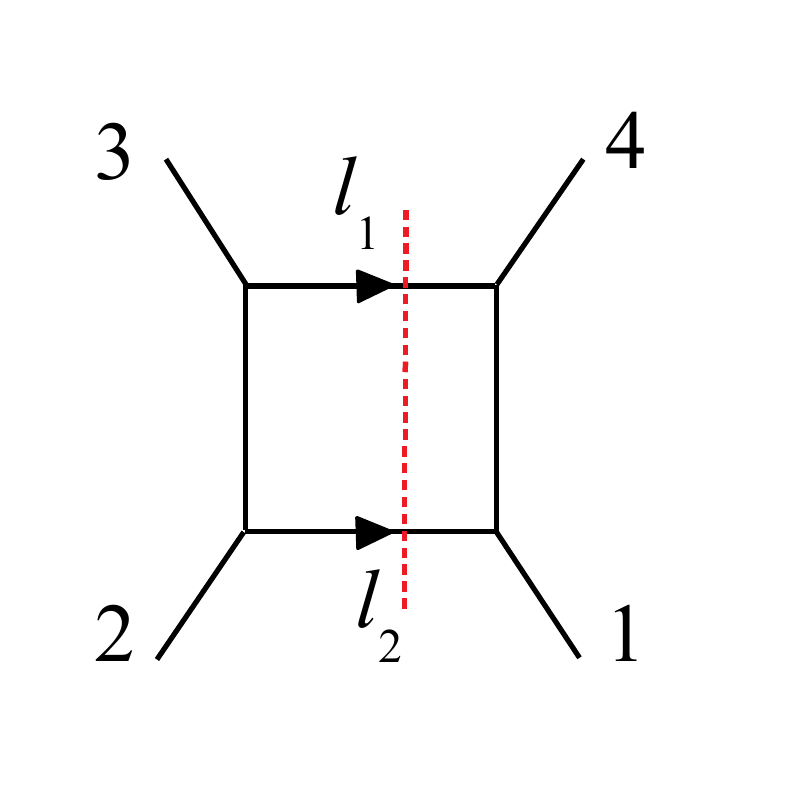} + 
\pic{0.35}{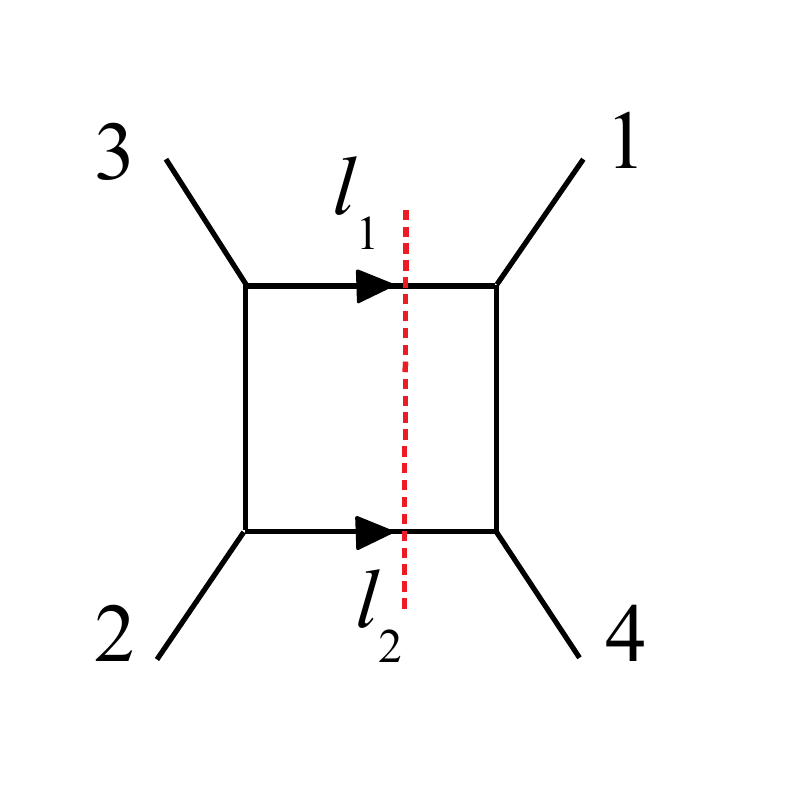}  + 
\pic{0.35}{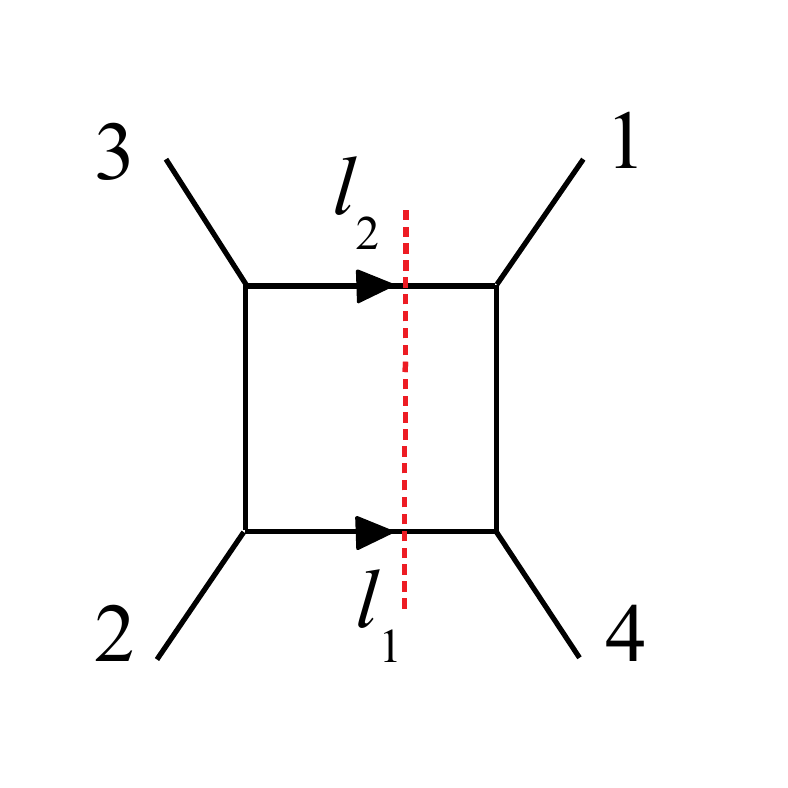} \Big]\ , 
\end{split}
\end{align}
obtained from $ A(4^{++},1^{-},l_{1,\phi},l_{2,\bar{\phi}})A(2^{+},3^{++},-l_{2,\phi},-l_{1,\bar{\phi}})$. Finally, 
\begin{align}
\begin{split}
\textrm{{\it u}-cut:}&\pic{0.35}{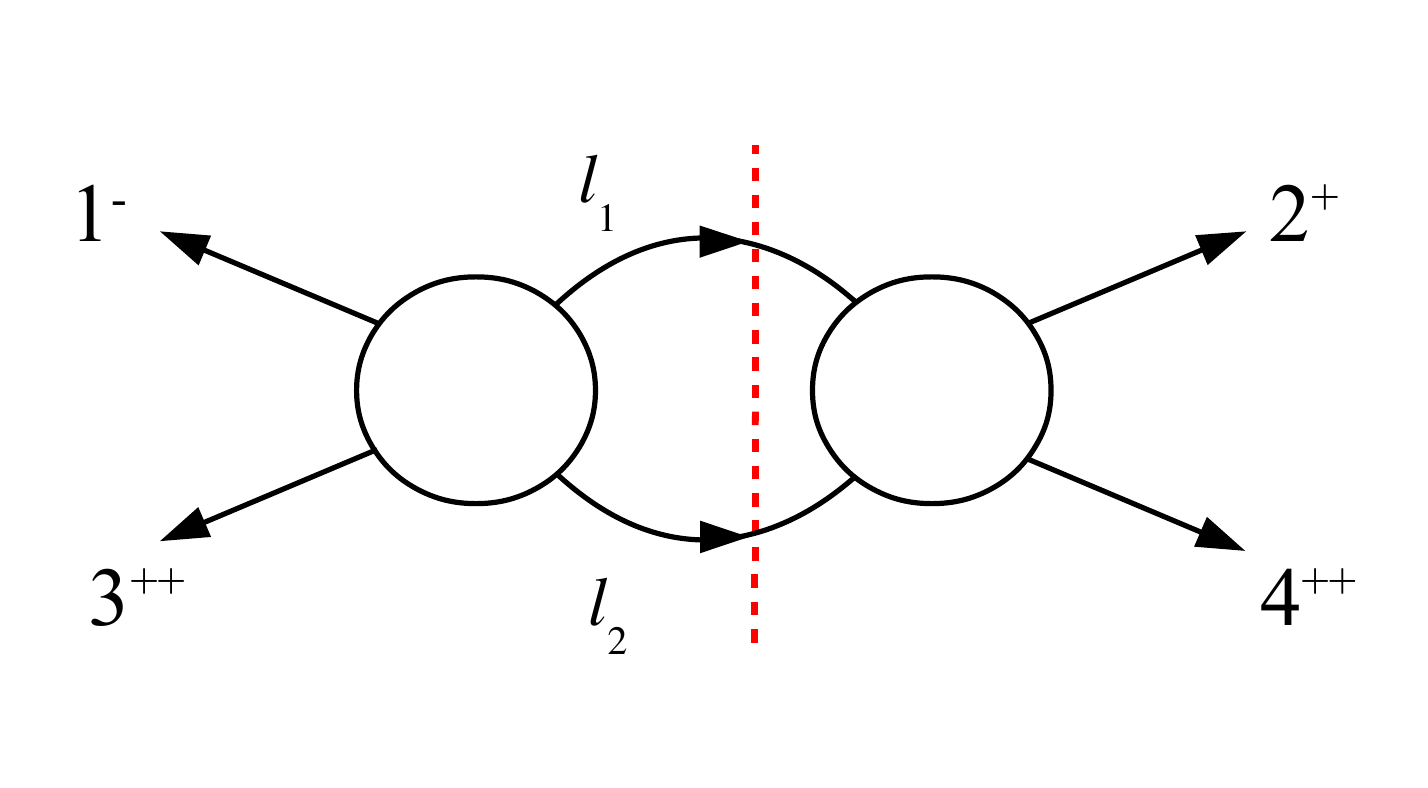} = \ 
\  2 \mu^2 {[42]\over \langle 42 \rangle^2 \langle 13 \rangle }
{\langle 1 |l_1 |3]^3
	\langle 2 |l_1 |4] \over u}
\\
&\cdot \Big[
\pic{0.35}{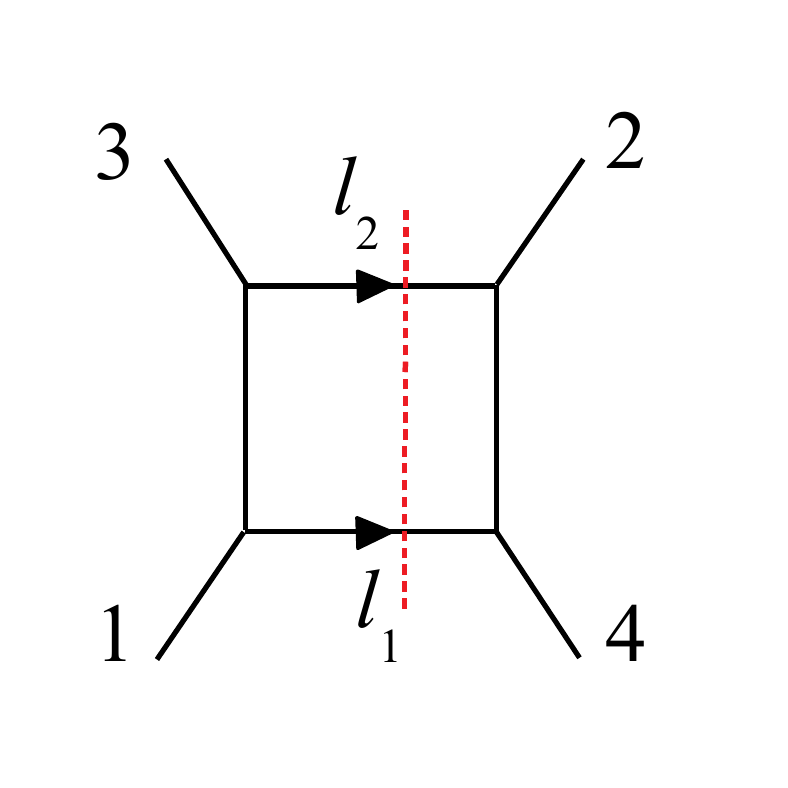} + 
\pic{0.35}{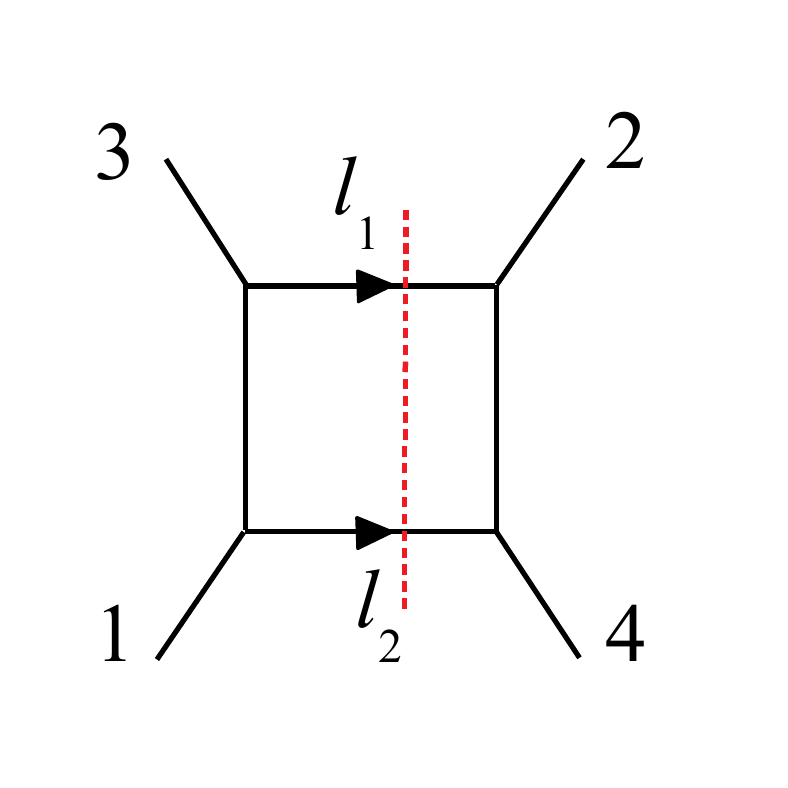} + 
\pic{0.35}{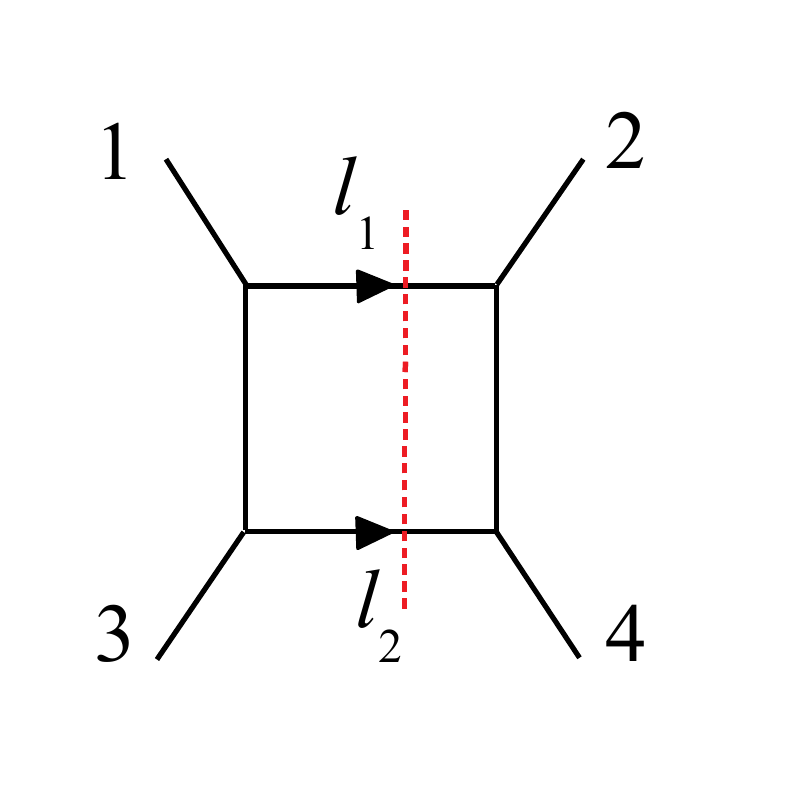}  + 
\pic{0.35}{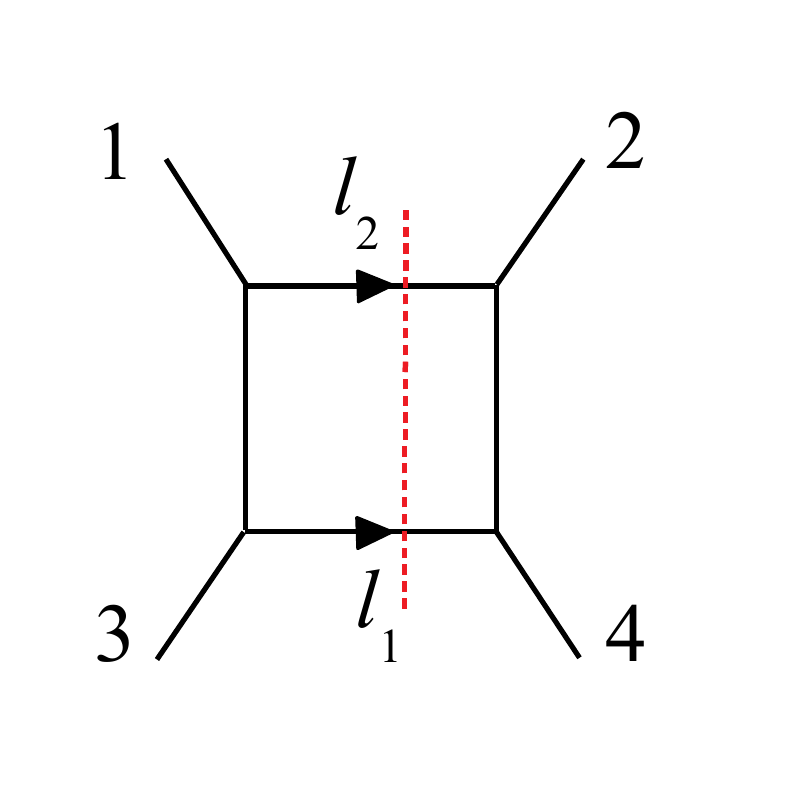} \Big]\ ,
\end{split} 
\end{align}
from $ A(3^{++},1^{-},l_{1,\phi},l_{2,\bar{\phi}})A(2^{+},4^{++},-l_{2,\phi},-l_{1,\bar{\phi}})$.
We also  define a  convenient spinor prefactor which has the correct spinor weights for the given amplitude:
\be
\label{J2}
\cJ=\frac{\bra(2,4)^2\bra(3,4)^2\ket(1,4)^2}{\ket(3,4)^2}\ .
\ee
We are now ready to  merge the different cuts. 
From the topology (1234) we  get: 
\begin{align}
\begin{split}
(1234) &:\ \  {4 i \over (4\pi)^{2-\eps}} \, \cJ \, 
\Biggl \{ - I_{4}[\mu^{6};s,t]\Bigl(\frac{1}{ut}\Bigr) -I_{4}[\mu^{4};s,t]\Bigl(\frac{s}{2u^2}\Bigr) \\
&+  I_{3}[\mu^{4}; t]\Bigl(\frac{s^2(2s+3t)}{u^2t^3}\Bigr) +I_{3}[\mu^{4}; s]\Bigl(\frac{-(2s+t)}{su^2}\Bigr)\\
 &+  I_{2}[\mu^{4}; t]\Bigl(\frac{(t-2s)(4s+3t)}{3ut^4}\Bigr) +I_{2}[\mu^{4}; s]\Bigl(\frac{(s+2t)}{uts^2}\Bigr)+  I_{2}[\mu^{2}; t]\Bigl(\frac{s}{3t^3}\Bigr)
 \Biggr \}.
 \label{eq:T1234}
 \end{split}
\end{align}
The box topology (1243) is simply obtained from the topology (1234) in \eqref{eq:T1234} by swapping $3\leftrightarrow 4$, or  $(s, t, u)\to (s, u, t)$.
 Note that $\cJ$ is invariant under this swap, hence  the  result for the $(1243)$ topology  is immediately found to be: 
\begin{align}
\begin{split}
(1243) &:  {4 i \over (4\pi)^{2-\eps}} \, \cJ \, 
\Biggl \{ - I_{4}[\mu^{6};s,u]\Bigl(\frac{1}{ut}\Bigr) -I_{4}[\mu^{4};s,u]\Bigl(\frac{s}{2t^2}\Bigr) \\
&+  I_{3}[\mu^{4}; u]\Bigl(\frac{s^2(2s+3u)}{t^2u^3}\Bigr) +I_{3}[\mu^{4}; s]\Bigl(\frac{-(2s+u)}{st^2}\Bigr)\\
 &+  I_{2}[\mu^{4}; u]\Bigl(\frac{(u-2s)(4s+3u)}{3tu^4}\Bigr) +I_{2}[\mu^{4}; s]\Bigl(\frac{(s+2u)}{uts^2}\Bigr)+  I_{2}[\mu^{2}; u]\Bigl(\frac{s}{3u^3}\Bigr)
 \Biggr \}.
 \label{eq:T1243}
 \end{split}
\end{align}
Note that in \eqref{eq:T1234} and \eqref{eq:T1243} the $I_{2}[\mu^{2}]$ functions only appear   in the $u$- and $t$-channel.

The last  topology is  (1324), for which we obtain 
\begin{align}
\begin{split}
(1324) &:  {4 i \over (4\pi)^{2-\eps}} \, \cJ \,  
\Biggl \{ - I_{4}[\mu^{6};u,t]\Bigl(\frac{1}{ut}\Bigr) -I_{4}[\mu^{4};u,t]\Bigl(\frac{2}{s}\Bigr) 
\\ &
+  I_{3}[\mu^{4}; t]\Bigl(\frac{-2(3t^2+3ut+u^2)}{st^3}\Bigr) +I_{3}[\mu^{4}; u]\Bigl(\frac{-2(3u^2+3ut+t^2)}{su^3}\Bigr)
 \\
 &+  I_{2}[\mu^{4}; t]\Bigl(\frac{(t+4u)(3t+2u)}{3ut^4}\Bigr) +I_{2}[\mu^{4}; u]\Bigl(\frac{(4t+u)(2t+3u)}{3tu^4}\Bigr)\\
 & -I_{4}[\mu^{2};u,t]\Bigl(\frac{ut}{2s^2}\Bigr) 
+  I_{3}[\mu^{2}; u]\Bigl(\frac{u}{s^2}\Bigr)+  I_{3}[\mu^{2}; t]\Bigl(\frac{t}{s^2}\Bigr)-  I_{2}[\mu^{2}; t]\Bigl(\frac{11t^2+7ut+2u^2}{6st^3}\Bigr)\\
& -I_{2}[\mu^{2}; u]\Bigl(\frac{2t^2+7ut+11u^2}{6su^3}\Bigr)
 \Biggr \}.
 \label{eq:T1324}
 \end{split}
\end{align}
The expression \eqref{eq:T1324} is symmetric in $u\leftrightarrow t$.

Finally we take the four-dimensional limit of  \eqref{eq:T1234},  \eqref{eq:T1243} and  \eqref{eq:T1324} using \eqref{hdint}, thus getting 
\begin{align} 
{i \over (4\pi)^{2}} \, \cJ\, {(t+2u )\over 30 \, t^2}\, , \qquad 
 {i \over (4\pi)^{2}} \, \cJ\,  {(u+2t) \over 30 \, u^2}\, , \qquad 
{i \over (4\pi)^{2}} \, \cJ\,  {s^3 \over 15\,  u^2\, t^2}\, , 
\end{align}
respectively. 
Thus, we arrive at the final result for the four-dimensional limit of the amplitude (using the expression of $\cJ$ in \eqref{J2}):
\beq
A^{(1)}(1^-, 2^+; 3^{++}, 4^{++}) \ = \   {i \over (4 \pi)^2} \,\frac{\bra(2,4)^2\bra(3,4)^2\ket(1,4)^2}{\ket(3,4)^2} \, 
{s \over 6\,  t\, u}\,
\ . 
\eeq
The $D$-dimensional answer is easily obtained  by adding \eqref{eq:T1234}, \eqref{eq:T1243} and \eqref{eq:T1324}.

\section{The  {\normalfont$\langle 1^+\, 2^{+} \,3^{++} \,4^{--}\rangle$} amplitude}
\label{sec:8}

We proceed similarly to the previous sections and  study all two-particle cuts of this amplitude.  As in earlier examples, we find three box topologies with tensor numerators. In this case, an appropriate  spinor prefactor which has the correct spinor weights for the given amplitude is
\be
\label{JL}
\cJ=\frac{\bra(1,2)\bra(1,3)^4\ket(1,4)^4}{\ket(1,2)} \ .
\ee
We construct the two-particle cuts of this amplitude using the tree-level expressions in  Section \ref{relevanttrees}.
The corresponding cuts  will again give rise to  three tensor boxes with different particle orderings and numerators which are  now quartic in the loop momenta. The expression of the relevant cut diagrams  are:
\begin{align}
\begin{split}
\text{$s$-cut}: \quad& A(3^{++},4^{--},l_{1,\phi},l_{2,\bar{\phi}})\ \big[ A(1^{+},2^{+},-l_{2,\phi},-l_{1,\bar{\phi}}) \, + 
\, 1\leftrightarrow 2\big]\, ,\\
\text{$t$-cut}:\quad& A(4^{--},1^{+},l_{1,\phi},l_{2,\bar{\phi}})\ A(2^{+},3^{++},-l_{2,\phi},-l_{1,\bar{\phi}})\, , \\
\text{$u$-cut}:\quad& A(3^{++},1^{+},l_{1,\phi},l_{2,\bar{\phi}})\ A(2^{+},4^{--},-l_{2,\phi},-l_{1,\bar{\phi}}) \ .
\end{split}
\end{align}
As in the cases studied in  Sections \ref{sec:6} and \ref{sec:7}, the $s$-cut integrand  includes the sum of two  color-ordered tree amplitudes on the right-hand side of the cut, which contribute to the same color-ordered amplitude, given that the external state on the left-hand side of the cut is colorless.  
Using the expressions of the relevant tree-level amplitudes and including a factor of two from the two possible assignments from the internal scalar fields, we obtain the following expressions for the cuts: 
\begin{align}
\begin{split}
\textrm{{\it s}-cut:}&\pic{0.35}{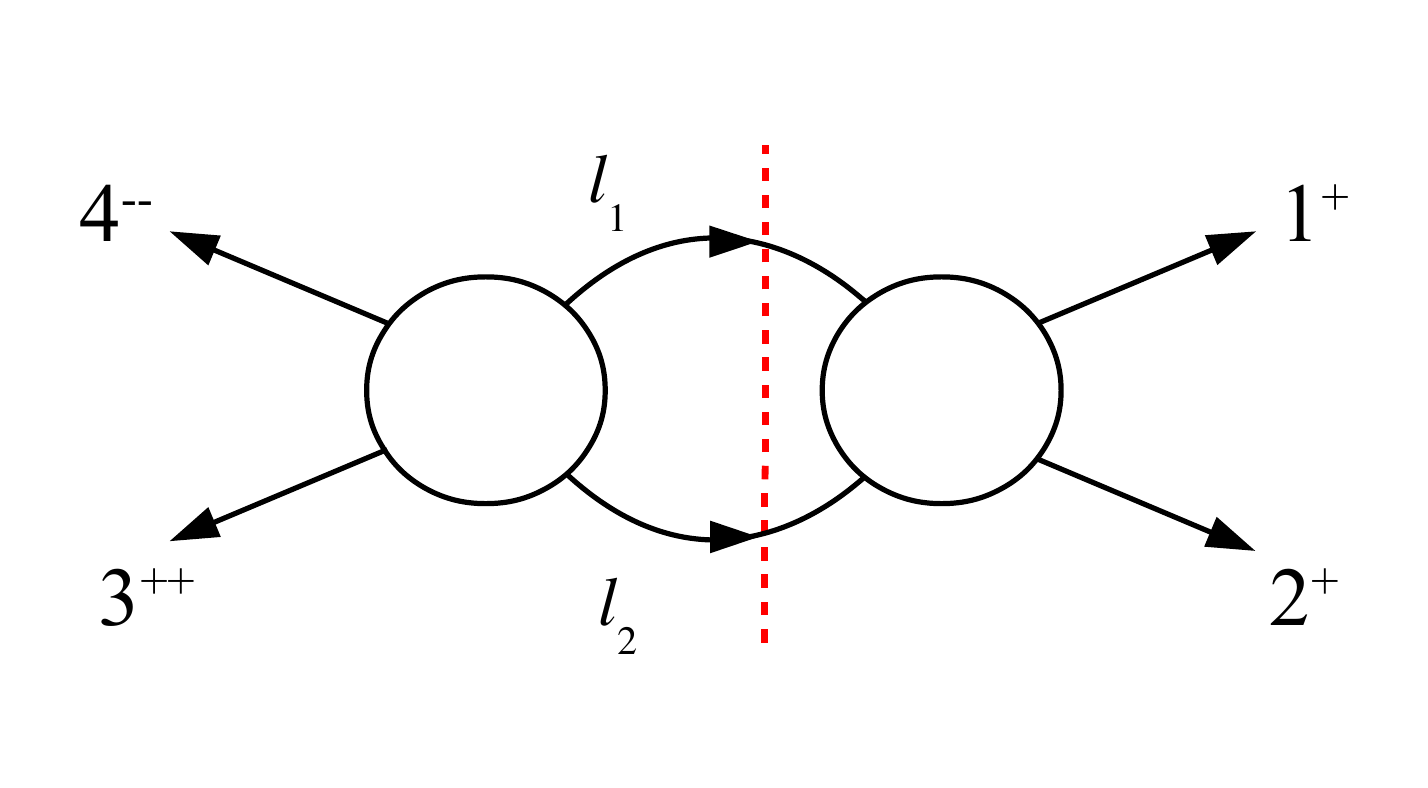} = \ 
 2 \mu^4 {[12]\over \langle 12 \rangle} {\langle 4|l_1|3]^4\over s^2}
 \\ 
& \cdot \Big[
\pic{0.35}{box1234-s-cut}  +
\pic{0.35}{box1234-s-cut-swapped} + 
\pic{0.35}{box1243-s-cut}+
\pic{0.35}{box1243-s-cut-swapped} \Big]
\ , 
\end{split}
\end{align}
\begin{align}
\begin{split}
\textrm{{\it t}-cut:}&\pic{0.35}{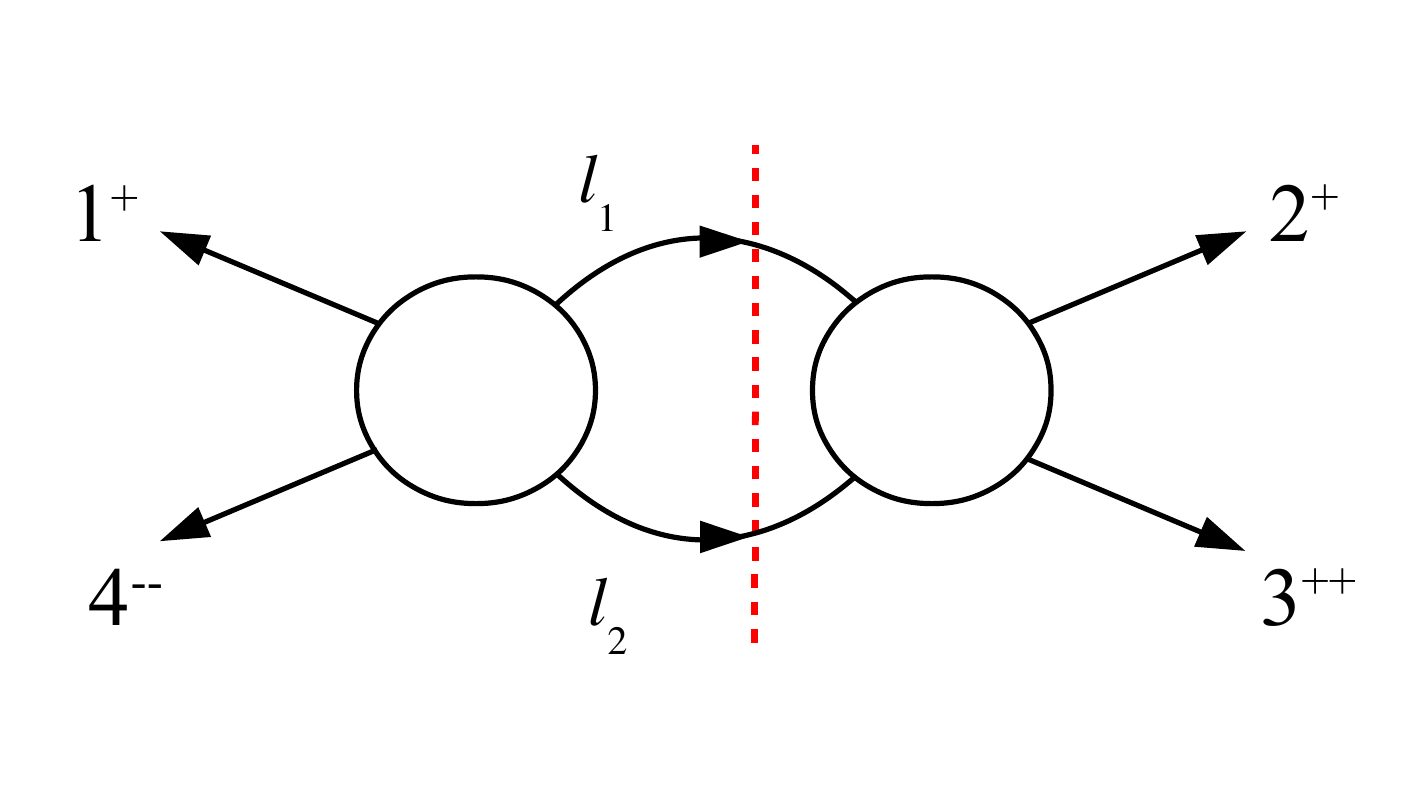} = \ 
\  2 \mu^2 {[32]\over \langle 32 \rangle^2 [41] }
{\langle 4 |l_1 |1]^3
	\langle 2 |l_1 |3] \over t}
\\ 
& \cdot \Big[
\pic{0.35}{sec7-tcut-d1} + 
\pic{0.35}{sec7-tcut-d2} + 
\pic{0.35}{sec7-tcut-d3}  + 
\pic{0.35}{sec7-tcut-d4} \Big]\ , 
\end{split}
\end{align}
and finally, 
\begin{align}
\begin{split}
\textrm{{\it u}-cut:}&\pic{0.35}{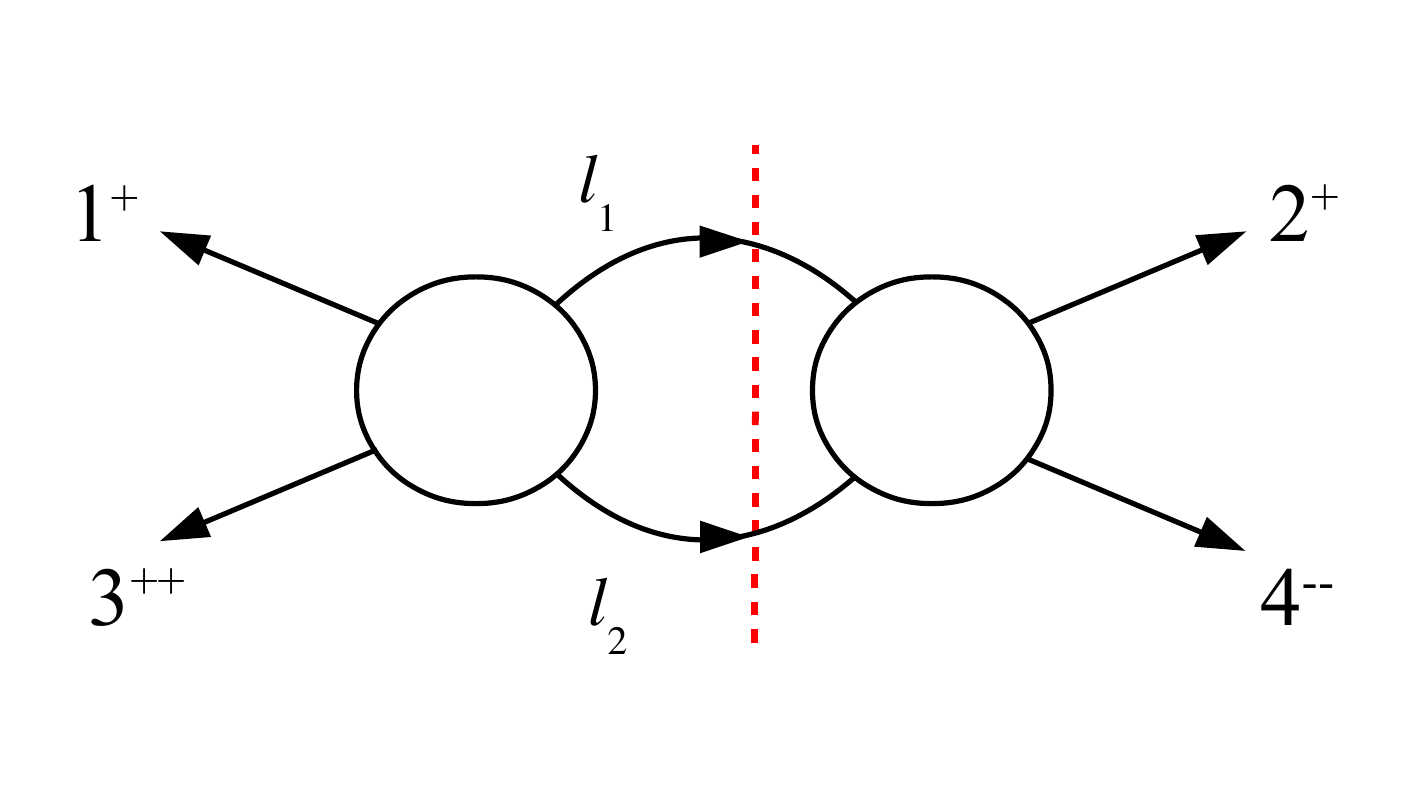} = \ 
\ 2 \mu^2 {[31]\over \langle 31 \rangle^2 [42] }
{\langle 4 |l_1 |2]^3
	\langle 1 |l_1 |3] \over u}
\\
&\cdot \Big[
\pic{0.35}{sec7-ucut-d1} + 
\pic{0.35}{sec7-ucut-d2} + 
\pic{0.35}{sec7-ucut-d3}  + 
\pic{0.35}{sec7-ucut-d4} \Big]\ .
\end{split} 
\end{align}
As usual, we now merge the cuts focusing separately on the three different box integrals. 
Merging the  $s$- and $t$-cut for the topology (1234) we get: 
\begin{align}
\begin{split}
(1234) 
&: 
\  {4 i \over (4\pi)^{2-\eps}} \, \cJ \, 
\Biggl \{ - I_{4}[\mu^{6};s,t]\Bigl(\frac{1}{u^2t^2}\Bigr) -I_{4}[\mu^{4};s,t]\Bigl(\frac{2s}{tu^3}\Bigr) 
\\
&+  I_{3}[\mu^{4}; t]\Bigl(\frac{2(t^3+u^3)}{t^4u^3}\Bigr) +I_{3}[\mu^{4}; s]\Bigl(\frac{2(6s^2+8st+3t^2)}{s^3u^3}\Bigr)
\\
&+  I_{2}[\mu^{4}; t]\Bigl(\frac{(2u-t)(4u+3t)}{3t^5u^2}\Bigr) +I_{2}[\mu^{4}; s]\Bigl(\frac{(s+2t)(3s^2-8st-8t^2)}{3s^4t^2u^2}\Bigr)
\\
&- I_{4}[\mu^{2};s,t]\Bigl(\frac{s^2}{2u^4}\Bigr)
\\
&
+I_{3}[\mu^{2}; t]\Bigl(\frac{s}{u^4}\Bigr)-I_{3}[\mu^{2}; s]\Bigl(\frac{(2s+t)(2s^2+2st+t^2)}{s^2u^4}\Bigr)
\\ 
&
-I_{2}[\mu^{2}; t]\Bigl(\frac{s(6t^2-3tu+2u^2)}{6t^4u^3}\Bigr)
+I_{2}[\mu^{2}; s]\Bigl(\frac{11s^3+59s^2t+64st^2+22t^3}{6s^3tu^3}\Bigr)
\Biggr \}.
\label{eq:T1234B}
\end{split}
\end{align}
The topology (1243) can  be  obtained  by swapping $3\leftrightarrow 4$ in \eqref{eq:T1234B}, or  $(s,t,u) \to (s, u, t)$. Noting  that $\cJ$ is invariant under this swap we get: 
\begin{align}
\begin{split}
(1243) &: \  {4 i \over (4\pi)^{2-\eps}} \, \cJ \, 
\Biggl \{ - I_{4}[\mu^{6};s,u]\Bigl(\frac{1}{u^2t^2}\Bigr) -I_{4}[\mu^{4};s,u]\Bigl(\frac{2s}{ut^3}\Bigr) 
\\
&+  I_{3}[\mu^{4}; u]\Bigl(\frac{2(t^3+u^3)}{u^4t^3}\Bigr) +I_{3}[\mu^{4}; s]\Bigl(\frac{2(6s^2+8su+3u^2)}{s^3t^3}\Bigr)\\
&+  I_{2}[\mu^{4}; u]\Bigl(\frac{(2t-u)(4t+3u)}{3u^5t^2}\Bigr) +I_{2}[\mu^{4}; s]\Bigl(\frac{(s+2u)(3s^2-8su-8u^2)}{3s^4t^2u^2}\Bigr)
\\
&
- I_{4}[\mu^{2};s,u]\Bigl(\frac{s^2}{2t^4}\Bigr)
+I_{3}[\mu^{2}; u]\Bigl(\frac{s}{t^4}\Bigr)-I_{3}[\mu^{2}; s]\Bigl(\frac{(2s+u)(2s^2+2su+u^2)}{s^2t^4}\Bigr)
\\
&-I_{2}[\mu^{2}; u]\Bigl(\frac{s(6u^2-3tu+2t^2)}{6u^4t^3}\Bigr)\
+\ I_{2}[\mu^{2}; s]\Bigl(\frac{11s^3+59s^2u+64su^2+22u^3}{6s^3ut^3}\Bigr)
\Biggr \}.
\label{eq:T1243B}
\end{split}
\end{align}
%
%
Next,  we merge the $u$- and $t$-cuts for  the topology (1324): 
\begin{align}
\begin{split}
(1324) &:\  {4 i \over (4\pi)^{2-\eps}} \, \cJ \, 
\Biggl \{ - I_{4}[\mu^{6};u,t]\Bigl(\frac{1}{u^2t^2}\Bigr) -I_{4}[\mu^{4};u,t]\Bigl(\frac{1}{2sut}\Bigr) 
+  I_{3}[\mu^{4}; t]\Bigl(\frac{2u+3t}{st^4}\Bigr) \\
&+I_{3}[\mu^{4}; u]\Bigl(\frac{2t+3u}{su^4}\Bigr)
+  I_{2}[\mu^{4}; t]\Bigl(\frac{(t-2u)(3t+4u)}{3u^2t^5}\Bigr) +I_{2}[\mu^{4}; u]\Bigl(\frac{(u-2t)(4t+3u)}{3t^2u^5}\Bigr)\\
&
+  I_{2}[\mu^{2}; u]\Bigl(\frac{s}{3u^4t}\Bigr)+  I_{2}[\mu^{2}; t]\Bigl(\frac{s}{3ut^4}\Bigr)
\Biggr \}.
\label{eq:T1324B}
\end{split}
\end{align}
As expected, the expression \eqref{eq:T1324B} is symmetric in $u\leftrightarrow t$.

Finally we take the four-dimensional limit of \eqref{eq:T1234B}, \eqref{eq:T1243B} and  \eqref{eq:T1324B}. These are given by 
\begin{align}
-{i\over (4\pi)^2}\, \cJ\, {(3t^2+ut+u^2 )\over 15 \,  s^2\, t^3}\, , \qquad 
-{i\over (4\pi)^2}\, \cJ\,{(3u^2+ut+t^2) \over 15 \,  s^2\, u^3}\, , \qquad 
-{i\over (4\pi)^2}\, \cJ\, {s(2t^2+ut+2u^2) \over 30\,  u^3\, t^3}\, , 
\end{align}
respectively. 
Thus, we arrive at the final result for the four-dimensional limit of the  amplitude, using the expression for $\cJ$ in \eqref{JL}, 
\beq
A^{(1)}(1^+, 2^+; 3^{++}, 4^{--}) \ = \   {i \over (4 \pi)^2} \,\frac{\bra(1,2)\bra(1,3)^4\ket(1,4)^4}{\ket(1,2)}\, 
{t^2+u^2 \over 6\,s \,  t^2\, u^2}\,
\ . 
\eeq
The $D$-dimensional result is obtained by adding \eqref{eq:T1234B},  \eqref{eq:T1243B} and  \eqref{eq:T1324B}.

\section{The  {\normalfont$\langle 1^{++} \,2^{++} \,3^{++}\, 4^\pm\, \rangle$} amplitudes}
\label{sec:3plusG1pmg}

We now move on to consider the one-loop amplitudes with three gravitons and a gluon, beginning with the amplitude with  three same-helicity gravitons and one gluon. 
It is easy to show that this amplitude vanishes upon integration. 
Consider for instance the $s$-cut diagram of the $\langle 1^{++} \,2^{++} \,3^{++}\, 4^+\rangle$ amplitude. Its expression is 
\begin{align}
\begin{split}
 \mu^2 {[34]\over \langle 34 \rangle^2} 
\langle 4 | l_2 |3] \ \Big[ {i\over (l_2 + p_3)^2 - \mu^2} + l_1 \leftrightarrow l_2 \Big]
 \
 \mu^4 {[12]^2 \over \langle 12 \rangle^2} 
\Big[ {-i\over (l_1 - p_1)^2 - \mu^2} + p_1 \leftrightarrow p_2 \Big]\, , 
\end{split}
\end{align}
or 
\begin{align}
\begin{split}
\textrm{{\it s}-cut:}&\pic{0.35}{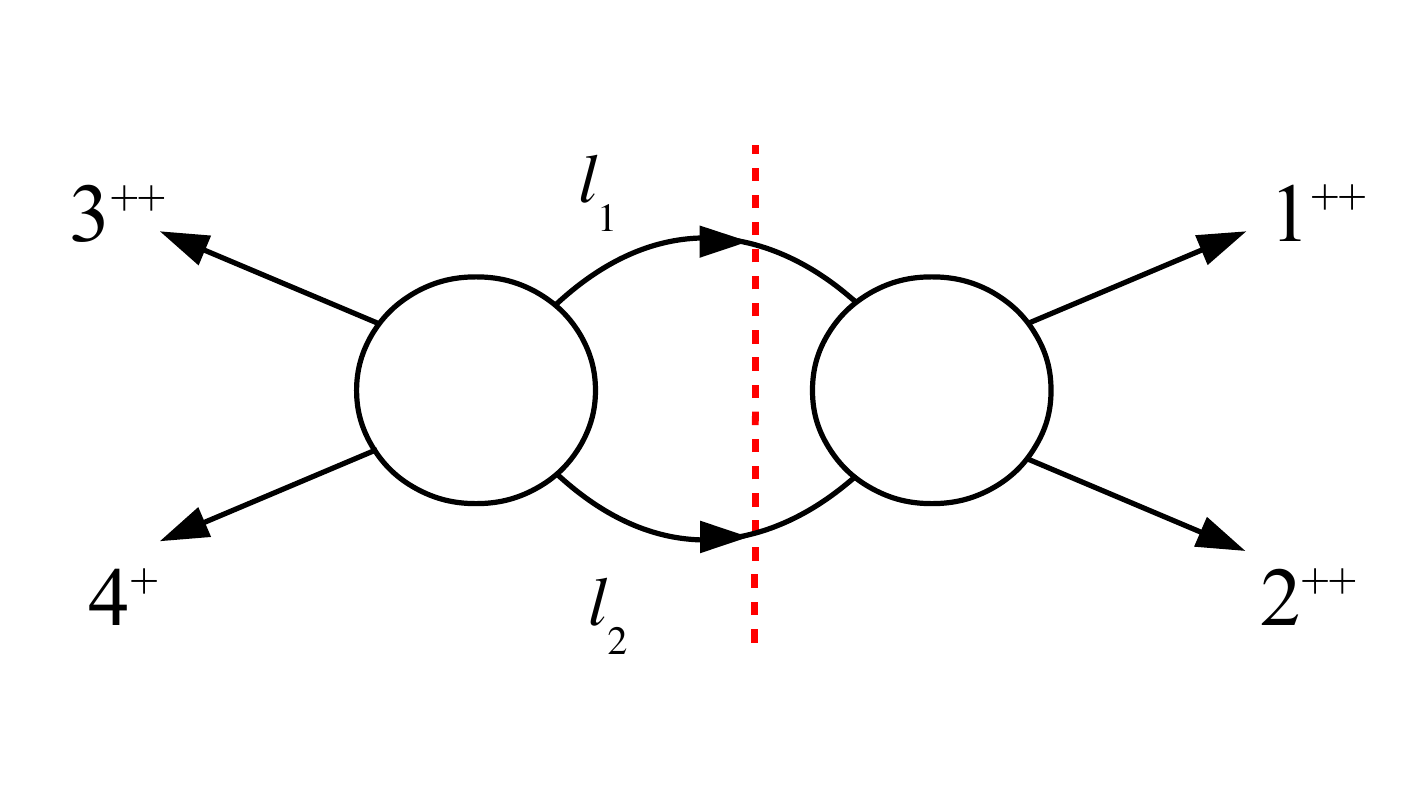} = \ 
 - \mu^2 {[34]\over \langle 34 \rangle^2} 
\langle 4 | l_2 |3] \  \mu^4 {[12]^2 \over \langle 12 \rangle^2} \ \cdot 
\\
&\Big[
\pic{0.35}{box1234-s-cut} \ + 
\pic{0.35}{box1234-s-cut-swapped} \ + 
\pic{0.35}{box1243-s-cut}
\pic{0.35}{box1243-s-cut-swapped} 
 \Big]
\ .
\end{split}
\end{align}
Although the integrand does not vanish, the integrated expression does because it is an odd function under the exchange of $l_1 \leftrightarrow l_2$. The $t$- and $u$-cut are simply given by permutations of the $s$-cut and hence combining the three cuts one obtains a vanishing integrated expression. 
Finally, using \eqref{1g1h} it is immediate to see that also the $\langle 1^{++} \,2^{++} \,3^{++}\, 4^-\rangle$ amplitude vanishes for the same reason. In conclusion, 
\beq
A^{(1)}( 1^{++} ,2^{++} ,3^{++}; 4^\pm) \ = \ 0\ . 
\eeq

\section{The  {\normalfont$\langle 1^+\, 2^{++} \,3^{++} \,4^{--}\rangle$} amplitude}
\label{sec:10}

Similarly to the previous section, we can easily show that the amplitude $\langle 1^+\, 2^{++} \,3^{++} \,4^{--}\rangle$ vanishes upon integration. 
Consider for instance its $s$-channel cut. This is given by 
\begin{align}
&\,\textrm{{\it s}-cut:}
\pic{0.40}{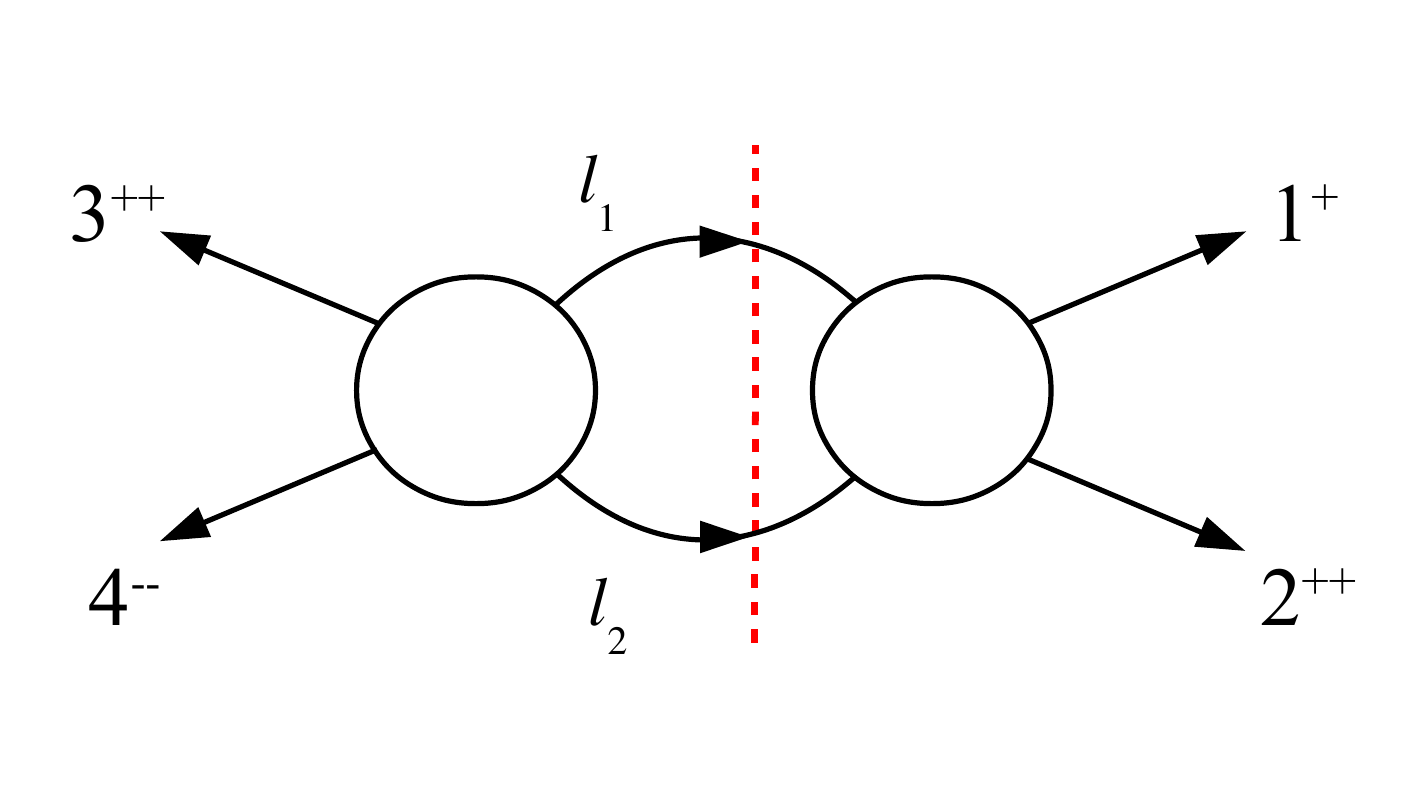} 
= A(3^{++},4^{--},l_{1,\phi},l_{2,\bar{\phi}})A(1^{+},2^{++},-l_{2,\phi},-l_{1,\bar{\phi}})\nonumber\\
&=\Biggl[{-\langle 4 | l_2 |3]^4 \over s^2}
\ \Big[ {i\over (l_2 + p_3)^2 - \mu^2} + l_1 \leftrightarrow l_2 \Big]\Biggr]
 \Biggl[\mu^2 {[21] \over \langle 21 \rangle^2}  \langle 1 |- l_1 |2]
\Big[ {i\over (l_1 - p_2)^2 - \mu^2} + l_1 \leftrightarrow l_2 \Big]\Biggr]\nonumber\\
&=\ 
  \mu^2 {[21]\over \langle 21 \rangle^2} 
 {\langle 4 | l_2 |3]^4 \langle 1 | l_2 |2]\over s^2} \ \cdot 
\nonumber\\
&\Big[
\pic{0.35}{box1234-s-cut} \ + 
\pic{0.35}{box1234-s-cut-swapped} \ + 
\pic{0.35}{box1243-s-cut}
\pic{0.35}{box1243-s-cut-swapped} 
 \Big]
\ .
\end{align}
Again, the integrated expression is an odd function under $l_1 \leftrightarrow l_2$ and hence it vanishes. The same holds true for the $t$- and $u$- channel cuts. In summary,  we get
\beq
A^{(1)}(1^+; 2^{++} ,3^{++} ,4^{--})=0\, .
\eeq

\section{The  {\normalfont$\langle 1^+\, 2^{+} \,3^{+} \,4^{++}\rangle$} amplitude from the double copy} \label{sec:11}

The color-kinematic duality or double copy \cite{Bern:2008qj,Bern:2010ue} was extended in the works
\cite{Chiodaroli:2014xia,Chiodaroli:2015rdg,Chiodaroli:2016jqw,Chiodaroli:2017ngp} also to the domain of mixed graviton-gluon amplitudes in the Einstein-Yang-Mills theory. In particular \cite{Chiodaroli:2017ngp} 
exposed explicitly how to construct an Einstein-Yang-Mills amplitude  through
a double copy from Yang-Mills and Yang-Mills $+$ $\phi^{3}$ theory:
\be
A_{\text{EYM}} = A_{\text{YM}} \otimes A_{\text{YM}+\phi^{3}} \, .
\label{DCLogic}
\ee
The latter Yang-Mills-Scalar theory contains biadjoint scalars $\phi^{A\hat a}$ next
to the gluons $A^{\hat a}_{\mu}$ and is defined through the Lagrangian
\begin{align}
\begin{split}
\mathcal{L}_{\text{YM}+\phi^{3}}=& -{1\over 4}F_{\mu\nu}^{\hat a}F^{\mu\nu\, \hat a}
+{1\over 2} (D_{\mu}\phi^{A})^{\hat a}(D^{\mu}\phi^{A})^{\hat a} +{ 1\over  3!}\,  \lambda \, g
\, F^{ABC}\, f^{\hat a \hat b\hat c}\, \phi^{A\hat a}\, \phi^{B\hat b}\, \phi^{C\hat c} 
\\ 
& -\frac{g^{2}}{4} \, f^{\hat a \hat b\hat d}\, f^{\hat d \hat c\hat d}\, 
\phi^{A\hat a}\, \phi^{B\hat b}\, \phi^{A\hat c} \, \phi^{B\hat d} \, .
\end{split}
\end{align}
As a one-loop application of \eqn{DCLogic},  we  wish  to derive the vanishing of the  
 {\normalfont$\langle 1^+\, 2^{+} \,3^{+} \,4^{++}\rangle$} amplitude, which we observed with a direct computation in Section \ref{allplus}. Thus we need
  to construct integrands for  the two amplitudes $A^{(1)}(1^{+},2^{+},3^{+},4^{+})$ and $A^{(1)}(1^{A_{1}}_{\phi},
 2^{A_{2}}_{\phi},3^{A_{3}}_{\phi}, 4^{+})$ where color ordering is performed in both cases
 with respect to the hatted  gauge group index. The first one, the all-plus helicity four-gluon
 amplitude, is well-known  and takes the form
 \be
  A^{(1)}(1^{+},2^{+},3^{+},4^{+}) = 
  \pic{0.44}{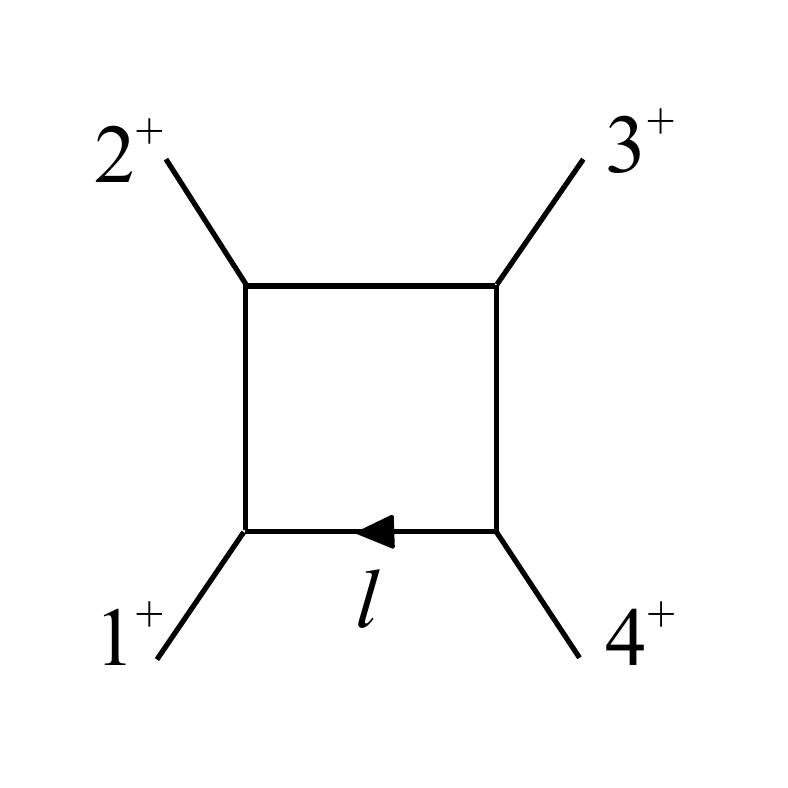}\ \  = \frac{[12][34]}{\vev{12}\vev{34}}\,
 \int\!{d^4l\over (2 \pi)^4}{d^{-2\eps} \mu  \over (2 \pi)^{-2 \eps}}
 \ \frac{\mu^{4}}{D_{0}D_{1}D_{2}D_{3}}\, .
 \ee
 As this is a pure box-integral, in the construction of the one-loop YM $+$ $\phi^{3}$ amplitude integrand
 we only need to construct the box-contribution to the $A^{(1)}(1^{A}_{\phi},
 2^{B}_{\phi},3^{C}_{\phi}, 4^{+})$ amplitude as well:
 \begin{align}
 A^{(1)}(1^{A_{1}}_{\phi}, 2^{A_{2}}_{\phi},3^{A_{3}}_{\phi}, 4^{+})\Bigr |_{\text{boxes}}= 
 \pic{0.44}{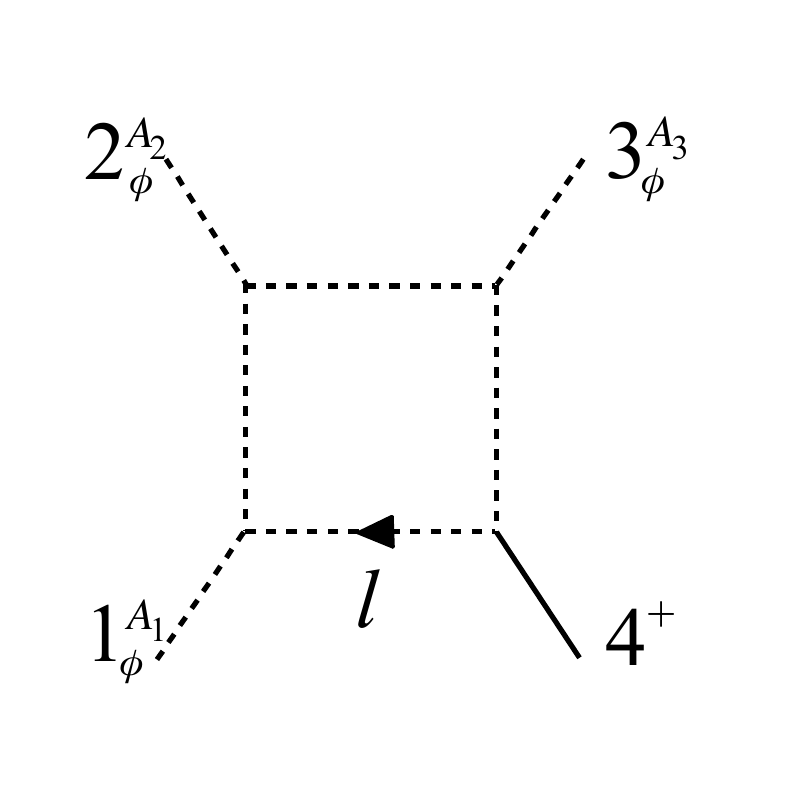} &=  \
i \int\!{d^4l\over (2 \pi)^4}{d^{-2\eps} \mu  \over (2 \pi)^{-2 \eps}}\, 
 \frac{ f^{A_{1}A_{2}A_{3}}}{D_{0}D_{1}D_{2}D_{3}}
  \frac{\langle q |l|4]}{\vev{q4}} 
  \nonumber \\
  &+ \ \text{cycl(1,2,3)}\, .
 \end{align}
Here we have  simply inserted the scalar-scalar-on-shell-gluon vertex of \eqn{gssb} in the
south-east corner with a reference
spinor~$\lambda_{q}$. The numerator emerging from this integrand respects color-kinematics duality
as it is built entirely from three-valent graphs. Employing the double-copy prescription
 \cite{Chiodaroli:2017ngp}  of \eqn{DCLogic} we are therefore led to the following representation
 of the all-plus single-gluon EYM-amplitude
 \be
 A^{(1)}(1_{A_{1}}^{+},2_{A_{2}}^{+},3_{A_{3}}^{+},4^{++}) = 
  if^{A_{1}A_{2}A_{3}}\, \frac{[12][34]}{\vev{12}\vev{34}}\,
 \int\!{d^4l\over (2 \pi)^4}{d^{-2\eps} \mu  \over (2 \pi)^{-2 \eps}}
 \frac{\mu^{4}}{D_{0}D_{1}D_{2}D_{3}}\,
 \frac{\langle q |l|4]}{\vev{q4}} + \text{cycl(1,2,3)}\,  .
 \ee
Passarino-Veltman reducing the integral one arrives at the $D$-dimensional expression
\begin{align}
A^{(1)}(1_{A_{1}}^{+},2_{A_{2}}^{+},3_{A_{3}}^{+},4^{++}) = 
  if^{A_{1}A_{2}A_{3}}\,& \frac{[12][34]}{\vev{12}\vev{34}}\,
  \frac{1}{\vev{q4}}\, \frac{1}{u} \Bigl \{ \ft 1 2 (t \langle q|3|4]-s\langle q|1|4])
  I_{4}[\mu^{4};s,t] \nonumber \\ & + \langle q|2|4]\, ( I_{3}[\mu^{4};s]- I_{3}[\mu^{4};t])
  \Big \}+ \text{cycl(1,2,3)}\, .
  \end{align}
Going to four dimensions  simplifies this result considerably, and one arrives at
\be
A^{(1)}(1_{A_{1}}^{+},2_{A_{2}}^{+},3_{A_{3}}^{+},4^{++}) = 
  if^{A_{1}A_{2}A_{3}}\, \frac{[12][34]}{\vev{12}\vev{34}}\,
  \frac{1}{\vev{q4}}\, \frac{1}{12} \Bigl \{\langle q | 3 |4] + \ft 1 2 \langle q|2|4]\Bigr \}
  + \text{cycl(1,2,3)} =0\, .
\ee
The expression above vanishes as the prefactor is invariant under cyclic shifts in $(1,2,3)$ 
and obviously the bracketed terms sum to zero, as $\langle q | 3|4] + \text{cycl(1,2,3)}\!=\!\langle q |\,  p_{1}+p_{2}+p_{3}\, |4]\!=\!0$.
Hence, we have reproduced the vanishing result of Section \ref{allplus}.

 Finally, we  comment on the question whether the amplitude relations of 
Stieberger and Taylor \cite{Stieberger:2016lng,Stieberger:2015qja} relating pairs
of collinear gluons to gravitons extend to the one-loop level for the one-loop rational  amplitudes we  have considered in this paper. 

We will test this for the simplest case of the all-plus amplitude with one graviton. For such a relation to
be true, the vanishing four-dimensional result must follow from  the specific collinear limit proposed by Stieberger and Taylor on the five-point all-plus rational amplitude in pure Yang-Mills. In  analogy to the tree-level relation, in four dimensions we expect to have:
\beqa
A^{(1)}_{\textsl{\rm EYM;ST}}(1^+,2^+,3^+,P^{++})
\stackrel{?}{=}
 {\kappa\over g^2}\, \mathcal{G}(x) \ \lim_{p_{4}\parallel p_{5}}
 s_{24}A^{(1)}_{\textsl{\rm YM}}(1^+,5^+,2^+,4^+,3^+) \, + \, {\rm cycl(1,2,3)}\, ,
\label{STallplusconjecture}
 \eeqa
where the equality would hold in the collinear limit $ \{p_4\rightarrow x P, \, p_5\rightarrow (1-x)P\} $ on the right-hand side of \eqref{STallplusconjecture},  and 
$\mathcal{G}(x)$ is an  undetermined function of the momentum splitting fraction $x$ which is expected to be independent of the helicities of the particles. Note that $\cG(x)$ has been determined for tree amplitudes in 
\cite{Stieberger:2015qja}. We have also added cyclic permutations of the three gluons to secure cyclic symmetry in these particles. 
 Using the well-known expression for the all-plus five-point rational amplitude in Yang-Mills
\cite{Bern:1993mq}, we see that  the right-hand side  of \eqref{STallplusconjecture} contains the factor 
\beqa
s_{24} A_{\rm YM}^{(1)}(1^+,5^+,2^+,4^+,3^+)=s_{24}\frac{i}{48 \pi^2}\frac{-s_{15}s_{52}-s_{13}s_{43}+\ket (5,2)\ket (4,3)\bra(2,4)\bra (3,5)}{\ket (1,5)\ket (5,2)\ket (2,4)\ket (4,3)\ket (3,1)}\, .
\label{STrelation5points}
\eeqa
Performing the above-mentioned collinear limit on \eqref{STrelation5points}, followed by a cyclic permutation of the three gluon legs in order to reflect the anticipated color structure,  
and  relabelling   $ P\rightarrow p_4 $ (with $p_4$ being the momentum of  the graviton leg),  we arrive at 
\beqa
 \lim_{p_{4}\parallel p_{5}}  s_{24} \, A_{\rm YM}^{(1)}(1^+,5^+,2^+,4^+,3^+)\,+ \, {\rm cycl(1,2,3)}\, = \, 
 \Bigl[\frac{1}{(1-x)}-2x\Bigr]\, \ {1\over 2} (st +ut+su)\, \cA_0\,  ,
 \label{STRHScollinear}
\eeqa
with 
\beqa
\cA_0\, :=\, \frac{i}{48 \pi^2}\frac{\langle 2|1|4]}{\ket(2,4)}\frac{1}{\ket(1,2) \ket(2,3)\ket(3,4)\ket(4,1)}\, .
\label{STcoefficientA0}
\eeqa
Clearly this does not vanish and hence invalidates the conjecture \eqref{STallplusconjecture}. 
However we note the following rather intriguing similarity:  Consider again our 
 $D$-dimensional result for this amplitude as obtained in \eqref{FinalAllPlus}, and focus only on  the pure  box contribution; evaluated in the $D\!\to\!4$ limit, it gives
   \begin{align}
6 \cA_0 
 \,  \Big[\frac{st}{2} I_4[\mu^{4}; s,t] \Big]\Bigg\vert_{D=4} \, + \, \mathrm{perms}\ = \  - {1\over 2} ( s t +  u t  + s u ) \cA_0\, . 
  \label{allplusBoxes}
   \end{align}
This is curiously   proportional to the $x$-independent part of the right-hand side  of \eqref{STRHScollinear}, which was  obtained from the Stieberger-Taylor collinear limit.
Given the vanishing of our final  result in four dimensions, also the triangle contribution in \eqref{FinalAllPlus} can be written in a similar way: 
\beq
\label{triaaa}
6 \, \cA_0 
 \,  \Big[\,s I_3[\mu^{4}; t]\,+\,t I_3[\mu^{4}; s] \Big]\Bigg\vert_{D=4} \, + \, \mathrm{perms}\, =\, {1\over 2} (st +ut+su)\, \cA_0\,  .
\eeq
In conclusion, even though the amplitude \eqref{FinalAllPlus}  vanishes in four dimensions, we find the similarities between 
\eqref{allplusBoxes} (or \eqref{triaaa}) and \eqref{STRHScollinear}
intriguing,  and worth further investigation. 

  
\section*{Acknowledgements}
	
We would like to thank Zvi Bern, Andi Brandhuber, Marco Chiodaroli, Henrik Johansson, Radu Roiban and  Oliver Schlotterer for interesting discussions. 
This research was supported in part by the Munich Institute for Astro and Particle Physics (MIAPP) of the DFG cluster of excellence ``Origin and Structure of the Universe".
The work of  GT was supported by the Science and Technology Facilities Council (STFC) Consolidated Grant 
ST/P000754/1  
\textit{``String theory, gauge theory \& duality"}.  GT is grateful to the Alexander von Humboldt Foundation for support through a Friedrich Wilhelm Bessel Research Award, and to the Institute for Physics and IRIS Adlershof at Humboldt University, Berlin, for their warm hospitality. JP would like to thank the Theory Department at CERN where this work
was completed for hospitality.  DN is supported by
the STFC consolidated grant ``Particle Physics at the Higgs Centre", by the National Science
Foundation.

	
	\appendix
	\section{Integrals}
	\label{Integrals}
	
		The integral functions used in this paper are defined as:
		\begin{align}
		\begin{split}
	\label{intdef}
	\int\!{d^{4-2\eps}L\over (2 \pi)^{4-2\eps}} {\mu^m \over {L^2 \cdots \big[ (L- \sum_{i=1}^{n-1} p_i)^2]}}\ &= \ 
	\int\!{d^4l\over (2 \pi)^4}{d^{-2\eps} \mu  \over (2 \pi)^{-2 \eps}} {\mu^m \over {(l^2-\mu^2) \cdots \big[ (l- \sum_{i=1}^{n-1} p_i)^2-\mu^2\big]}} \\
	&:=\ 
{i\over (4 \pi)^{2-\eps}} \, 	I_n [\mu^m ]   \ , 
	\end{split}
	\end{align}
	where  $L^2 = L_{(4)}^2 + L_{(-2\eps)}^2 := l^2 - \mu^2$.%
\footnote{Our  definition \eqref{intdef} differs from \cite{Bern:1995db} in that we do not include a factor of $(-)^n$ on the right-hand side of this equation. Hence, note the minus sign on the right-hand side of \eqref{tri}, in contradistinction with e.g.~(I.4) of \cite{Bern:1994cg}.}
The exact expressions for  the bubble,  one-mass triangle and zero-mass box integral functions in $4-2\eps$ dimensions 
following from the definition \eqref{intdef} are
\beq
\label{bub}
I_2 [1; s] \  = \ r_\Gamma \frac{(-s)^{-\eps} 
 }{ \eps (1-2 \eps) 
  }\ , 	
  \eeq
	\beq
	\label{tri}
I_3 [1; s]  \ = \ -  {r_\Gamma\over \eps^2}  (-s)^{-1-\eps}\ , 
\eeq
for the bubble and one-mass triangle, while for the zero-mass box function one has \cite{Green:1982sw, Bern:1993kr}
\beq
\label{boxallD}
I_4 [1; s,t]\ =  \  r_{\Gamma} {2\over {s t}}  \left[{(-s)^{-\eps}\over \eps^2 } \,
   _2F_1\left(1,-\eps,1-\eps;1+\frac{s}{t}\right)+{(-t)^{-\eps} \over \eps^2 }\,
   _2F_1\left(1,-\eps,1-\eps;1+\frac{t}{s}\right)\right]
   \ , 
   \eeq
where  
\beq
	r_\Gamma := \frac{\Gamma(1+\eps) \Gamma^2 (1-\eps )
	}{\Gamma (1-2 \eps)} \ .
		\eeq
The results \eqref{bub}, \eqref{tri}  and \eqref{boxallD} are exact to all orders in $\eps$, and the expression of the corresponding  integral 	functions in a different number of dimensions can be obtained by simply replacing  $\eps$ to the appropriate  value, for instance $\eps \to \eps -1$ and $\eps \to \eps -2$ for $D=6-2\eps$ and $D=8-2\eps$, respectively.   The dependence on the relevant kinematic invariants is indicated in brackets along with the power of $\mu$.
Using  \cite{Bern:1995db} 
\beq
I_n^{D=4-2\eps} [ (\mu^2)^p]  \ = \  - \eps(1-\eps)(2-\eps)\cdots (p-1-\eps)
 \, I_n^{D=2p+4-2\eps}
 \ , 
 \eeq
 along with the expressions \eqref{bub}, \eqref{tri} and \eqref{boxallD}, which are correct in any number of dimensions, 
one easily arrives at the following result, used widely in this paper: 
 \begin{align}
 \label{hdint}
 \begin{split}
 I_2[ \mu^2; s]  &= \ -{s\over 6} + \mathcal{O}(\eps) \ ,
 \qquad
  I_2[ \mu^4; s]  = \ -{s^2\over 60} + \mathcal{O}(\eps) \ , 
  \\
  I_3[ \mu^2; s]  &= \  {1\over 2} + \mathcal{O}(\eps) \ , \qquad 
  I_3[ \mu^4; s]  \ = \  {s\over 24} + \mathcal{O}(\eps) \ , \\
  I_4[ \mu^2; s, t] & = \  \mathcal{O}(\eps)\, , \qquad \qquad
   I_4[ \mu^4; s, t]   \ = \ - {1\over 6} + \mathcal{O}(\eps) , \\
    I_4[ \mu^6; s, t]   & = \ - {s+t\over 60} + \mathcal{O}(\eps)  \ ,\\
   I_4[ \mu^8; s, t]  &  = \ -\frac{1}{840} \left(2 s^2+s t+2 t^2\right)  + \mathcal{O}(\eps) \
    \ , 
  \end{split}
  \end{align}
in complete agreement with results of \cite{Bern:1995db,Bern:1998sv} (after taking into account the opposite sign in the definition of triangle functions compared to those papers). 

\section{Tree-level amplitudes via recursion relations}
\label{recrel}	
In this appendix we derive the relevant  tree amplitudes involving gravitons, gluons and massive scalars which enter the one-loop calculations in EYM performed in earlier sections. \\

\noindent
{\bf The $A(4^{++}, 1^{+}, 2_{\phi}, 3_{\bar\phi})$ amplitude} 

\noindent
We use a BCFW recursion relation with a $\langle 4 \, 1]$ shift, i.e.~we perform a shift
\beq
\label{bcfwshifts}
\hat{\lambda}_4 = \lambda_4 + z \lambda_1\, , \qquad \quad \hat{\tilde\lambda}_1 = \tilde\lambda_1 - z \tilde\lambda_4\ . 
\eeq
There are two  recursion diagrams to compute, $A$ and $B$. The first one is 
\beq
A_A(4^{++}, 1^{+}, 2_{\phi}, 3_{\bar\phi}) \ = \ A( \hat{4}^{++}, \hat{P}_{\phi}, 3_{\bar\phi}) \, 
{i\over (p_3 + p_4)^2 - \mu^2} \, 
A (\hat{1}^+, 2_{\phi}, - \hat{P}_{\bar\phi})
\ .
\eeq
 In accordance with  \eqref{gssb} and \eqref{hssb} we have 
 \begin{align}
 \begin{split}
 A( \hat{4}^{++}, \hat{P}_{\phi}, 3_{\bar\phi})  & = -i\, \ {\langle q_1 | 3 | 4]^2 \over \langle q_1 \hat{4}\rangle^2} \ , 
 \\
 A (\hat{1}^+, 2_{\phi}, - \hat{P}_{\bar\phi}) &= i \ {\langle q_2 | -\hat{P} | \hat{1} ]
 \over \langle q_2 1\rangle}
 \ , 
 \end{split}
 \end{align}
 with $\hat{P} = \hat{p}_1+ p_2$. 
 The reference spinors  $q_1$ and $q_2$ can be conveniently chosen to be $q_2 = \hat{4}$ and $q_1 = 1$. 
 Using 
 \beq
  \langle 1 | 3 | 4]^2 \ \langle \hat{4} |-  \hat{P} | \hat{1}] =  - \mu^2 \, s_{14} \, \langle 1 | 3 | 4]
  \ , 
  \eeq 
one quickly arrives at the result 
  \beq
 A_A(4^{++}, 1^{+}, 2_{\phi}, 3_{\bar\phi})  \ = \ 
 -i^2 \, \mu^2 \, {[41]\over \langle 41\rangle^2} \langle 1 | 3 | 4]
  \ {i\over (p_3 + p_4)^2 - \mu^2}
 \ . 
 \eeq
 The second diagram  corresponds to  swapping the position of the graviton with the gluon, to account for the fact that the graviton is colour blind. We have 
 \beq
 A_B(4^{++}, 1^{+}, 2_{\phi}, 3_{\bar\phi}) \ = \ A (\hat{1}^+, \hat{P}_{\phi}, 3_{\bar\phi})
 {i\over (p_2 + p_4)^2 - \mu^2} \,
  A( \hat{4}^{++}, 2_{\phi}, - \hat{P}_{\bar\phi}) \, 
\ . 
\eeq
With the same choice of reference spinors, we get
\beq
A_B(4^{++}, 1^{+}, 2_{\phi}, 3_{\bar\phi})  \ = \ 
 -i^2 \, \mu^2 \, {[41]\over \langle 41\rangle^2} \langle 1 | 3 | 4]
  \ {i\over (p_2 + p_4)^2 - \mu^2}
  \, , 
  \eeq
  and hence the result for the complete amplitude is 
 \beq
 A (4^{++}, 1^{+}, 2_{\phi}, 3_{\bar\phi}) \ = \ \, \mu^2 \, {[41]\over \langle 41\rangle^2} \langle 1 | 3 | 4]
  \ \Big[ {i\over (p_3+ p_4)^2 - \mu^2} \, + \, {i\over (p_2 + p_4)^2 - \mu^2}
 \Big]
  \, .
  \eeq 
  Note that this amplitude vanishes for $\mu^2=0$. \\

\noindent
{\bf Soft limits of the $A(4^{++}, 1^{+}, 2_{\phi}, 3_{\bar\phi})$ amplitude}

\noindent
It is an interesting check to confirm that the amplitude obtained in this way has the correct soft limits. To this end we consider the case with gluon $1^{+}$ becoming soft. We then expect the amplitude to factorize as 
\beq
A(4^{++}, 1^{+}, 2_{\phi}, 3_{\bar\phi})    \xrightarrow[p_1\to 0]{} 
S_1^{(0)} A( 4^{++}; 2_\phi , 3_{\bar\phi})
\ , 
\eeq
where the soft function is 
\beq
S_1^{(0)} \ = \ {p_2 \cdot \varepsilon (p_1)\over \sqrt{2} ( p_2 \cdot p_1) }\,  -\, 
{p_3 \cdot \varepsilon (p_1)\over \sqrt{2} ( p_3 \cdot p_1) } \ .
\eeq
Using $\varepsilon^{(+)}_\nu (p_1) = \langle \xi |\nu| 1] / ( \sqrt{2} \langle \xi \, 1\rangle)$, where $|\xi\rangle$ is a reference spinor, and choosing for convenience $\xi = 4$, we get 
\beq
S_1^{(0)} A( 4^{++}; 2_\phi , 3_{\bar\phi}) \ = \ i\, {\langle 4|3|1]\over \langle 41  \rangle}\Big[ {1\over 2 (p_2 \cdot p_1)} + {1\over 2 (p_3 \cdot p_1)} \Big]{\langle q |3|4]^2\over \langle q\, 4\rangle^2}
\ . 
\eeq
In the soft limit, one easily finds that 
\beq
\langle 4|3|1]\langle q |3|4]  \xrightarrow[p_1\to 0]{}  -\mu^2 \langle q4\rangle [41]\ , 
\eeq
and choosing the arbitrary spinor $q$ to be equal to $1$, we finally get 
\beq
S_1^{(0)} A( 4^{++}; 2_\phi , 3_{\bar\phi})  \xrightarrow[p_1\to 0]{}\ i\, \mu^2 {[41]\over \langle 41 \rangle^2}\langle 1 | 3 | 4] \Big[  {1\over 2 (p_2 \cdot p_1)} + {1\over 2 (p_3 \cdot p_1)} \Big]\ , 
\eeq
which is identical to the result for $A(4^{++}, 1^{+}, 2_{\phi}, 3_{\bar\phi})$.\\

\noindent
{\bf The $A(4^{++}, 1^{-}, 2_{\phi}, 3_{\bar\phi})$ amplitude}

\noindent
We will use the same  BCFW  shift as in \eqref{bcfwshifts}. Again, there are  two recursion diagrams to compute,  A, and B. The first one  is 
\beq
A_A(4^{++}, 1^{-}, 2_{\phi}, 3_{\bar\phi}) \ = \ A( \hat{4}^{++}, \hat{P}_{\phi}, 3_{\bar\phi}) \, 
{i\over (p_3 + p_4)^2 - \mu^2} \, 
A (\hat{1}^-, 2_{\phi}, - \hat{P}_{\bar\phi})
\ ,
\eeq
 with 
 \begin{align}
 \begin{split}
 A( \hat{4}^{++}, \hat{P}_{\phi}, 3_{\bar\phi})  & = \ - i\,  {\langle q_1 | 3 | 4]^2 \over \langle q_1 \hat{4}\rangle^2} \ , 
 \\
 A (\hat{1}^-, 2_{\phi}, - \hat{P}_{\bar\phi}) &= \ i\, {\langle 1 | -\hat{P}| q_2 ]
 \over [ \hat{1} q_2]}
 \ , 
 \end{split}
 \end{align}
and  with $\hat{P} = \hat{p}_1+ p_2$. 
 A convenient choice for the reference spinors  $q_1$ and $q_2$ is again $q_2 = \hat{4}$ and $q_1 = 1$, which immediately leads to  
 \beq
 A_A(4^{++}, 1^{-}, 2_{\phi}, 3_{\bar\phi})  \ = \  -i^2  \,  {\langle 1 | 2 | 4]^3 \over  \langle 1 4 \rangle\,  s_{14}} \ {i\over (p_3 + p_4)^2 - \mu^2}
 \ . 
 \eeq
Similarly 
 \beq
A_B(4^{++}, 1^{-}, 2_{\phi}, 3_{\bar\phi}) \ = \ 
A (\hat{1}^-, \hat{P}_{\phi},  3_{\bar\phi})
 \, 
{i\over (p_2 + p_4)^2 - \mu^2}
\, 
A( \hat{4}^{++}, 2_{\phi}, - \hat{P}_{\bar\phi}) \, 
\ ,
\eeq
which leads to  
\beq
 A_B(4^{++}, 1^{-}, 2_{\phi}, 3_{\bar\phi})  \ = \  -i^2  \,  {\langle 1 | 2 | 4]^3 \over  \langle 1 4 \rangle\,  s_{14}} \ {i\over (p_2 + p_4)^2 - \mu^2}
 \ . 
 \eeq
Adding the two contributions, we get 
  \beq
A (4^{++}, 1^{-}, 2_{\phi}, 3_{\bar\phi}) \ = \ 
 \,  {\langle 1 | 2 | 4]^3 \over  \langle 1 4 \rangle\,  s_{14}} \ \Big[ {i\over (p_3 + p_4)^2 - \mu^2}\, +\, 
 {i\over (p_2 + p_4)^2 - \mu^2}\Big]
 \ . 
 \eeq
Note that this amplitude does not vanish for $\mu^2=0$.  \\

\noindent
{\bf The $A(1^{++}, 2^{++}, 3_{\phi}, 4_{\bar\phi})$ amplitude}

\noindent
We now consider the case of two gravitons and two scalars. The simplest case to consider is that of two 
gravitons of the same helicity, already considered in \cite{Bern:1998sv} in the computation of all-plus graviton amplitudes.
We will use the shifts
\beq
\hat{\lambda}_1 = \lambda_1 + z \lambda_2\, , \qquad \quad \hat{\tilde\lambda}_2 = \tilde\lambda_2 - z \tilde\lambda_1\ . 
\eeq
As usual, there are two diagrams to consider. The first one is 
\beq
A_A(1^{++}, 2^{++}, 3_{\phi}, 4_{\bar\phi}) \ = \ A( \hat{1}^{++}, \hat{P}_{\phi}, 4_{\bar\phi}) \, 
{i\over (p_4 + p_1)^2 - \mu^2} \, 
A (\hat{2}^{++}, 3_{\phi}, - \hat{P}_{\bar\phi})
\ ,
\eeq
while 
$A_B(1^{++}, 2^{++}, 3_{\phi}, 4_{\bar\phi})  = \big[A_A(1^{++}, 2^{++}, 3_{\phi}, 4_{\bar\phi}) \big]_{1\leftrightarrow 2}$.
Thus, using  \eqref{gssb} we get
\beq
A_A = (-i\,) {\langle q_1 | 4 | 1]^2 \over \langle q_1 \hat{1}\rangle^2} {i\over {(p_4 + p_1)^2- \mu^2}} (-i\,) 
{{\langle q_2 | -\hat{P} | \hat{2}]^2 \over \langle q_2 2\rangle^2}}\ . 
\eeq
Choosing $q_2 = \hat{1}$, $q_1 = 2$ and using 
$
\langle q_1 | 4 | 1] \langle q_2 | -\hat{P} | \hat{2}]  = -\mu^2 s_{12}$, we finally arrive at 
\beq
\label{agreesBDPR}
A(1^{++}, 2^{++}, 3_{\phi}, 4_{\bar\phi}) \ = \ 
-\mu^4 { [12]^2\over \langle 12\rangle^2} 
\Big[ {i \over (p_4 + p_1 )^2 - \mu^2} +  {i\over (p_3 + p_4 )^2 - \mu^2} \Big] \ . 
\eeq
Note that \eqref{agreesBDPR} agrees with (4.10) of \cite{Bern:1998sv}.

	\newpage

\phantomsection
\addcontentsline{toc}{section}{References}
\bibliographystyle{nb}
\bibliography{bibliothek}

\end{document}